\documentclass[11pt]{article}
\usepackage{epsfig}
\renewcommand{\arraystretch}{1.3}

\catcode`\@=11
\def\marginnote#1{}

\newcount\hour
\newcount\minute
\newtoks\amorpm
\hour=\time\divide\hour by60
\minute=\time{\multiply\hour by60 \global\advance\minute by-\hour}
\edef\standardtime{{\ifnum\hour<12 \global\amorpm={am}%
        \else\global\amorpm={pm}\advance\hour by-12 \fi
        \ifnum\hour=0 \hour=12 \fi
        \number\hour:\ifnum\minute<10 0\fi\number\minute\the\amorpm}}
\edef\militarytime{\number\hour:\ifnum\minute<10 0\fi\number\minute}

\def\draftlabel#1{{\@bsphack\if@filesw {\let\thepage\relax
      \xdef\@gtempa{\write\@auxout{\string
          \newlabel{#1}{{\@currentlabel}{\thepage}}}}}\@gtempa \if@nobreak
    \ifvmode\nobreak\fi\fi\fi\@esphack} \gdef\@eqnlabel{#1}}
    \def\@eqnlabel{}
\def\@vacuum{}
\def\draftmarginnote#1{\marginpar{\raggedright\scriptsize\tt#1}}

\def\draft{
  \oddsidemargin -.5truein
  \def\@oddfoot{\footnotesize \sl preliminary draft \hfil
    \rm\thepage\hfil\sl\today\quad\militarytime}
  \let\@evenfoot\@oddfoot \overfullrule 3pt
    \let\label=\draftlabel
    \let\marginnote=\draftmarginnote
  \def\@eqnnum{(\theequation)\rlap{\kern\marginparsep\tt\@eqnlabel}%
    \global\let\@eqnlabel\@vacuum}

  }

\makeatletter
\newdimen\normalarrayskip              
\newdimen\minarrayskip                 
\normalarrayskip\baselineskip
\minarrayskip\jot
\newif\ifold             \oldtrue            \def\new{\oldfalse}
\def\arraymode{\ifold\relax\else\displaystyle\fi} 
\def\eqnumphantom{\phantom{(\theequation)}}     
\def\@arrayskip{\ifold\baselineskip\z@\lineskip\z@
     \else
     \baselineskip\minarrayskip\lineskip2\minarrayskip\fi}
\def\@arrayclassz{\ifcase \@lastchclass \@acolampacol \or
\@ampacol \or \or \or \@addamp \or
   \@acolampacol \or \@firstampfalse \@acol \fi
\edef\@preamble{\@preamble
  \ifcase \@chnum
     \hfil$\relax\arraymode\@sharp$\hfil
     \or $\relax\arraymode\@sharp$\hfil
     \or \hfil$\relax\arraymode\@sharp$\fi}}
\def\@array[#1]#2{\setbox\@arstrutbox=\hbox{\vrule
     height\arraystretch \ht\strutbox
     depth\arraystretch \dp\strutbox
     width\z@}\@mkpream{#2}\edef\@preamble{\halign
\noexpand\@halignto
\bgroup \tabskip\z@ \@arstrut \@preamble \tabskip\z@ \cr}%
\let\@startpbox\@@startpbox \let\@endpbox\@@endpbox
  \if #1t\vtop \else \if#1b\vbox \else \vcenter \fi\fi
  \bgroup \let\par\relax
  \let\@sharp##\let\protect\relax
  \@arrayskip\@preamble}
\def\eqnarray{\stepcounter{equation}%
              \let\@currentlabel=\theequation
              \global\@eqnswtrue
              \global\@eqcnt\z@
              \tabskip\@centering
              \let\\=\@eqncr
 \halign to \displaywidth\bgroup
    \eqnumphantom\@eqnsel\hskip\@centering
    $\displaystyle \tabskip\z@ {##}$%
    \global\@eqcnt\@ne \hskip 2\arraycolsep
         $\displaystyle\arraymode{##}$\hfil
    \global\@eqcnt\tw@ \hskip 2\arraycolsep
         $\displaystyle\tabskip\z@{##}$\hfil
         \tabskip\@centering
    &{##}\tabskip\z@\cr}
\begingroup\ifx\undefined\newsymbol \else\def\input#1 {\endgroup}\fi
\newfont{\hr}{msbm10}
\newfont{\ams}{msam10}

\voffset= - 1.2in
\hoffset= - 1.0in         

\def\beq{\begin{equation}}
\def\eeq{\end{equation}}
\def\ba{\beq\new\begin{array}{c}}
\def\ea{\end{array}\eeq}
\def\be{\ba}
\def\ee{\ea}

\def\N2{${\cal N}=2$}

\def\1N{${\cal N}=1$}
\def\4N{${\cal N}=4$}
\def\nn{\nonumber}

\def\p{\partial}

\def\la{\left\langle}
\def\ra{\right\rangle}

\newcommand{\Tr}{\mathop{\rm Tr}\nolimits}

\newcommand\vol{\mathop{\rm Vol} \nolimits}

\def\theequation{\Roman{part}.\arabic{section}.\arabic{equation}}

\def\lla{\la\la}
\def\rra{\ra\ra}
\def\sd{(S\partial)}
\def\cdd{(c\partial\partial)}

\title{{\bf
Partition Functions of Matrix Models
\\
as the First Special Functions of String Theory
\\
I. Finite Size Hermitean 1-Matrix Model}
\vspace{.5cm}}
\author{{\bf A. Alexandrov}\thanks{E-mail: \ alex@gate.itep.ru}
\date{ } \\ {\small {\it MIPT} and
{\it ITEP, Moscow, Russia}}\\ \\
{\bf A. Mironov}\thanks{E-mail:
\ mironov@itep.ru; mironov@lpi.ac.ru}
\date{ } \\
{\small {\it Theory Department, Lebedev Physics Institute}
and {\it ITEP, Moscow, Russia}}\\ \\
{\bf A.Morozov}\thanks{E-mail: \ morozov@itep.ru}
\date{ } \\ {\small
{\it ITEP, Moscow, Russia}}}

\begin{document}

\setcounter{footnote}{3}

\maketitle

\vspace{-12cm}

\begin{center}
\hfill FIAN/TD-17/03\\
\hfill ITEP/TH-50/03\\
\end{center}

\vspace{9.5cm}

\begin{abstract}
Even though
matrix model partition functions do not exhaust
the entire set of $\tau$-functions relevant for string theory,
they seem to be elementary building blocks for many others
and they seem to properly capture the fundamental symplicial
nature of quantum gravity and string theory.
We propose to consider matrix model partition functions
as new special functions.
This means they should be investigated and put into some
standard form, with no reference to particular applications.
At the same time, the tables and lists of properties should
be full enough to avoid discoveries of unexpected peculiarities
in new applications. This is a big job, and the present paper
is just a step in this direction.
Here we restrict our consideration
to the finite-size Hermitean 1-matrix model and concentrate
mostly on its phase/branch structure arising when the
partition function is considered as a D-module. We discuss
the role of the CIV-DV prepotential
(as generating a possible basis in the linear space of solutions
to the Virasoro constraints, but with a lack of understanding of
why and how this basis is distinguished)
and evaluate first few multiloop correlators,
which generalize semicircular distribution
to the case of multitrace and non-planar correlators.
\end{abstract}
\newpage
\tableofcontents

\newpage
\setcounter{footnote}{0}
\part{A Piece of Theory: Partition Functions and their Properties}

\section{Introduction \label{intro}}


$\bullet$
One of the goals of {\it generic string theory} \cite{ST} is to identify the
properties of {\it partition functions} of various {\it string models}.
These are defined as generating functions of all the
correlators in a given quantum theory.
Associated with the three existing formulations of quantum mechanics
(linear algebra of operators in the Hilbert and Fock spaces,
wave equations, path integral), there
are three possible descriptions/definitions of the partition function:
as a matrix element, as a solution to a system of linear differential
equations (as an element of a $D$-module), as a (functional) integral
over trajectories in configuration and/or phase space (over field
configurations).
These
very different definitions emphasize different
properties of partition functions, and their equivalence implies
deep and non-trivial relations and symmetries.
Among these implications, there are integrability properties of
partition functions, placing them into
the class of {\it generalized $\tau$-functions} \cite{GKLMM},
which satisfy infinite sets of compatible {\it non-linear}
difference-differential equations, ({\it generalized
Hirota equations}).
Moreover, partition functions of different models are often
related by {\it dualities} and/or {\it mirror-like symmetries}.
Despite their general definitions and symmetries,
partition functions are rarely expressible through
conventional {\it special functions} and possess complicated
analytical properties, with all kinds of {\it singularities} and
{\it branchings}.

All this -- universality, rich symmetry properties and
impossibility to reduce the problem to known functions --
means that the partition functions ($\tau$-functions) of string models
are natural candidates for the role of the
next generation of special functions: they should be studied without
reference to any particular applications, their properties should be
searched for and listed, particular interesting cases (at specific
values of parameters, in asymptotics, on particular branches)
should be enumerated, listed and tabulated --
and finally collected in the special-functions reference-books.
This task is even more natural, because, similarly to conventional
special functions (of hypergeometric, elliptic and Riemann theta-functions
families), all the $\tau$-functions are closely related to representation
theory of Lie algebras and groups \cite{GKLMM}.
It is important to emphasize only that, while partition functions are
usually some $\tau$-functions, inverse is not true: partition
functions form a very special subclass in the set of $\tau$-functions --
with additional structures and deeper properties (the situation
is well modeled by the relation between Riemann and generic
theta-functions: the former, representing truncated partition functions of
free fields on Riemann surfaces, possess additional properties
of crucial importance, e.g. the Fay identities, implied by the Wick theorem
for the free fields).
In other words, integrability theory is a natural part of the theory
of partition functions, but only a small part, by no means exhaustive.

In this paper we present a first iteration of the above-proposed
analysis/listing
for the simplest and very important string-theory special function:
the partition function of the finite-size {\it Hermitean 1-matrix model}
(FSHMM) \cite{HMM}-\cite{revs}. The general scheme of
reasoning is equally applicable to any other matrix model.
The obvious next candidates to be analyzed from this perspective are
quiver (conformal) matrix models \cite{KMMMP}-\cite{kost}
(see also the recent progress in the two-matrix (normal matrix) models
\cite{zab}),
and the Kontsevich $\tau$-function \cite{K}-\cite{GKMfollowup}.
After this is done, one would be capable to proceed
to the {\it geometric} $\tau$-functions,
associated with the topological sigma-models and models of quantum
gravity in different backgrounds.
However, of all partition functions, the matrix model ones are
the natural candidates to begin with (see various aspects of matrix models
in \cite{Wig}-\cite{MMs}).
Not only are they the simplest example,
one can expect they are actually elementary
building blocks for many other $\tau$-functions.
An example of decomposition of a
{\it geometric} partition function
(the one for the $CP_n$ topological sigma-model) into a multilinear
composition of the elementary building blocks ($n+1$ Kontsevich $\tau$-functions
in this particular case) has been found by A.Givental
\cite{Giv}-\cite{Givfollowup} (see also \cite{kost}).
A simpler decomposition formula for the FSHMM
(which, actually, appeared already in \cite{MMDV})
will be analyzed in the present paper.

\bigskip

$\bullet$
In the theory of conventional special functions one often
distinguishes between the two levels of generality:
one can look at {\it generic} solution to a differential equation
and one can instead concentrate on {\it particular} solutions
(branches), normally associated with specific integral representations
(and/or particular representations of an underlying Lie algebra).
To give a trivial example we can refer to the cylindric function \cite{GR}
($D$-module, associated with the operator $z^2 \partial^2_z +
z\partial_z + (z^2-\nu^2)$) and its particular branches:
Bessel and Neumann functions or Hankel functions. Any pair of these
branches forms a basis in the space of solutions (in the $D$-module) and can
be fixed by choosing integration contours.

The same is true for the string-theory special functions, particular
(functional) integrals provide particular branches of generic
partition function that is defined by a set of Ward identities
(Shwinger-Dyson equations) -- only the difference between branches
can be somewhat more pronounced than in the elementary case.
We shall elaborate on this in detail in the case of the FSHMM,
beginning with a naive matrix integral in s.\ref{pffi},
generalizing to arbitrary solution to the Virasoro constraints in
ss.\ref{pfDm}--\ref{nsol} (though today we do not have much to say
when the genus expansion does not hold), and coming back to
particular solutions: the Gaussian $\tau$-function $Z^M_G(t)$
in s.\ref{Gpf} and the CIV-DV $\tau$-function $Z_{DV}[T|S](t)$
in s.\ref{Giv}. With the help of the Givental-style {\it decomposition
formulas} (s.\ref{Giv}), the CIV-DV $\tau$-function is constructed
from a multilinear combination of the Gaussian ones, and, according
to the arguments of s.\ref{uop}, it can be considered as providing a
basis in the linear space of all solutions to the Virasoro constraints
(somewhat analogous to $e^{ipx}$ in the space of all the functions
of $x$; in other words, just like $e^{ipx}$, $Z_{DV}[T|S]$
provides a kernel of an integral transform from functions of $T$
into functions of $S$).
Unfortunately, no feature is known yet which distinguishes this basis
among other possibilities, although it has clear advantages of being
provided by Givental-style decomposition formulas and
being related to Whitham-Seiberg-Witten theory.

\bigskip

$\bullet$
The goal of this paper is four-fold:

-- We formulate the task of tabulating the properties of
{\it matrix-model} and
{\it geometric} $\tau$-functions, defined as
$D$-modules, i.e. as generic
formal-series solutions to a system of Virasoro-like constraints,
and make a preliminary list of those for the FSHMM.

-- We claim that description of the complicated phase (branch) structure
of the $\tau$-function is a very important -- and so far under-estimated
and under-investigated -- ingredient of this program,
and make an attempt of such description for the FSHMM.

-- We introduce a hierarchical description of phases.
First, with the help of {\it the string coupling} $g$, we introduce the
gradation in powers of $t$-variables
({\it coupling constants} or {\it times}), and use it to
select the phases possessing the {\it genus expansion},
where logarithm of the partition function ({\it the prepotential})
admits a {\it semi}-infinite expansion in powers of $g$.
Second, a shift of finitely-many $t$-variables, $t \rightarrow t-T$,
allows us to introduce different kinds of $t$-expansions (phases)
with different quantities (actually, the differences $\alpha_{ij}=
\alpha_i - \alpha_j$ between the roots of the polynomial of degree $n$,
$W'(z) = \sum kT_kz^{k-1} = \prod_{i=1}^n (z-\alpha_i)$) emerged in
denominators. The polynomial $W(\phi)$ of degree $n+1$ plays role
of the {\it tree superpotential} in the
context of Dijkgraaf-Vafa theory \cite{DV}.
Third, each of these phases is further split into infinite-parametric
set with different positions of singularities (branching points)
of the {\it multidensities}\footnote{These quantities can be associated with
the multiloop correlators (correlators of Wilson loops), $\la\la \prod_{k=1}^m
{\rm Tr}\frac{1}{z_k-\phi}
\ra\ra_W$ \cite{loopeqn}
(where the double brackets denote {\it connected correlators}),
when a matrix integral representation with some
diagram perturbation expansion exists.
The imaginary part of these correlators
is naturally associated with multidensities of
eigenvalue distributions in this case. This is why we usually call
$\rho_W(z_1,\ldots,z_m)$ as
multidensities. We also call them multiresolvent, since $\rho (z)$ is a
resolvent.}
\be
\rho_W(z_1,\ldots,z_m) = \prod_{i=1}^m \left(\sum_{k \ge 0}
{1\over z_i^{k+1}}{\partial
\over \partial t_{k}}\right){\cal F}_W (t)
\ee
where $Z_W(t)=e^{{1\over g^2}{\cal F}_W(t)}$ is the FSHMM partition function.
This family of phases is parameterized by an arbitrary
function $F_W(g) =
F[g,T]={\cal F}_W(t=0,g)$ of
$g$ and $n$ variables $T_0,\ldots,T_{n-1}$.

-- We argue that an alternative approach to description of branches can be
provided by the Givental-style decomposition formulas,
representing a matrix-model/geometric $\tau$-function
through an operator, acting on a product of ``elementary $\tau$-functions''.
In the case of the FSHMM, the ``elementary $\tau$-functions'' are the
Gaussian ones, and the decomposition formulas
give rise to the CIV-DV $\tau$-functions $Z_{DV}[T|S](t)$.
Given $W(z)$, the generic partition function, which is built from
an arbitrary function
$Z[T] = e^{F[T]/g^2}$ with the help of the evolution operator
$\check U_W(t)$ (see s.\ref{uop}),
\be
Z_W(t) = \check U_W(t) Z[T]
\ee
can be expressed through the CIV-DV $\tau$-functions,
\be
Z_W(t) = \int Z_{DV}[T|S](t) C[S] dS
\ee
where the coefficients $C[S]$ are defined by the expansion of
``initial" $Z[T]$ into a linear combination of
$Z_{DV}[T|S] = e^{F_{DV}[T|S]/g^2} = Z_{DV}[T|S](t=0)$,
\be
Z[T] = \int Z_{DV}[T|S] C[S] dS
\ee

\section{Partition function as functional integral \label{pffi}}
\setcounter{equation}{0}
\subsection{The notion of partition function}


$\bullet$
{\it String theory} studies {\it partition functions}
\be
Z_{{\cal M},S_0}\{t|\phi_0\} =
\int_{\cal M} d\phi\
\exp{\frac{S(\phi_0 + \phi)}{g}},
\nn \\
S(\phi) = S_0(\phi) + \sum_{\vec k} t_{\vec k}V_{\vec k}(\phi)
\label{pafun}
\ee
(with complete set of {\it vertex operators} $V_{\vec k}(\phi)$)
as functions of:

(i) the space of fields ${\cal M}$, parameterized by $\phi$,
and equipped with some ``natural" measure $d\phi$,

(ii) the measure $d\mu(\phi) = d\phi\ e^{\frac{1}{g}S_0(\phi)}$ on ${\cal M}$
(composed from $d\phi$,
the {\it bare action} $S_0(\phi)$ and the
{\it string coupling}/Planck constant $g$),

(iii) the infinitely many {\it coupling constants} $t_{\vec k}$ and

(iv) {\it the background state} (vacuum) $\phi_0$.

Usually {\it the background field} $\phi_0$ is a solution of
{\it the equations of motion}
\be
\left.\frac{\delta S_0(\phi)}{\delta \phi}\right|_{\phi_0} = 0
\label{eqm}
\ee
but sometime their quantum counterpart,
\be
\left.\frac{\delta d\mu(\phi)}{\delta \phi}\right|_{\phi_0} = 0
\label{eqmq}
\ee
can be used, and/or a deformed action $S(\phi)$ substituted
instead of $S_0(\phi)$ into (\ref{eqm}) and (\ref{eqmq})
(thus mixing the dependencies on $\phi_0$, $g$ and $t_{\vec k}$).

The space ${\cal M}$, together with possible restrictions on the choice
of measures $d\mu(\phi)$, defines a particular {\it string model}.

The task of string theory is also to provide a unified construction of all
interesting spaces ${\cal M}$ from some elementary building blocks
(for example, from elementary simplexes).

\bigskip

$\bullet$
Typically,
\be
{\cal M} = {\rm Hom}(W,T)/{\cal G}
\label{mapsspace}
\ee
is the space of mappings from the {\it world sheet} $W$ into the {\it target space}
$T$, factorized by some {\it symmetry group} ${\cal G}$.

\bigskip

$\bullet$
For every ${\cal M}$
one can construct the new spaces ${\cal M}_G \equiv {\cal M}\otimes G_R/U$,
where $G_R$ is a group $G$
in some representation $R$, factored out by
action of its maximal subgroup $U$.
The characteristic example is $G_R = M_N$, the space of square $N\times N$
matrices (the $N$-dimensional representation of $GL(N)$  -- the space of
linear mappings of a vector space $V_N$ into itself), on which the
unitary group $U(N)$ acts by conjugations $M \rightarrow U^{-1} MU$.
This ``branization" prescription
(substituting numbers by matrices) is familiar from the
studies of brane stacks \cite{Wbr}.

For ${\cal M}$ of type (\ref{mapsspace}), one
can also switch to mappings between more general bundles, sheaves
and {\it complexes} over $W$
and $T$ (substituting numbers by
{\it mappings of complexes}, with complexes described by auxiliary variables
like, for example, those in BW or BFW  formalisms
\cite{BV}-\cite{WBVfollowup}).

\bigskip

$\bullet$
For a given vacuum $\phi_0$, it often makes sense to redefine
the set of coupling constants in the action $S(\phi)$ in (\ref{pafun}):
this is always possible if the vertex operators $V_{\vec k}(\phi)$
form a {\it complete system}
(either {\it strong} or {\it weak}, see \cite{MMRG} for more details):
\be
S(\phi_0+\phi) = S_0(\phi_0+\phi) +
\sum_{\vec k} t_{\vec k}V_{\vec k}(\phi_0+\phi) = \nn \\ =
S_0(\phi_0+\phi) + \sum_{\vec k} t_{\vec k}^{\{\phi_0\}}V_{\vec k}(\phi)
\ee
The couplings $t_{k}^{\{\phi_0\}}$ are more adequate than $t_k$ for building
up the perturbation theory around a particular background $\phi_0$, but they
themselves depend on $\phi_0$, and the Ward identities (like the
Virasoro constraints
below) look more sophisticated if expressed through them.
Of course, all $t_{k}^{\{\phi_0\}}=0$ whenever $t_k=0$.

\subsection{Matrix Model as the simplest String Model}


$\bullet$
The very starting example of string model is $W = pt$, $T= pt$ (i.e. both
the world sheet and the target space are just single points). The simplest
associated brane stack provides ${\cal M} = M_N/U(N)$,
and the partition function is the
one of {\it Hermitean 1-matrix model} of {\it the eigenvalue type}
\cite{UFN3}:
\be
{\cal Z}_N^{(matr)}(t) = \frac{1}{\vol_{U(N)}}\int_{M_N} D\phi\
\exp \left( \frac{1}{g}\sum_{\vec k} t_{\vec k}\
{\rm Tr}\ \phi^{k_1}\cdot\ldots \cdot {\rm Tr}\ \phi^{k_m}\right)
\label{fullZN}
\ee
where the sum is over all non-negative integer-valued vectors
$\vec k = (k_1,\ldots, k_m)$ of all possible lengths $m$.
The measure
\be
D\phi = \prod_{i,j=1}^N d\phi_{ij}
\ee
is $U(N)$ invariant, and, if $\phi = U^{-1} \phi_D U$, where
$\phi_D = {\rm diag}(\phi_i)$,
\be
D\phi = DU \prod_{i=1}^N d\phi_i \prod_{i<j}^N(\phi_i - \phi_j)^2
\label{VdM}
\ee
$DU$ is the Haar measure on $U_N$, induced by the norm
$||\delta U||^2 = {\rm Tr} (U^{-1} \delta U)^2$,
and the volume of the unitary group is
\be\label{volume}
\vol_{U(N)} = \int_{U(N)}DU \sim \prod_{k=1}^{N-1} k!
\ee
The model is of {\it eigenvalue} type, because it is essentially
reproduced by an $N$-fold integral over the eigenvalues $\phi_i$ only.
Integration contours for $\phi_i$ are not obligatory restricted to
the real line (this can be considered as
a sort of analytical continuation, see footnote \ref{bas}).

\bigskip

$\bullet$
Often instead of ${\cal Z}_N(t)$ the truncated partition function $Z_N(t)$
is considered, with $t_{\vec k}$ non-vanishing only for the vectors $\vec k$
of unit length,
\be
Z_N^{(matr)}(t) = \frac{1}{\vol_{U(N)}}\int_{M_N} D\phi\
\exp \left(\frac{1}{g} \sum_{k\in Z_+} t_{k}\ {\rm Tr}\ \phi^{k}\right)
\label{ZN}
\ee
This is what we will use as a prototype of partition function of
the finite-size (i.e. finite $N$) Hermitean matrix model (FSHMM).

\bigskip

$\bullet$
A number of powerful methods exists for investigation of the
truncated model (\ref{ZN}) as an eigenvalue model \cite{UFN3,revs}.
The most important two are the method of orthogonal polynomials
\cite{HMM} and the free-fermion formalism \cite{detrep,versus}.
They can be used to derive {\it the determinant representation}
\cite{detrep,UFN3,revs}
for $Z_N$,
\be
Z_N^{(matr)}(t) = {\rm det}_{\leq i,j < N} {\cal H}_{i+j}(t)
\label{detf}
\ee
where
\be
{\cal H}_j = \int \phi^j \exp\left(
\frac{1}{g}\sum_{k=0}^{\infty} t_k \phi^k
\right)   d\phi
\ee
In turn, it can be used to prove \cite{GMMMO}
that $Z_N$ is a Toda-chain $\tau$-function, i.e. satisfies an
infinite system of bilinear differential equations like
\be
Z_N^{(matr)}\frac{\partial^2Z_N^{(matr)}}{\partial t_1^2} -
\frac{\partial Z_N^{(matr)}}{\partial t_1 }\frac{\partial Z_N^{(matr)}}{\partial t_1 }
= Z_{N-1}^{(matr)}Z_{N+1}^{(matr)}
\label{Toda}
\ee

\bigskip

$\bullet$
In (\ref{ZN}) the bare action is absent.
As in every string model, it can be introduced by a shift
of the coupling constants, $t_k \rightarrow t_k-T_k$:
\be
Z_{N,W}^{(matr)}(t) = \frac{1}{\vol_{U(N)}}
\int d\phi e^{-\frac{1}{g}{\rm Tr}W(\phi)}
e^{\frac{1}{g}\sum_k t_k{\rm Tr}\phi^k}
\label{ZNW1}
\ee
where $W(z) = \sum_{k=1}^{n+1} T_kz^k$. Despite obtained by a shift
of variables, this partition function essentially depends on
$W(z)$, especially on its power $n+1$, since changing $n$ completely changes
structure of the perturbation series.
In other words, the system undergoes a phase
transition (the partition function is singular)
at zero values of the coefficient of
the leading ($z^{n+1}$) term in $W(z)$.\footnote{This
situation is much similar to matrix models of the Kontsevich type,
which also drastically depends on the matrix model potential,
especially on its power \cite{LGM}.}

\section{Partition function as a $D$-module \label{pfDm}}
\setcounter{equation}{0}

\subsection{The notion of partition function}


$\bullet$
Partition function ($\tau$-function) is, in fact, a highly sophisticated
quantity. Actually, it is not a function, but a formal $D$-module, i.e.
the entire collection of power series (in $t$-variables), satisfying a
system of consistent linear
equations.
Solution to the equations does not need to be unique, however,
an appropriate analytical continuation in $t$-variables transforms
one solution to another, and, on a large enough moduli space
(of {\it coupling constants} $t$), the whole entity can be considered,
at least, formally as a single object: this is what we call
{\it the partition function}.
Naively different solutions are interpreted as different {\it branches}
of the partition function, associated with different {\it phases} of the theory.
Further, solutions to the linear differential
equations can be often represented as integrals (over {\it spectral
varieties}), but integration ``contours'' remain unspecified: they can
be generic {\it chains} with complex coefficients (in the case of integer
coefficients this is often described in terms of {\it monodromies}, but
in the case of partition functions the coefficients are not restricted
to be integer).
{\it A model} of partition function is an integral formula which has
enough many free parameters to represent the generic solution of the
differential equations in question.\footnote{\label{bas}
Several familiar examples can be helpful to clarify these notions.

First, a {\it flat} $SU(2)$  Seiberg-Witten modulus $a(\lambda)$ \cite{SW}
satisfies
\be
\left(4\lambda(\lambda-1)\partial_\lambda^2 + 1\right) a(\lambda) = 0
\label{Ex1.1}
\ee
possesses integral representation,
\be
a(\lambda) = \int_C \sqrt{\frac{x-\lambda}{x(x-1)}}dx
\label{Ex1.2}
\ee
and {\it the model} is provided by the generic contour
\be
C = \alpha[0,1] + \beta[1,\lambda]
\label{Ex1.3}
\ee
which is the generic linear combination
(with complex(!) coefficients) of $A$ and $B$ cycles on the toric spectral
curve.
Alternative description of the same object (the toy partition function)
is in terms of the Seiberg-Witten 1-form
\be
dS(x,\lambda) = \sqrt{\frac{x-\lambda}{x(x-1)}}dx
\ee
related to the original equation (\ref{Ex1.1}) through
\be
\left(4\lambda(\lambda-1)\partial_\lambda^2 + 1\right)dS(x,\lambda)
= d\sqrt{\frac{x(x-1)}{x-\lambda}} \cong 0
\ee
(where $\cong$ means cohomologically equivalent).

Another example already discussed in the introduction is the cylindric
functions. Their defining equation is
\be
\left[\lambda^2\partial_{\lambda}^2 +\lambda\partial_{\lambda}+
\left(\lambda^2-\nu^2 \right)\right]Z_{\nu}(\lambda)=0
\ee
and an integral representation is
\be\label{cylint}
Z_{\nu}(x)={1\over 2\pi}\int_C e^{-ix\sin\theta+i\nu\theta}d\theta
\ee
{\it The model} is given by the generic
linear combination of two contours, say,
chosen as in 8.423 of \cite{GR} (this choice fixes as the basis the Hankel
functions).

One more example is the Airy function (a simple example of Kontsevich
model \cite{K}-\cite{GKMfollowup}), with the defining equation
\be
\left(\partial_\lambda^2 + \lambda\right)A(\lambda) = 0
\label{Ay}
\ee
and an integral representation
\be
A(\lambda) = \int_C \exp \left(\frac{x^3}{3} + \lambda x\right) dx
\ee
This $D$-module is a particular case of the cylindric functions with
$\nu={1\over 3}$.
}

\subsection{Virasoro constraints for the FSHMM}


$\bullet$
Coming back  to the FSHMM, its original definition
has been done in terms of a matrix
integral (\ref{ZN}).
This integral
depends on (infinitely many) parameters $t_k$ and,
according to above arguments, implies another
{\it definition} of the partition function:
as solution to the (infinite) system of consistent linear
(differential) equations
(Ward identities or Picard-Fucks equations), i.e. as
{\it a D-module}.
In the particular case of FSHMM, these equations,\footnote{
The origin of these equations
is invariance of integral (\ref{ZN}) under the change of integration variable
$\phi \rightarrow \phi + \epsilon_m\phi^{m+1}$ \cite{MM}.
Similarly, integral (\ref{fullZN}) is invariant under the changes
$\phi \rightarrow \phi + \epsilon_{m,\vec k} \phi^{m+1}
{\rm Tr}\ \phi^{k_1}\cdot\ldots \cdot {\rm Tr}\ \phi^{k_l}$. The resulting
set of constraints is a corollary of (\ref{vircon})
and the obvious relations
$$
\frac{\partial{\cal Z}}{\partial t_{\vec k}} =
\frac{\partial^m{\cal Z}}{\partial t_{k_1}\ldots \partial t_{k_m}}
$$
The inverse way, from the equations (\ref{vircon}) to the integral
representation (\ref{ZN}), is discussed in \cite{KMMMP}.
}
\be
\hat L_m Z(t) = 0, \ \ m\geq -1
\label{vircon}
\ee
\be
\hat L_m = \sum_{k\geq 0} kt_k \frac{\partial}{\partial t_{k+m}} + g^2
\sum_{\stackrel{a+b=m}{a,b\geq 0}} \frac{\partial^2}{\partial t_a\partial t_b}
\ee
are called \cite{vircref} {\it Virasoro constraints}, since
$\left[\hat L_m, \hat L_n\right] = (m-n) \hat L_{m+n}$ (in fact, this is only
the Borel part of the Virasoro algebra).

From now on we {\it define} the FSHMM partition function as the generic
solution to the system (\ref{vircon}), so that the matrix integral
(\ref{ZN}) is its particular integral representation. One of our tasks
below is to analyze how general this integral representation is and
whether it can be treated as {\it a model} of the FSHMM partition
function.

\bigskip

$\bullet$
Since
\be
\left[\frac{\partial}{\partial t_0},\hat L_m\right] = 0
\ee
one can diagonalize the operator $\partial/\partial t_0$ and impose
an additional constraint on the partition function
\be\label{t0}
\frac{\partial}{\partial t_0}Z_S = SZ_S
\ee
with constant $S$.
The integral representations (\ref{ZN}) and (\ref{ZNW1}) obviously
satisfy this relation with the integer $S=N$.
The generic solution to the Virasoro constraints can be decomposed into
{\it a linear
combination} of $Z_S$'s with different {\it constant} $S$'s,
but, in practice, this
decomposition does not look very natural.\footnote{What is worse,
while the constraints $N_i=const$
in the Givental-style decomposition formulas sect.II.2 are clearly
generalizations of (\ref{t0}) (and, moreover,
$\sum_{i=1}^n N_i = N=S$),
no operators are known that have $N_i$ as their eigenvalues.
This causes problems with justification of linear decomposition
of every branch of partition function into the Givental functions
(\ref{factfor}) with constant $N_i$'s.}

Note that, for particular values of
$N$, the partition function $Z_N^{(matr)}$ satisfies
additional differential equations, impled by the fact that ${\rm Tr}\phi^k$
with $k>N$ can be expressed through an algebraic combination of ${\rm Tr}\phi^k$
with $k\leq N$. For example, for $N=1$,
\be
\left(\frac{\partial}{\partial t_k} - \left(\frac{\partial}{\partial
t_1}\right)^k\right) Z_{N=1}^{(matr)} = 0
\ee
We do {\it not} impose these constraints in our definition of the generic
FSHMM partition function: the only requirements are (\ref{vircon})
(and (\ref{addcon}), to be defined below).

\bigskip

$\bullet$
Solutions to the system (\ref{vircon}) can be looked for
among the formal series in powers of $t_k$, and these series
are different, depending on what is allowed to appear in
denominators. Different such possibilities are interpreted as
different {\it phases} of the matrix model or as different
{\it branches} of its partition $\tau$-function $Z(t)$.
Formally, they can be regarded as different approximations and
analytical continuations of the integral (\ref{ZN}).
These different phases are labeled, in particular, by the {\it bare actions},
i.e. by consideration of (\ref{ZNW1}) instead of (\ref{ZN}).

\bigskip

$\bullet$
Introduction of the bare action $W(\phi) = \sum_{k=1}^{n+1} T_k\phi^k$
modifies the Virasoro constraints (\ref{vircon}) in a trivial way,
by shifting
\be
\hat L_m \rightarrow \hat L_m^W = \hat L_m -
\sum_{k\geq 0} kT_k \frac{\partial}{\partial t_{k+m}}
\label{viropW}
\ee
but the solutions $Z_W(t) = e^{{\cal F}_W(t)/g^2}$ can be now looked for among the formal
series in positive powers of the $t$-variables with $T$'s
emerging in denominators.
Also, now the set of the Virasoro equations should be supplemented
by additional constraints
\be
\frac{\partial Z_W}{\partial T_k} +
\frac{\partial Z_W}{\partial t_k} = 0, \ \
\forall k=0,\ldots,n+1
\label{addcon}
\ee

\subsection{Virasoro constraints and loop operators}


$\bullet$
A powerful technique for dealing with the
Virasoro equations (\ref{vircon}) is provided by {\it the loop operator}
\be
\hat\nabla(z) = \sum_{k\geq 0}\frac{1}{z^{k+1}} \frac{\partial}{\partial
t_k}
\ee
and the two complementary projectors $\hat P^{\mp}_z$ onto
negative and non-negative powers of $z$,
\be
\hat P^-_z\left[\sum_{k=-\infty}^{+\infty}\xi_k z^k \right] =
\sum_{k<0} \xi_k z^{k}, \nn \\
\hat P^+_z\left[\sum_{k=-\infty}^{+\infty}\xi_k z^k \right] =
\sum_{k\geq 0} \xi_k z^{k}
\ee
With the help of these operators  and the generating function
\be
v(z) = \sum_{k\geq 0}^\infty t_kz^k
\ee
one can write down the entire set of the Virasoro constraints
(\ref{vircon}) in the form of a single {\it loop equation} \cite{loopeqn}
\be
\hat P^-_z\left[v'(z)\rho(z)\right] + \rho^2(z) + g^2\hat\nabla(z)\rho(z) = 0
\label{virconf}
\ee
for the resolvent ({\it density})\footnote{The Virasoro constraints
(\ref{vircon}) can be also realized in free field terms. Indeed, consider
the $U(1)$-current $\partial_z\phi(z)$ in the theory of one free scalar
field realized in terms of Heisenberg algebra acting on the space of
functions of time-variables $t_k$,
$$
\partial_z\phi(z) ={1\over g} v'(z)+ g {\hat\nabla (z)}
$$
Then, the Virasoro generators are given by the standard expansion of the
energy-momentum tensor
$$
T(z)\equiv {1\over 2}:\!\left(\partial_z\phi(z)\right)^2\!:\ =
\sum {L_n\over z^{n+2}}
$$
Now the loop equation is just the requirement that the partition function as
a function of time variables
realizes the vacuum w.r.t. the Virasoro algebra:
$$
\hat P^-_z T(z) Z=0
$$
This language proved to be very effective in dealing with different quiver
models and with continuum limits, see \cite{MMMM,KMMMP}.
}
\be
\rho(z) \equiv \hat\nabla(z) {\cal F} \equiv
\sum_{k\geq 0}\frac{1}{z^{k+1}}
\frac{\partial {\cal F}}{\partial t_k},
\ \ \ {\cal F} = g^2\log Z
\ee
Eq.(\ref{virconf}) should be
complemented with the zero-curvature condition
\be
\hat\nabla(x)\rho(z) = \hat\nabla(z)\rho(x)
\label{zcc}
\ee

\bigskip

$\bullet$
After the shift (\ref{viropW}), the modified
Virasoro relations for
\be
\rho_W(z) = \hat\nabla(z) {\cal F}_W
\ee
are
\be
\hat P^-_z\left[W'(z)\rho_W(z)\right]
=\
\rho_W^2(z) + g^2\hat\nabla(z)\rho_W(z) +
\hat P^-_z\left[v'(z)\rho_W(z)\right]
\label{virconfW0}
\ee

\bigskip

$\bullet$
In considerations below we make use of one more loop
operator, defined for arbitrary function $u(z)$,
\be
\hat R_u(z) = \sum_{l\geq 0} \hat P^+_z\left[\frac{u'(z)}
{z^{l+1}}\right] \frac{\partial}{\partial t_l}
\ee
Characteristic property of this operator is that, for
arbitrary function $g(t)$ of time-variables, we have
\be
\hat P^+_z\left[u'(z)\hat\nabla(z) g(t)\right] =
\hat R_u(z) g(t)
\ee
In particular, if $u(z)$ is the potentials $v(z)$ or $W(z)$,
and $g(t)$ is ${\cal F}(t) = g^2\log Z$ or
${\cal F}_W(t) = {\cal F}(T,t) = g^2\log Z_W$,
one obtains for the corresponding pieces in the loop equations
(\ref{virconf}) and (\ref{virconfW0}),
\be
\hat P^-_z\left[v'(z)\rho(z)\right] = v'(z)\rho(z) -
\hat P^+_z\left[v'(z)\rho(z)\right] = v'(z)\rho(z) -
\hat R_v(z) {\cal F}
\ee
and
\be
\hat P^-_z\left[W'(z)\rho_W(z)\right] = W'(z)\rho_W(z) -
\hat P^+_z\left[W'(z)\rho_W(z)\right] = W'(z)\rho_W(z) -
\hat R_W(z) {\cal F}_W
\ee
For polynomial $W(z)$, the last term at the r.h.s. is just a polynomial
of degree $n-1$ in $z$.
Moreover, due to (\ref{addcon}), one can also write
\be
\check R_W(z) {\cal F}_W = - \sum_{k,l\geq 0} (k+l+2)T_{k+l+2}z^k
\frac{\partial {\cal F}_W}{\partial T_l}
\ee
Hereafter, we use $\ \check{}\ $ to denote operators containing
$T$-derivatives, while those containing
$t$-derivatives are labeled by $\ \hat{}\ $.

With the help of $\hat R_W(z)$, we can rewrite (\ref{virconfW0})
in the most convenient form
\be
W'(z)\rho_W(z) =\
\rho_W^2(z) + f_W(z) + g^2\hat\nabla(z)\rho_W(z) +
\hat P^-_z\left[v'(z)\rho_W(z)\right]
\label{virconfW}
\ee
where
\be
f_W(z) = \hat P^+_z\left[W'(z)\rho_W(z)\right]
= \hat R_W(z) \log Z_W
\ee
Note also that
\be
\hat\nabla(x) v(z) = \hat P^-_x\left[\frac{1}{z-x}\right], \ \
\hat\nabla(x) v'(z) = -\hat P^-_x\left[\frac{1}{(z-x)^2}\right]
\label{derWprime}
\ee
and for any function $h(z)$ such that $\hat P^+_z [h(z)] = 0$,
\be
\hat\nabla(x) P^-_z \left[v'(z)h(z)\right] =
- P^-_z\left[P^-_x\left[\frac{h(z)}{(z-x)^2}\right]\right]
= \partial_x \frac{h(z)-h(x)}{z-x}
\label{PPrel}
\ee
where $\partial_x = \partial/\partial x$.
The commutator
\be
\left[ \hat\nabla(x), \hat R_v(z)\right] = \partial_x \hat
P^+_z\left[\frac{\hat\nabla(z)}{z-x}\right]
\ee

\subsection{Solving Virasoro constraints recurrently
\label{NGW}}


$\bullet$
In this section we assume that $Z_W(t)$ possesses {\it genus expansion},
i.e. that
\be
\log Z_W(t) = g^{-2}{\cal F}_W(t) = \sum_{p\geq 0} g^{2p-2} {\cal F}_W^{(p)}(t)
\ee
(see s.\ref{gdep} for discussion of this assumption).

Actually, the partition functions can be obtained by one
of the two procedures:

\bigskip

-- {\it Virasoro-constraints method}, i.e. by solving the loop equations
(\ref{virconfW}), and

-- {\it Givental method}, i.e. by making use of decomposition formulas
(\ref{factfor}), to be introduced below in s.II.\ref{Giv}.

\bigskip

In the framework of the Givental method, an
arbitrary phase of the FSHMM with given $W(\phi)$ of
degree $n+1$ is rather described in terms of $n$ Gaussian models,
according to the decomposition formula (\ref{factfor}).
This means that the averages
depend, in fact, on the additional set of variables $N_i$,
characterizing a particular saddle-point (vacuum) of the FSHMM.
We go into more details along this line in s.II.\ref{defo} below,
while here we concentrate on analysis of the Virasoro
constraints (\ref{virconfW}), which provides an
{\it a priori} richer variety of possible solutions/phases (though they
would still have to be linear combinations of the Givental-style ones).

\bigskip

$\bullet$
In order to convert
(\ref{virconfW}) into a solvable set of recurrent relations, we expand
\be
\rho_W(z|t) = g^2\hat\nabla(z) \log Z_W(t) = \hat\nabla(z){\cal F}_W(t)
\ee
in powers of $g^2$ and $t$'s
\be
\rho_W(z|t) = \sum_{p,m\geq 0} \frac{g^{2p}}{m!}
\oint \ldots \oint v(z_1)\ldots v(z_m) \rho_W^{(p|m+1)}(z,z_1,\ldots,z_m),
\nn \\
f_W(z|t) = \sum_{p,m\geq 0} \frac{g^{2p}}{m!}
\oint \ldots \oint v(z_1)\ldots v(z_m)
f_W^{(p|m+1)}(z|z_1,\ldots,z_m)
\ee
In this way, we introduce the full set
of multidensities
\be
\rho_W^{(p|m)}(z_1,\ldots,z_m) =
\left.\hat\nabla(z_1)\ldots\hat\nabla(z_m) {\cal F}^{(p)}_W\right|_{t=0}
\ee
and auxiliary polynomials
(which distinguishes between different phases for a given $W(z)$)
\be
f_W^{(p|m+1)}(z|z_1,\ldots,z_m) =
\check R_W(z)\rho_W^{(p|m)}(z_1,\ldots,z_m)
\label{fthroughrho}
\ee
In fact, eq.(\ref{fthroughrho}) guarantees that the constraints
(\ref{addcon}) are fulfilled.

Acting on eq.(\ref{virconfW}) with the operator
$\hat\nabla(z_1)\ldots\hat\nabla(z_m)$, making use of (\ref{PPrel})
and putting all $t_k=0$ afterwards,
we obtain a double-recurrent (in $p$ and $m$) relation for
the multidensities $\rho_W^{(p|m)}$
\be
W'(z)\rho_W^{(p|m+1)}(z,z_1,\ldots,z_m) -
f_W^{(p|m+1)}(z|z_1,\ldots,z_m) = \nn \\ =
\sum_q \sum_{m_1+m_2=m}
\rho_W^{(q|m_1+1)}(z,z_{i_1},\ldots,z_{i_{m_1}})
\rho_W^{(p-q|m_2+1)}(z,z_{j_1},\ldots,z_{j_{m_2}}) + \nn \\ +
\sum_{i=1}^m \frac{\partial}{\partial z_i}
\frac{\rho_W^{(p|m)}(z,z_1,\ldots,\check z_i,\ldots,z_m) - \rho_W^{(p|m)}(z_1,\ldots,z_m)}{z-z_i}
+ \nn \\ +
\hat\nabla(z)\rho_W^{(p-1|m+1)}(z,z_1,\ldots,z_m)
\label{recrel1}
\ee
Together with (\ref{fthroughrho}) this relation is enough
to find explicit expressions for all the multidensities through
$W(z)$ and $f^{(p|1)}_W(z)$. In fact, the latter polynomials
(all of degree $n-1$) are not
independent, since (\ref{fthroughrho}) for $m=0$,
\be\label{fthrF}
f^{(p|1)}_W(z) = \check R_W(z) F^{(p)}[T]
\ee
holds only because of the zero-curvature constraint (\ref{zcc}) and
expresses all the $f$-polynomials
through a single function of $T$ (i.e. of $W(z)$) and $g$,
the prepotential at $t=0$,
\be
{\cal F}_W(t=0,g) = F[g,T] = \sum_{p=0}^\infty g^{2p} F^{(p)}[T]
\ee

\bigskip

$\bullet$
The $T_n$ and $T_{n+1}$ dependencies of $F[g,T]$ are not arbitrary,
since the $\hat L^W_{-1}$ and $\hat L^W_0$ Virasoro constraints can be
consistently truncated to $t=0$ and then allow one to express
$\partial F/\partial T_{n+1}$ and
$\partial F/\partial T_{n}$ through $\partial F/\partial T_{l}$
with $l = 0,\ldots, n-1$,
\be
\frac{\partial F^{(0)}}{\partial T_n} = -\frac{1}{(n+1)T_{n+1}}
\sum_{l=0}^{n-1}
(l+1)T_{l+1}
\frac{\partial F^{(0)}}{\partial T_l},
\nn \\
\frac{\partial F^{(0)}}{\partial T_{n+1}} =
\frac{1}{(n+1)T_{n+1}}\left(\frac{\partial F^{(0)}}{\partial T_0}\right)^2 -
\nn \\
-\frac{1}{(n+1)T_{n+1}}
\sum_{l=0}^{n-1}
lT_l\frac{\partial F^{(0)}}{\partial T_l}
 -\frac{nT_n}{(n+1)^2T_{n+1}^2}
\sum_{l=0}^{n-1}
(l+1)T_{l+1}\frac{\partial F^{(0)}}{\partial T_l}
\label{Tnn1}
\ee

\bigskip

$\bullet$
As an immediate corollary of (\ref{recrel1}),
we obtain for $p=0$ and $m=0$
\be
\rho_W^{(0|1)}(z) = \frac{W'(z) - y_W(z)}{2}
\label{sdenW}
\ee
with
\be
y^2_W(z) = (W'(z))^2 - 4 f^{(0|1)}_W(z)
\ee

\bigskip

$\bullet$
Sometime it is useful to represent the polynomial $f^{(0|1)}_W(z)$ through
peculiar $n$ parameters $\tilde S_i$ \cite{IM7}
\be
4 f^{(0|1)}_W(z) = W'(z) \sum_{i=1}^n \frac{\tilde S_i}{z - \alpha_i}
\label{tildeS}
\ee
where $\alpha_i$  are the roots of $W'(z) = \prod_{i=1}^n (z-\alpha_i)$
(so that there are no poles at the r.h.s. of (\ref{tildeS})).
If $\tilde S$'s do not depend on $T$'s, the corresponding truncated
prepotential is especially simple,
\be
\tilde F^{(0)}[T|\tilde S] = -\sum_{i=1}^n \tilde S_i W(\alpha_i)
\ee
Indeed, since $W'(\alpha_i) = 0$, the second term at the r.h.s. of
\be
\frac{\partial \tilde F^{(0)}[T|\tilde S]}{\partial T_k} =
-\sum_{i=1}^n \tilde S_i \alpha_i^k +
\sum_{i=1}^n \frac{k\alpha_i^{k}}{M_i}
\frac{\partial \tilde F^{(0)}[T|\tilde S]}{\partial \alpha_i},\ \ \ \
M_i\equiv W''(\alpha_i)
\ee
vanishes, and the action of $\check R_W(z)$ straightforwardly converts the first
term into the r.h.s. of (\ref{tildeS}).

See Tables below for more explicit expressions for
the multidensities.

\section{Comments on the number of solutions  \label{nsol}}
\setcounter{equation}{0}

In this section we discuss an {\it a priori} way to introduce
phases, and naturality of their labeling introduced in the previous
section.
We also emphasize that
an interesting task is to construct a matrix-integral {\it model} of the partition
function defined by the Virasoro constraints (\ref{vircon}) or
(\ref{virconf}), and briefly discuss in what sense the Givental-style decomposition
formulas (which are related to matrix integrals and give rise to the CIV-DV
partition functions) do provide a solution to this problem.

As we already saw above,
an important ingredient\footnote{This is new as compared to the simple example
of (\ref{Ex1.1}), but can be already observed in the integral representation
for the cylindric functions (\ref{cylint}). Indeed, the integration contour,
say, for the Hankel functions is quite fancy so that the shift of integration
variable generates from any one Hankel function their linear combination.
}
in the theory of matrix model integrals,
is relevance of {\it the shifts} of $t$-variables: $Z(t-T)$
and $Z(T'+t)$ with different $T$ and $T'$ can be different branches
of the same partition function, moreover, they are further ramified
into (generically infinitely many) {\it sub-branches}. Below we explain the
meaning of this claim and describe the
pattern of branches and their properties.\footnote{
After this is done, one can switch to studies
of {\it the continuum limits} of these branches, which arise when
$N \rightarrow \infty$ and other parameters behave somehow, thus
providing (depending on behaviour of these  infinitely many parameters)
$\infty^\infty$ many types of continuum limits (there can be rather few
on the lowest branches -- and these are all so far described in the
literature). However, below we stop at the level of finite $N$:
classification of branches is a precondition for classification of
continuum limits, and it is already enough sophisticated.
}
The first question to address
is to enumerate/label the
branches, i.e. different formal series solutions to the
system (\ref{vircon}).
As already mentioned, an origin of complications is that
the series depends on ratios of $t$-variables and
different branches occur when different combinations
of $t$'s are allowed to appear in the denominators.
One can distinguish three levels in classification
of branches, characterized by the parameter
$g$, $W(z)=\sum T_kz^k$ and $f_W^{(p|m)}(z_1,\ldots,z_m|g,T)$.

\subsection{Types of $g$ dependencies \label{gdep}}


$\bullet$
The $g$-dependence
characterizes the degree of singularity when all $t_k \rightarrow 0$
or all $t_k \rightarrow \infty$.
From the very beginning one can distinguish between three possibilities:

-- Expansion of $Z$ in non-negative powers of $g$.
The leading term of expansion in this
phase describes the ``naive $g=0$ limit" and
satisfies the equations
\be
\sum_{k\geq 0} kt_k\frac{\partial Z}{\partial t_{k+m}} = 0,
\ \ m\geq -1
\label{g0Vir}
\ee

-- Expansion of $Z$ in non-positive powers of $g$. In this
phase the leading term of expansion corresponds to the ``naive $g=\infty$ limit"
and satisfies the equations
\be
\sum_{\stackrel{a,b\geq 0}{a+b=m}}
\frac{\partial^2 Z}{\partial t_{a}\partial t_b} = 0,
\ \ m\geq -1
\label{ginfVir}
\ee

-- Alternative to these two cases is a Laurent expansion in powers
of $g$, infinite in both positive and negative powers of $g$
(if the set of powers is semi-infinite, multiplication of $Z$ by
appropriate power of $g$ puts it into one of the two above classes).
Infinite Laurent series can be sometimes further separated into classes,
if, instead of $Z$, one considers the {\it prepotential}
${\cal F} = g^2\log Z$. Then
\be
{\cal F}_{q,s} = \sum_{q\leq p \leq s} g^{2p}{\cal F}^{(p)}_{q,s}
\label{Fexpang}
\ee
gives rise to an infinite Laurent series for $Z$ whenever $q\leq 0$
and $s\geq 2$. Different pairs $(q,s)$ give rise to different
branches of $Z$.
In fact, solutions to (\ref{vircon}) exist only if $q=-\infty$
or $s=\infty$. If both $q=-\infty$ and $s=\infty$, one can repeat the
trick and take, say, $\log{\cal F}$ instead of ${\cal F}$ and  split the
``point" $(-\infty,\infty)$ further and so on. Also $p$ in
(\ref{Fexpang}) can be, at least, half-integer (more general
expansions in non-integer powers of $t_k$, though quite interesting,
are even more exotic).
Of all allowed ${\cal F}_{q,s}$, the most attention in the literature is
paid to ${\cal F} \equiv {\cal F}_{0,\infty}$, usually with integer
(rather than half-integer)
$p$. In this case (\ref{Fexpang}) is called {\it genus expansion}.
${\cal F}^{(0)}$ is the
{\it spherical} contribution to the prepotential ${\cal F} =
{\cal F}_{0,\infty}$, $p$ labels the genus. The half-integer $p$'s,
if included, are associated with Riemann surfaces with boundaries
or with non-orientable surfaces (both have {\it doubles}, which
are ordinary Riemann surfaces with integer genus).

\bigskip

$\bullet$
In the most of phases ${\cal F}_{q,s}$, for given $q$
and $s$ (in particular, ${\cal F}_{0,\infty}$) possesses
additional free parameters, which are not fixed by the
Virasoro constraints (\ref{vircon}). In s.3.4 we labeled these parameters
by $f$ (actually by $f(z)$ made from an arbitrary function of $T$-variables
$F[g,T]$, (\ref{fthrF})).
Every linear combination
\be
Z\{c\} = \sum_{q,s;f} c_{q,s;f} \exp {\cal F}_{q,s;f}(g|t)
\ee
with $t$-independent (but, probably, $g$-dependent)
coefficients $c_{q,s;f}$ is still a solution to the linear system
(\ref{vircon}), and $c_{q,s;f}$ can not be absorbed into redefinition
of $f$. These coefficients $c_{q,s;f}$ are analogues of $\alpha$ and
$\beta$ in (\ref{Ex1.3}) and can be interpreted as substitution of the
integration contours by cycles with complex coefficients.
In what follows we pay no more attention to the freedom associated with
$c_{q,s;f}$ and concentrate on the {\it D-module generators}
$Z_{q,s;f} = \exp {\cal F}_{q,s;f}(g|t)$. Our next task is to explain the
origin of the free parameters $f$.\footnote{The separation of
``degrees of freedom'' into $f$ and $c_f$ is not absolutely obvious. We described
a possible approach to the separation with the help of basis
in the Hilbert space of solutions
in the very end of introduction, and more details about the corresponding
evolution operator are given in s.\ref{uop} below.
}

\subsection{The shifts of $t$ and the origin of $f$-parameters:
an oversimplified example \label{g0vir}}


$\bullet$
Once the $g$-dependence is specified, one can start enumerating
solutions to the system (\ref{vircon}).
To describe different phases, we shift the variables $t_k \rightarrow
-T_k + t_k$
and assume that only the first $n+1$ of $T_k$'s can
be non-vanishing.

In order to understand how the new free parameters appear for
non-trivial $W(z)$, it is instructive to
start from the ``naive $g=0$ limit" , i.e. from the partition function
satisfying eq.(\ref{g0Vir}). Note that it has nothing to do
with the partition function in ``spherical" (genus $p=0$) approximation,
$\exp {\cal F}^{(0)}$. Instead of (\ref{virconf}), we have for
\be
\rho_0(z) = \hat\nabla(z)\log Z
\label{rhozZ}
\ee
(note the absence of the factor $g^2$ in this definition)
evaluated at $t_k=0$ (i.e. when $v(z) = 0$)
\be
\hat P^-_z\left[ W'(z) \rho_0(z) \right] = 0,
\ \ {\rm i.e.} \ \ \rho_0(z) = \frac{f(z)}{W'(z)}
\ee
where $deg(f(z)) < deg{W'(z)}$,
since the operator $\nabla(z)$ and $\rho_0(z)$ in (\ref{rhozZ})
are expanded in negative powers of $z$ only.
This means that $f(z)$ is an arbitrary
polynomial of degree $n-1$. In other words, we see that, for
given $W(z)$ of degree $n+1$, we get an $n$-parametric set of
solutions, even if all $t_k = 0$. Switching on $t_k$'s is
equivalent to raising the degree of $W(z)$, and thus could lead to a
new arbitrariness, to new arbitrary polynomials like $f(z)$.
However, the coefficients of these polynomials are actually functions
of $T$-variables restricted by additional constraints (\ref{zcc}).
The formalism, developed in s.\ref{NGW} below, allows one to demonstrate
that all the $f$-functions are made from a single arbitrary
function $F(T)$ of $n$ variables $T_0,\ldots,T_{n-1}$.

In our oversimplified example this can, of course, be also seen without
any special formalism, just by looking at the truncated Virasoro
constraints
\be
\sum_{k=1}^{n+1} kT_k\frac{\partial{\cal F}_W(t)}{\partial t_{k+m}} =
\sum_{k=1}^{\infty} kt_k\frac{\partial{\cal F}_W(t)}{\partial t_{k+m}},
\ \ m\geq -1
\label{tv1}
\ee
Putting all $t_k=0$, we express all the derivatives
$\left.\frac{\partial{\cal F}_W}{\partial t_k}\right|_{t=0}$ with $k\geq n$
through those with $k<n$.
Now, take the $t_l$-derivative of (\ref{tv1}),
\be
\sum_{k=1}^{n+1} kT_k\frac{\partial^2{\cal F}_W(t)}
{\partial t_{k+m}\partial t_l} =
l\frac{\partial{\cal F}_W(t)}{\partial t_{l+m}} +
\sum_{k=1}^{\infty} kt_k\frac{\partial{\cal F}_W(t)}
{\partial t_{k+m}\partial t_l},
\ \ m\geq -1
\label{tv2}
\ee
Putting all $t_k=0$, we express all the second derivatives
$\left.\frac{\partial^2{\cal F}_W}
{\partial t_k\partial t_l}\right|_{t=0}$ with $k\geq n$
through those with $k<n$. Using the symmetry between $k$ and $l$ we can
conclude that, of all the second-derivatives at $t=0$, only
$\left.\frac{\partial^2{\cal F}_W}
{\partial t_k\partial t_l}\right|_{t=0}$ with $k,l<n$
are left unrestricted by the Virasoro constraints.
Continuing in the same fashion, one proves that the truncated Virasoro constraints
fix everything except for
$\left.\frac{\partial^p{\cal F}_W}
{\partial t_{k_1}\ldots\partial t_{k_p}}\right|_{t=0}$
with $k_1,\ldots,k_p < n$, with any $p$. This means that, indeed, the only freedom left
after the truncated Virasoro constraints are imposed is in the choice of
arbitrary function $F[T] = {\cal F}_W(t=0)$ of $n$ variables $T_1,\ldots,
T_{n-1}$ (since the dependence on $T_n$ and $T_{n+1}$ is fixed by the
constraints $\hat L_{-1}^W(t=0)Z_W(t=0) = \hat L_0^W(t=0)Z_W(t=0) = 0$,
see (\ref{Tnn1})).

\bigskip

$\bullet$
Just the same is going to happen when $g\neq 0$ with more interesting
prepotentials ${\cal F}^{(p)}_W$. They depend on the choice of
$W(z)$ and on additional functions like $f(z)$.
All these functions, which are getting more and more
complicated with increasing genus $p$ and with increasing complexity of correlators
(what is equivalent to increasing the number of non-vanishing,
but infinitesimal variables $t_k$), are,
in fact, made from the $T$-derivatives of just a single arbitrary function
of $T_0,\ldots,T_{n-1}$, that is,
the truncated (evaluated at $t_k=0$) prepotential $F(T|g)$.

\subsection{Evolution operator and basis in the Hilbert space of solutions
\label{uop}}

In this section we claim that, for any $W(z)$, i.e. for any
choice of $T$-variables, there is an operator $\check U_W(t)$,
which converts {\it any} function $Z[T]$ of $T_0,\ldots,T_{n-1}$
(with prescribed dependence on $T_n$ and $T_{n+1}$) into
$Z_W(t) = \check U_W(t)Z[T]$, which satisfies the Virasoro constraints,
$L^W_m(t)Z_W(t) = 0, \ \ m\geq -1$. Moreover, if
$Z[T] = \sum_a c_a Z^{(a)}[T]$, then $Z_W(t) = \sum_a c_aZ^{(a)}_W(t)$.
The construction is, in fact, a straightforward generalization of that
in s.\ref{g0vir} to the case of $g\neq 0$.

Namely, for given $T$'s , we make use of the Virasoro constraints
$\hat L^W_m Z(t) = 0$, i.e.
\be
\sum_{k=1}^{n+1} kT_k\frac{\partial Z}{\partial t_{k+m}} =
\sum_{k=1}^\infty kt_k\frac{\partial Z}{\partial t_{k+m}} +
\sum_{a+b=m} \frac{\partial^2 Z}{\partial t_a\partial t_b},
\ \ \ m\geq -1
\ee
and their multiple $t$-derivatives to recurrently express
$\left.\frac{\partial^p Z}{\partial t_{k_1}\ldots \partial
t_{k_p}}\right|_{t=0}$ with all $0 \leq k_i < \infty$ through
$g^{2s}\left.\frac{\partial^{p+s} Z}{\partial T_{l_1}\ldots \partial
T_{l_{p+s}}}\right|_{t=0}$ with all $0 \leq l_j < n$
\be
\left.\frac{\partial^p Z}{\partial t_{k_1}\ldots \partial
t_{k_p}}\right|_{t=0}
 = \sum_{\stackrel{s}{l_1,\ldots,l_{p+s}}}
g^{2s}{\cal D}^{l_1\ldots l_{p+s}}_{k_1\ldots k_p}
\left.\frac{\partial^{p+s} Z}{\partial T_{l_1}\ldots \partial
T_{l_{p+s}}}\right|_{t=0}
\ee
This is a little more tedious procedure than in the case of $g=0$,
but still it is straightforward, and the sum over $s$ is finite,
from $0$ to the integer part of $\left\{\frac{k}{n-1}\right\}$:
the expression for
$\left.\frac{\partial Z}{\partial t_k}\right|_{t=0}$ contains
$\left.\frac{\partial^2 Z}{\partial t_a\partial t_b}\right|_{t=0}$,
but with $a,b \leq k-n-1$, further, the expression for
$\left.\frac{\partial^2 Z}{\partial t_a\partial t_b}\right|_{t=0}$
contains
$\left.\frac{\partial^3 Z}{\partial t_a\partial t_b\partial
t_c}\right|_{t=0}$,
this time with $a,b,c \leq k-2n-2$ and so on.

Now we can define the operators
\be
\check D^{(p)}_{k_1\ldots k_p} =
\sum_{\stackrel{s}{0\leq l_1,\ldots,l_{p+s} \leq n-1}}
g^{2s}{\cal D}^{l_1\ldots l_{p+s}}_{k_1\ldots k_p}
\left.\frac{\partial^{p+s}}{\partial T_{l_1}\ldots \partial
T_{l_{p+s}}}\right|_{t=0}
\ee
Because of their origin in the Virasoro constraints, these operators satisfy
some obvious relations. Actually these are the eqs.(\ref{lincom}) below.
Now we construct the evolution operator $\check U_W(t)$ as a series in these
$\check D$-operators
\be
\check U_W(t) = 1 + t_k\check D_k^{(1)} +
\frac{1}{2}t_kt_l \check D_{kl}^{(2)} +
\frac{1}{6}t_kt_lt_m\check D_{klm}^{(3)} +
\ldots
\ee

The fact that for any $Z[T]$
\be
\hat L^W_m(t) Z_W(t) = \hat L^W_m(t)  \check U_W(t) Z[T] = 0
\ee
or, simply, that
\be
\hat L^W_m(t) \check U_W(t) =
\left(\sum_k kT_k \check D_{k+m}^{(1)} +
\sum_{a+b = m} \check D_{ab}^{(2)}\right)  +
\sum_l t_l \left(l\check D_{l+m}^{(1)} + \sum_k kT_k \check D_{k+m,l}^{(2)} +
\sum_{a+b = m} \check D_{abl}^{(3)}\right) +\\+
\frac{1}{2}\sum_{l_1,l_2} t_{l_1}t_{l_2}
\left(l_1\check D_{l_1+m,l_2}^{(2)} + l_2\check D_{l_2+m,l_1}^{(2)} +
\sum_k kT_k \check D_{k+m,l_1,l_2}^{(3)} +
\sum_{a+b = m} \check D_{abl_1l_2}^{(4)}\right) + \ldots =0
\label{lincom}
\ee
is equivalent to vanishing of all the linear combinations of operators
in brackets, and these are the characteristic equations for the $\check D$-operators.
Note that we proved that the evolution operator is the same for any values
of the arbitrary parameters $f$ (or for any function $Z[T]$) once $W(z)$ is
fixed. This proves that if
$Z[T] = \sum_a c_a Z^{(a)}[T]$, then $Z_W(t) = \sum_a c_aZ^{(a)}_W(t)$, and
means that ``orbits" of the evolution operators are completely parameterized
by $W(z)$.

\newpage

\part{Particular Special Functions}
\setcounter{equation}{0}
\setcounter{section}{0}
\section{Gaussian partition function \label{Gpf}}
\setcounter{equation}{0}


$\bullet$
The Gaussian partition function (phase) is distinguished by existence
of {\it diagrammatic representation} for correlators and thus by the
possibility of interpreting all the formulas in terms of combinatorics of
fat graphs. This makes the Gaussian partition function an important special
function by itself, even without a reference to the generic FSHMM partition
function, of which it is a particular branch. Moreover, the Givental-style
decomposition formulas (see s.II.\ref{Giv}) represent
various branches as polilinear combinations of the Gaussian
partition functions. For these reasons, the Gaussian partition function
$Z_G^M(t|N)$ deserves a special attention, and we consider it separately,
though a significant part of the formalism below is a particular case of
the generic formalism for non-Gaussian models.

\bigskip

$\bullet$
The {\it Gaussian matrix model (phase)} is the one with
$W(x) = \frac{M }{2}x^2$ (i.e. $T_k = \frac{M}{2}\delta_{k,2}$,
where $M$ is a number), and the saddle point (see s.I.2.1) $\phi_0 = 0$,
\be
Z_G^M(t|N) =
\frac{1}{\vol_{U(N)}}
{\int D\phi \exp\left(-\frac{M}{2g}{\rm Tr}\phi^2
+ \frac{1}{g}\sum_{k=0}^\infty t_k{\rm Tr}\phi^k\right)}
= \nn \\ =
\left(\frac{2\pi g}{M}\right)^{N^2/2}
\sum_{\vec k}
\frac{t_{k_1}\ldots t_{k_m}}{g^m m!}
\la {\rm Tr}\phi^{k_1}\cdot\ldots\cdot{\rm Tr}\phi^{k_m}\ra_G
\ee
Here the brackets denote averaging over the random matrices with the Gaussian measure,
$\exp \left( -\frac{M}{2g}{\rm Tr}\phi^2\right)$.

Each correlator at the r.h.s. is a sum over the Feynman-t'Hooft diagrams
({\it fat graphs}) with $m$ vertices of valencies $k_1,\ldots, k_m$.
Each vertex contributes the factor of $g^{-1}$, each propagator,
denoted by a double-line (strip of the fat graph), contributes the factor
of $g/M$, and each closed line (boundary component of the fat graph)
contributes the factor of $N = S/g$, where $S \equiv gN$ is the
t'Hooft's coupling constant\footnote{
We remind that it can make sense to
assume that $S$ can still depend on $g$
$$S = \sum_{k=0}^\infty S_k g^k$$
However, the expansion in (\ref{genexpa}) is defined for $S=gN$,
otherwise one could write
$${\cal F}_G(t|S) = \sum_{p=0}^\infty g^{2p} \hat{\cal F}_G^{(p)}(t)$$
with
$$\hat{\cal F}_G^{(0)}(t) = {\cal F}_G^{(0)}(t|S_0)$$
$$\hat{\cal F}_G^{(1)}(t) = {\cal F}_G^{(1)}(t|S_0) +
S_1\frac{\partial}{\partial S_0}{\cal F}_G^{(0)}(t|S_0)$$
etc.
}.
The prepotential ${\cal F}_G = g^2\log Z_G^M$
is a similar sum over the {\it connected} diagrams only.
If we switch from the $N,g$ variables to $S,g$, then particular terms
of the expansion
\be
{\cal F}_G(t|S) = \sum_{p=0}^\infty g^{2p} {\cal F}_G^{(p)}(t|S)
\label{genexpa}
\ee
are sums over connected genus-$p$ fat graphs
\be
{\cal F}_G^{(p)}(t|S)
=
F^{(p)}(S) +
\sum_{\vec k} \frac{ t_{k_1}\ldots t_{k_m}}{m!}
\la\la {\rm Tr}\phi^{k_1}\cdot\ldots
{\rm Tr}\phi^{k_m}\ra\ra^{(p)} =
\nn \\ =
F_G^{(p)}(S) +
\sum_{m\geq 1} \frac{1}{m!}\oint\ldots\oint v(z_1)\ldots v(z_m)
\rho^{(p|m)}_G(z_1,\ldots,z_m)
\label{Gpre}
\ee
where the $t$-independent quasiclassical prepotential
\be
F_G^{(p)}(S) =-\frac{S^2}{2}\log\frac{M}{2\pi S}\delta_{p,0} -
\left(\log \vol_{U(N)}\right)^{(p)}(1-\delta_{p,0})
\ee
and the
Gaussian {\it multidensities} are
\be
\rho^{(p|m)}_G(z_1,\ldots,z_m) = \la\la
T(\phi|z_1)\ldots T(\phi|z_m)
\ra\ra^{(p)}_G
\ee
with
$T(\phi|z) \equiv {\rm Tr}\frac{1}{z-\phi}$,
and $\langle\langle \ldots \rangle\rangle^{(p)}$
denoting contribution of the {\it connected} Feynman-t'Hooft diagrams of
genus $p$ to the Gaussian average with the $g$ and {\it $N$-independent} weight
$\exp \left(- \frac{M}{2}{\rm tr}\phi^2\right)$. Formally,
the size of the matrices is now
$S$, not $N$:
${\rm tr} \ I = S$
(accordingly, we change the notation ${\rm Tr}$ for ${\rm tr}$).
This calculus is justified by the Euler's theorem stating that
\be
g^{-\#\ {\rm of\ vertices}}g^{\# \ {\rm of \ edges}}g^{-\#\ {\rm of\ loops}}
= g^{2\times{\rm genus} -2}
\ee

\bigskip

$\bullet$
The Gaussian multidensities $\rho^{(p|m)}_G$ can be obtained recurrently
in $m$ and then in $p$
by solving the Virasoro constraints (\ref{vircon}) rewritten
in terms of the prepotentials ${\cal F}$
\be
M\frac{\partial{\cal F}_G^{(p)}}{\partial t_{m+2}} =
\sum_{\stackrel{a+b=m}{a,b\geq 0}}
\sum_{\stackrel{q+r = p}{q,r\geq 0}}
\frac{\partial{\cal F}_G^{(q)}}{\partial t_{a}}
\frac{\partial{\cal F}_G^{(r)}}{\partial t_{b}}
+ \sum_{\stackrel{a+b=m}{a,b\geq 0}}
\frac{\partial^2{\cal F}_G^{(p-1)}}{\partial t_{a}\partial t_b}
+ \sum_{k\geq 0} kt_k\frac{\partial{\cal F}_G^{(p)}}{\partial t_{k+m}}
\nn \\
m\geq -1
\label{VirG}
\ee
Summing over $m$ with the weight $z^{-m-2}$, one obtains
for the generating function
\be
\rho_G(z|t) = \sum_{k\geq 0} \frac{1}{z^{k+1}}
\frac{\partial{\cal F}_G(t|\nu)}{\partial t_k}
= \la\la
T(\phi|z)
\ra\ra_G
= \nn \\ = \sum_{p=0}^\infty g^{2p}
\sum_{m\geq 0} \frac{1}{m!}\oint\ldots\oint v(z_1)\ldots v(z_m)
\rho^{(p|m+1)}_G(z,z_1,\ldots,z_m)
\label{genden}
\ee
the equation ({\it cf.} (\ref{virconfW}))
\be
M (z\rho_G(z|t) - S) =
\rho_G^2(z|t) + g^2\hat\nabla(z) \rho_G(z|t)
+\hat P_z^- \left[v'(z) \rho_G(z|t)\right]
\label{VirGgen}
\ee
where
\be
\hat P_z^- \left[v'(z) \rho_G(z|t)\right] = v'(z) \rho_G(z|t) -
\la\la {\rm tr}\frac{v'(\phi) - v'(z)}{\phi - z} \ra\ra_G
\ee
while
\be
\hat P_z^- \left[z \rho_G(z|t)\right] = z \rho_G(z|t) -
\la{\rm tr}\ I\ra = z \rho_G(z|t) - S
\ee
Putting $v(z) = 0$ in equation (\ref{VirGgen}), one obtains a recurrent
relation in $p\ $ for the densities.
Similar recurrent relations for {\it multidensities} with $m>1$ are obtained
by acting with $\hat\nabla(z_1)\ldots \hat\nabla(z_{m-1})$ on
(\ref{VirGgen}), making use of (\ref{PPrel})
and putting $v(z) = 0$ afterwards. This double recurrent procedure
(with recursions in $p$ and $m$) allows one to obtain every particular
$\rho^{(p|m)}$. See Tables below for some explicit expressions.

\bigskip

$\bullet$
The first (genus-zero, single-point) density, satisfying
\be
M (z\rho^{(0|1)}_G(z) - S) = \left(\rho^{(0|1)}_G(z)\right)^2
\ee
is the celebrated semicircular distribution \cite{Wig},
\be
\rho^{(0|1)}_G(z) = \frac{M}{2}\left(z - \sqrt{z^2 - \frac{4S}{M}}\right) =
\frac{Mz - y_G(z)}{2} =
\sum_{k=0}^\infty \frac{c_kS^{k+1}}{M^kz^{2k+1}} = \frac{S}{z} +
\frac{S^2}{Mz^3} + \ldots
\label{sdenG}
\ee
It is a generating function for the {\it Catalan numbers}
\be
c_k = \frac{4^k\Gamma(k+1/2)}{\Gamma(1/2)(k+1)!}
\ee
counting the numbers of planar leaf diagrams with $k$ leaves.

The two-point density satisfies
\be
\left(Mz - 2\rho^{(0|1)}_G(z)\right)\rho^{(0|2)}_G(z,x) =
\partial_x \frac{\rho^{(0|1)}_G(z) - \rho^{(0|1)}_G(x)}{z-x}
\ee
so that
\be
\rho^{(0|2)}_G(z,x) = \frac{1}{2(z-x)^2}\left( -1 +
\frac{4MS}{y_G(z)y_G(x)}\right) = \nn \\ =
\frac{1}{2x^2}\left(-1 + \frac{Mz}{y_G(z)}\right) +
\frac{1}{x^3}\frac{Mz^2-2MS - zy_G(z)}{y_G(z)} + O\left(\frac{1}{x^2}\right)
\label{ddenG}
\ee
The first two terms at the r.h.s. are the generating functions of the correlators
\be
\frac{1}{z^{k+1}}\la\la{\rm tr}\phi\ {\rm tr}\phi^k\ra\ra^{(0)}_G
\ \ {\rm and} \ \
\frac{1}{z^{k+1}}\la\la{\rm tr}\phi^2\ {\rm tr}\phi^k\ra\ra^{(0)}_G
\ee
The latter generating function can be alternatively obtained by taking the
derivative of $\rho^{(0|1)}_G$ over $T_2 = M/2$ provided that
\be
\frac{\partial S}{\partial T_2} = 0
\ \ {\rm for\ the\ Gaussian\ model}
\ee
(note that this is not true for the generic FSHMM of s.I.\ref{NGW},
even for the quadratic potential $W(z) = \frac{Mz^2}{2}$: whenever $S \neq const$
the diagram technique breaks down even for the quadratic potentials).

\bigskip

$\bullet$
Along with the Gaussian multidensities
\be
\rho_G^{(p|m)}(z_1,\ldots,z_m) = \la\la
{\rm Tr}\frac{1}{z_1-\phi}\ldots {\rm Tr}\frac{1}{z_m-\phi}
\ra\ra_G^{(p)}
\ee
one can consider their inverse Laplace transforms
\be
\eta_G^{(p|m)}(x_1,\ldots,x_m) =
\oint\ldots\oint e^{x_1z_1 + \ldots +x_mz_m}
\rho_G^{(p|m)}(z_1,\ldots,z_m)dz_1\ldots dz_m =
\la\la
{\rm Tr}\ e^{x_1\phi}\ldots {\rm Tr}\ e^{x_m\phi}
\ra\ra_G^{(p)}
\ee
and their generating functions like
\be
\eta_G(x|t) = \oint e^{xz}\rho(z|t) dz =
\sum_{p,m=0}^\infty
\frac{g^{2p}}{m!}\oint\ldots\oint dzdz_1\ldots dz_m
\int_0^{\infty}\ldots\int_0^{\infty}dx_1\ldots dx_m\times\\ \times
e^{-x_1z_1-\ldots-x_mz_m}
\bar v(z_1)\ldots \bar v(z_m)
\eta^{(p|m+1)}_G(x,x_1,\ldots,x_m)
\label{genden1}
\ee
The direct Laplace transform converts $\eta_G(x)$ back into $\rho_G(z)$
\be
\rho_G(z|t) = \int_0^\infty e^{-xz} \eta_G(x|t) dx
\ee
and
\be
\rho_G^{(p|m)}(z_1,\ldots,z_m) =
\int_0^\infty\ldots\int_0^\infty
e^{-x_1z_1-\ldots-x_mz_m} \eta_G^{(p|m)}(x_1,\ldots,x_m) dx_1\ldots dx_m
\ee

\bigskip

$\bullet$
The (one-point) Gaussian densities satisfy the {\it sum rule},
calculating the total number of ``flower" diagrams contributing to
$\langle {\rm Tr}\phi^{2k}\rangle$
\be
\sum_{p=0}^\infty \oint z^{2k}\rho^{(p|1)}(z) = (2k-1)!!
\left({g\over m}\right)^k
\label{sumrule1}
\ee

The sum rule (\ref{sumrule1}) is a particular case (at $N=1$) of a more general
identity \cite{sumru,sumru1},
providing a generating function for all (any genus)
Gaussian one-point functions (densities)
\be
\sum_{k=0}^\infty \frac{(x^2M/g)^k}{(2k-1)!!}
\langle {\rm Tr}\phi^{2k}\rangle_G =
\frac{1}{2x^2}\left(\left(
\frac{1+x^2}{1-x^2}\right)^N - 1\right) =
N + N^2x^2 + \frac{(2N^3+N)x^4}{3} +\\+
\frac{(5N^4+10N^2)x^6}{15} +
\frac{(14N^5+70N^3+21N)x^8}{105} +
\frac{(42N^6 + 420N^4 + 483N^2)x^{10}}{945} + \ldots
\label{sumrule2}
\ee
Here $(2k-1)!! = 2^k\frac{\Gamma(k+1/2)}{\Gamma(1/2)}$
so that $(-1)!! = 1$.

\bigskip

$\bullet$
The Gaussian partition function $Z_G^M(t|N)$ can be also represented
as a member of the family of Kontsevich models \cite{K}-\cite{GKMfollowup},
\cite{UFN3}:
it coincides with the Gaussian Kontsevich model with logarithmic potential
\cite{versus}
\be
\int dH_{N\times N}
e^{-\sum_k t_k\Tr M^k} = {(2\pi )^{N^2/2}}{
\int dX_{n\times n}\det (I-X/H) e^{-\Tr X^2/2}\over
\int dX_{n\times n}e^{-\Tr X^2/2}}
\ee
where the time variables $t_k$'s are expressed through the external matrix in the
Kontsevich integral, $t_k={1\over k}\Tr H^{-k}+{1\over 2}\delta_{k,2}$,
the size of matrix in the $r.h.s.$ is $n\times n$ and in the $l.h.s.$
is $N\times N$, and these parameters are absolutely independent.
The multidensities $\rho^{(p|m)}$
can be also reproduced from quadratic Kontsevich model (we postpone
their discussion for a
separate publication devoted to the Kontsevich type matrix models).

\bigskip

$\bullet$
The genus-zero three-point functions,
in particular $\rho^{(0|3)}(z_1,z_2,z_3)$,
are expected to satisfy the
WDVV equations \cite{WDVV}-\cite{WDVVf}.

\section{Givental-style decomposition formulas for non-Gaussian CIV-DV
partition functions \label{defo} \label{Giv}}
\setcounter{equation}{0}


$\bullet$
The shift of $t$-variables converts integral (\ref{ZN}) into
\be
Z_{N,W}^{(matr)}(t) = \frac{1}{\vol_{U(N)}}
\int d\phi e^{-\frac{1}{g}{\rm Tr}W(\phi)}
e^{\frac{1}{g}\sum_k t_k{\rm Tr}\phi^k}
\label{ZNW}
\ee
At the first glance, it can seem that there is no freedom in this
definition and no place for the free parameters to get in.
This is, however, incorrect, if one wants to allow any kind of analytical
continuation in $T$-variables (coefficients of $W(\phi)$).
One of the ways to construct formal series (in powers of $t_k$)
solutions to the linear system of the Virasoro constraints (\ref{vircon})
is provided by an analytically continued saddle-point
approximation. Different generators of the $D$-module are represented
by expansions around different saddle-points $\phi = \phi_0$ such that
$W'(\phi_0) = 0$. If the polynomial
\be
W'(x) = \prod_{i=1}^n(x-\alpha_i)
\ee
has roots $\alpha_i$, then, since $\phi_0$ are matrices
defined modulo $U(N)$-conjugations (which allow one
to diagonalize any matrix and permute its eigenvalues),
the different saddle points are represented by
\be
\phi_0 =
 diag(\alpha_1,\ldots,\alpha_1;
\alpha_2,\ldots,\alpha_2;\ \ldots\ ; \alpha_n,\ldots,\alpha_n)
\ee
with $\alpha_i$ appearing $N_i$ times, $\sum_{i=1}^n N_i = N$.

Since we are dealing with a $D$-module, there is no need to keep
these $N_i$ non-negative integers:
in final expressions (like formulas
for the multidensities and prepotentials)
they can be substituted by any complex
numbers. Moreover, $N_i$ can depend on $g$ and on $T_k$ (i.e. on the
shape of $W(\phi)$), the only restriction coming from additional
constraints (\ref{addcon}). The so restricted freedom to choose functions
$N_i(g,W)$ matches the free parameters $f$ in $\exp {\cal F}_{q,s;f}$.

\bigskip

$\bullet$
The contribution of a particular extremum $\phi_0$
(labeled by the set of $N_i$)
is given by the Givental-style decomposition formula,\footnote{
It is derived, as in \cite{MMDV}, by switching from an eigenvalue integral
over $N\times N$ matrix $\phi$ to $n$ eigenvalue integrals over $N_i\times N_i$
matrices $\phi_i$ (each obtained with the shift by $\alpha_i$: just
changing of variables in integral (\ref{ZNW})),
namely by rewriting the square of the Van-der-Monde determinant
in (\ref{VdM}) as
\be
\Delta_n^2(\phi) \equiv \prod_{a<b}^N(\phi_a-\phi_b)^2 =
\prod_{i=1}^n\left(\prod_{a<b}^{N_i}(\phi_{i,a}-\phi_{i,b})^2\right)
\prod_{i<j}^n\left(\prod_{a=1}^{N_i}\prod_{b=1}^{N_j}
(\alpha_{ij} + \phi_{i,a}-\phi_{j,b})^2\right) = \nn \\ =
\prod_{i=1}^n \Delta^2_{N_i}(\phi_i)\prod_{i<j}^n\alpha_{ij}^{2N_iN_j}
\exp \left(2\sum_{m=1}^\infty\frac{(-)^{m-1}}{m\alpha_{ij}^m}
{\rm Tr}_i{\rm Tr}_j (\phi_i\otimes I_j - I_i\otimes \phi_j)^m\right) = \nn \\
= \prod_{i=1}^n \Delta^2_{N_i}(\phi_i)\prod_{i<j}^n\alpha_{ij}^{2N_iN_j}
\exp \left(2\sum_{k,l=0}^\infty (-)^{k}\frac{(k+l-1)!}
{\alpha_{ij}^{k+l}k!l!}{\rm Tr}_i\phi_i^k{\rm Tr}_j\phi_j^l\right)
\ee
Despite the technical simplicity of this transformation, its real
quantum-\-field-\-theory meaning (exact embedding of the $SU(N)$
Cartanian $\tau$-function into the $S\left(\otimes_{i=1}^n U(N_i)\right)$
one) remains underinvestigated (in particular, no effective transformation
in terms of matrices, avoiding intermediate eigenvalue integrals,
is known), what makes generalizations to non-eigenvalue models obscure.
}
expressing it
through the product of $n$ Gaussian partition functions, each with
its own $N_i$ and $M_i = W^{\prime\prime}(\alpha_i) =
\prod_{j\neq i}(\alpha_i - \alpha_j)$
\be
Z_{N,W}^{(matr)}(t|\phi_0) =
\frac{\prod_{i=1}^n e^{-N_iW(\alpha_i)}\vol_{U(N_i)}}{\vol_{U(N)}}
\prod_{i<j}^n\alpha_{ij}^{2N_iN_j}
\prod_{i<j}^n \hat{\cal O}_{ij} \prod_{i=1}^n
\hat{\cal O}_i \prod_{i=1}^n Z_G^{M_i}(t^{(i)}|N_i)
\label{factfor}
\ee
with operators
\be
\hat{\cal O}_{ij} = \exp \left(2\sum_{k,l=0}^\infty (-)^{k}\frac{(k+l-1)!}
{\alpha_{ij}^{k+l}k!l!}
\frac{\partial}{\partial t^{(i)}_k}\frac{\partial}{\partial t^{(j)}_l}
\right),
\ee
$\alpha_{ij} = \alpha_i-\alpha_j$, and
\be
\hat{\cal O}_i = \exp \left(-\sum_{k\geq 3} \frac{W^{(k)}(\alpha_i)}{k!}
\frac{\partial}{\partial t^{(i)}_k}\right)
\ee
($W^{(k)}(x) = \partial_x^kW(x)$). The relation between the time-variables
at the two sides of eq.(\ref{factfor}) is given by
\be
\sum_{k=0}^\infty t_k \left(\sum_{i=1}^n {\rm Tr}_i(\alpha_i + \phi_i)^k\right) =
\sum_{i=1}^n \left(\sum_{k=0}^\infty t_k^{(i)} {\rm Tr}_i \phi_i^k\right)
\label{ttalpha}
\ee
with arbitrary $N_i\times N_i$ matrices $\phi_i$.

The operator $\hat\nabla(z) = \sum_{k\geq 0}\frac{1}{z^{k+1}}
\frac{\partial}{\partial t_k}$ is a linear combination of the operators
$\hat\nabla(z)^{(i)} = \sum_{k\geq 0}\frac{1}{z^{k+1}}
\frac{\partial}{\partial t_k^{(i)}}$,
\be
\hat\nabla(z) = \sum_{i=1}^n \hat\nabla(z-\alpha_i)^{(i)}
\label{nanai}
\ee

$\bullet$
Acting by operators $\hat\nabla(z)$ on the decomposition formula
(\ref{factfor}),
one can express the non-Gaussian multidensities through the Gaussian ones.
Note that $\hat\nabla(z) = \sum_{k\geq 0}z^{-k-1}\partial/\partial t_k$
commutes with $\hat {\cal O}_{ij}$ and
$\hat {\cal O}_i$ and it does not act on $M_i =
W^{\prime\prime}(\alpha_i) = \prod_{j\neq i}\alpha_{ij}$,
which do not depend on $t$'s.
According to (\ref{ttalpha}) and (\ref{nanai}), the
action of $\hat\nabla(z)$ on a particular Gaussian constituent,
\be
g^2\hat\nabla(z) \log Z_G^{[M_i]}(t^{(i)}|N_i) =
\rho_G^{[M_i]}(z-\alpha_i)
\ee
provides the corresponding Gaussian density, but with shifted arguments.
As a result, we have, for example, for
\be
\rho_W^{(matr)}(z) = \hat\nabla(z)\log Z_{W}^{(matr)} =
\frac{\prod_{i<j}^n \hat{\cal O}_{ij} \prod_{i=1}^n
\hat{\cal O}_i
\left(\sum_{i=1}^n \rho_G^{[M_i]}(z-\alpha_i)\right)
\prod_{i=1}^n Z_G^{[M_i]}(t^{(i)}|N_i)}
{\prod_{i<j}^n \hat{\cal O}_{ij} \prod_{i=1}^n
\hat{\cal O}_i \prod_{i=1}^n Z_G^{[M_i]}(t^{(i)}|N_i)}
\label{denGiv}
\ee
Of course, $\hat{\cal O}_i$ and $\hat{\cal O}_{ij}$ act both on $Z_G$'s and
$\rho_G$'s.
Similarly, expressions for the multidensities $\rho_W(z_1,\ldots,z_m) =
\hat\nabla(z_1)\ldots\hat\nabla(z_m)\log Z_{W}$ contains
as insertions in the numerator sums over splittings of the set
$\{z_1,\ldots,z_m\}$ into $n$ subsets (perhaps, some of them empty,
then the corresponding $\rho$ with no arguments is just unity)
\be
\sum_{\{m\} \rightarrow \{m_1\},\ldots,\{m_n\}}
\left(\prod_{i=1}^n\rho_G^{[M_i]}\left(z_{j_1},\ldots,z_{j_{m_i}}\right)\right)
\label{multidG}
\ee

\bigskip

$\bullet$
Expression (\ref{denGiv}) and its generalization with (\ref{multidG})
are rather sophisticated. Role of the denominator is, as usual, to
eliminate the disconnected Feynman-t'Hooft diagrams, including
disconnected contributions from different Gaussian constituents.
The first contributions to (\ref{denGiv}) look as follows
(we remind that $\rho_G(z) = \la\la T(z|\phi) \ra\ra_G$, and introduce
an obvious notation $z_i = z - \alpha_i$ for shifted $z$-variables and
$\phi_i$ for the $N_i\times N_i$-matrix fields)
\be
\rho_W(z)=\rho_W^{(matr)}(z),\\
\rho_W^{(matr)}(z)= \sum_i \la\la T(z_i|\phi_i)
e^{-\frac{1}{g}{\rm Tr}_i\tilde W_i(\phi_i)}\ra\ra_G^{[M_i,N_i]} +
\sum_{j\neq i} \frac{2N_j}{\alpha_{ij}} \la\la T(z_i|\phi_i) {\rm Tr}_i\phi_i
e^{-\frac{1}{g}{\rm Tr}_i\tilde W_i(\phi_i)}\ra\ra_G^{[M_i,N_i]} + \ldots
\label{denGiv1}
\ee
Here
\be
\tilde W_i(\phi_i) \equiv W(\alpha_i+\phi_i) -
W(\alpha_i) - \frac{M_i\phi_i^2}{2}
\ee
and the structure of other corrections can be understood since
every operator $\hat{\cal O}_{ij}$ inserts into the correlator a series
\be
\exp{ \sum_{k=1}^\infty\left( (-)^{k-1}\frac{2}{k\alpha_{ij}^k}
{\rm Tr}_i{\rm Tr}_j (\phi_i - \phi_j)^k\right)} =
1 + \frac{2}{\alpha_{ij}}(N_j{\rm Tr}_i\phi_i   - N_i{\rm Tr}_j\phi_j)
+ \ldots
\ee
The leading contributions to the l.h.s. and to
the r.h.s. of the first line of (\ref{denGiv1}) are respectively
\be
\rho_W(z) =
\frac{1}{2}\left(W'(z) - \sqrt{(W'(z))^2 - 4W'(z)\sum_i\frac{\tilde
S_i}{z_i}}\right) = \sum_i\frac{\tilde S_i}{z_i} + O(\tilde S^2)
\ee
(see (\ref{sdenW}) and (\ref{tildeS})) and
\be
\sum_i \rho_G^{[M_i,N_i]}(z_i) = \sum_i\frac{M_i}{2}\left(z_i -
\sqrt{z_i^2 - \frac{4gN_i}{M_i}}\right) =
\sum_i \frac{gN_i}{z_i} + O(g^2N^2)
\ee
(see (\ref{sdenG})), i.e. in the leading approximation
(comp.with \cite{IM7})
\be
\tilde S_i = gN_i + O(g^2N^2) = S_i + O(S^2)
\ee

\bigskip

$\bullet$
The quantities $\rho_W(z|t)$ and $f_W(z|t)$ are very similar
to the non-trivial {\it Gaussian} averages, respectively,
\be\label{pertf1}
\la\la T(\phi|z) \ra\ra_W =
\la\la T(\phi|z) e^{-{\rm tr} \tilde W(\phi)}\ra\ra_G
\ee
and
\be
\la\la R_W(\phi|z) \ra\ra_W =
\la\la R_W(\phi|z) e^{-{\rm tr} \tilde W(\phi)}\ra\ra_G
\label{pertf2}
\ee
with
\be
T(\phi|z) = {\rm Tr}\ \frac{1}{z-\phi}, \ \
R_W(\phi|z) = {\rm Tr}\ \frac{W'(z)-W'(\phi)}{z-\phi}
\ee
and $\tilde W(z) = W(z) - W(\alpha) - \frac{M}{2}(z-\alpha)^2$.
The problem is, however, that these averages are not well-defined,
because there are different saddle
points $\alpha$, which are roots of $W'(z)$
(since these are saddle points, they can be both minima and maxima of
$W(z)$), with different values of $M = W''(\alpha)$.
Expressions (\ref{pertf1})-(\ref{pertf2}) can make sense only for particular
$\tilde W$, associated with one of the roots.
Thus, they describe only particular, {\it perturbative} branches
of the partition function, when just a single $N_i=N$ and all other $N_j=0$.
To describe generic {\it non-perturbative} branches, one needs to
return to (\ref{denGiv}) and (\ref{multidG}).


As to {\it perturbative} phases, in any of them one can take the
expansion (\ref{Gpre}) for the Gaussian prepotential,
\be
{\cal F}_G^{(\cdot)} = F_G^{(\cdot)} + \oint v \rho^{(\cdot|1)}_G +
\frac{1}{2!}\oint\oint v v \rho^{(\cdot|2)}_G + \ldots
\ee
and define an expansion for the non-Gaussian one, by simply
performing a shift $v(z) \rightarrow -\tilde W(z) + v(z)$,
i.e. by resummation of the diagrams
\be
{\cal F}_W^{(\cdot)} = F_W^{(\cdot)} + \oint v \rho^{(\cdot|1)}_W +
\frac{1}{2!}\oint\oint v v \rho^{(\cdot|2)}_W + \ldots\
\ee
so that
\be
F_W^{(\cdot)} = F_G^{(\cdot)} + \oint\tilde W\rho^{(\cdot|1)}_G +
\frac{1}{2!}\oint\oint \tilde W \tilde W \rho^{(\cdot|2)}_G + \ldots\ , \nn \\
\rho^{(\cdot|1)}_W =  \rho^{(\cdot|1)}_G + \oint \tilde W \rho^{(\cdot|2)}_G
+ \frac{1}{2!}\oint \oint \tilde W \tilde W \rho^{(\cdot|3)}_G + \ldots\ =
\nn \\
= \sum_{k=0}^\infty \frac{1}{z^{k+1}}
\la\la {\rm Tr}\phi^k \exp \left(
-{\rm Tr}\tilde W(\phi)\right)\ra\ra_G
\equiv \sum_{k=0}^\infty \frac{1}{z^{k+1}}
\la\la {\rm Tr}\phi^k \ra\ra_W, \nn \\
\rho^{(\cdot|2)}_W =
\rho^{(\cdot|2)}_G + \oint \tilde W \rho^{(\cdot|3)}_G
+ \frac{1}{2!}\oint \oint \tilde W \tilde W \rho^{(\cdot|4)}_G + \ldots\
\equiv \sum_{k_1,k_2=0}^\infty \frac{1}{z_1^{k_1+1}z_2^{k_2+1}}
\la\la {\rm Tr}\phi^{k_1} {\rm Tr}\phi^{k_2} \ra\ra_W, \nn \\
\ldots
\label{preW}
\ee

\section{CIV-DV prepotential}
\setcounter{equation}{0}

As we saw in part I, for a given $W$ the solutions to Virasoro constraints,
which possess the genus expansion, are labeled by arbitrary function
$F[g,T]$ of the first $n$ time variables $T_0,\ldots,T_{n-1}$
and $g$ (dependencies on $T_n$ and $T_{N+1}$ are prescribed by
(\ref{Tnn1})). To match the generic $F[g,T]$, the variables $S_i$
in the decomposition formulas (\ref{factfor}) should also be made
$T$- and $g$-dependent (though their possible dependencies on
$T$ and $g$ are severely constrained).

The CIV-DV prepotential \cite{DV}-\cite{DVfollowup} is defined
as $F[g,T]$ with {\it constant}
\be
S_i = {\rm const}
\label{Sconst}
\ee
Then it is immediately given by (\ref{factfor}), but it remains
unclear, what kind of additional constraint should be added to
the Virasoro equations (\ref{vircon}) in order to reproduce this
condition (\ref{Sconst}). Instead, as was
discovered by R.Dijkgraaf and C.Vafa
\cite{DV} a new representation occurs, that is,
in terms of Seiberg-Witten theory \cite{SW},\cite{SWth}-\cite{SWthDV}.
See Tables below for explicit expressions for the CIV-DV prepotential
\be
{\cal F}_{DV}[T|S] =
\sum_{p\geq 0} g^{2p} {\cal F}^{(p)}_{DV}[T|S]
\ee

\bigskip

$\bullet$
In the framework of the Seiberg-\-Witten-\-Dijkgraaf-\-Vafa theory
the $T$-independent $S_i$ can be expressed through ${\cal F}_{DV}^{(0)}$
\be
S_i = \oint_{A_i} y_W(z)dz
\\
y_W^2(z) = \left[W'(z)\right]^2 - 4f_W^{DV}(z), \nn \\
f_W^{DV}(z) = \check R_W(z) {\cal F}_{DV}^{(0)}[T|S] =\sum_{k.l\geq 0}
(k+l+2)T_{k+l+2}z^k \frac{\partial {\cal F}_{DV}^{(0)}}{\partial T_l}\\
{\partial {\cal F}^{(0)}_{DV}\over\partial S_i}=
\oint_{B_i} y_W(z)dz
\ee

\bigskip

Now we are going to calculate the CIV-DV prepotentials using the decomposition
formulas (\ref{factfor}).

\bigskip

$\bullet$
We begin with defining the action of the operator
\be
\hat{\cal O}_i =
\exp\left(-\sum_{k=3}^{n+1}\frac{W^{(k)}(\alpha_i)}{k!}
\frac{\partial}{\partial t_k}\right) =
\exp\left(-M_i\sum_{k=1}^{n-1}\frac{\sigma^{(i)}_k}{(k+2)}
\frac{\partial}{\partial t_{k+2}}\right)
\ee
with
\be
M_i = W^{(2)}(\alpha_i) = W''(\alpha_i) =
(n+1)T_{n+1}\prod_{j\neq i}^n\alpha_{ij}, \nn \\
2M_i\sigma^{(i)}_1 = W^{(3)}(\alpha_i) = 2M_i\sum_{j\neq i}
\frac{1}{\alpha_{ij}}, \nn \\
6M_i\sigma^{(i)}_2 = W^{(4)}(\alpha_i) = 6M_i
\sum_{\stackrel{j<k}{j,k\neq i}}
\frac{1}{\alpha_{ij}\alpha_{ik}}, \nn \\
\ldots, \nn \\
(p+1)!M_i\sigma^{(i)}_p = W^{(p+2)}(\alpha_i) = (p+1)!M_i
\sum_{\stackrel{j_1<\ldots<j_p}{j_1,\ldots,j_p\neq i}}
\frac{1}{\alpha_{ij_1}\ldots\alpha_{ij_p}}
\ee
on a particular Gaussian partition function
$Z_G^{M_i}(t|N_i)$:
\be
\left.\hat{\cal O}_iZ_G^{M_i}(t|N_i)\right|_{t=0} =
\left[ 1 + \left(-\frac{W^{(4)}(\alpha_i)}{4!}\la {\rm tr} \phi^4\ra^{M_i}_G
+ \frac{\left[W^{(3)}(\alpha_i)\right]^2}{2!(3!)^2}
\la \left({\rm tr} \phi^3\right)^2\ra^{M_i}_G\right) + \right.\nn\\ \left.
+ \left(-\frac{W^{(6)}(\alpha_i)}{6!}\la {\rm tr} \phi^6\ra^{M_i}_G
+ \frac{W^{(3)}(\alpha_i)W^{(5)}(\alpha_i)}{3!5!}
\la {\rm tr} \phi^3\ {\rm tr} \phi^5\ra^{M_i}_G
+ \frac{\left[W^{(4)}(\alpha_i)\right]^2}{2!(4!)^2}
\la \left({\rm tr} \phi^4\right)^2 \ra^{M_i}_G
- \right.\right.\nn\\ \left.\left.-
\frac{\left[W^{(3)}(\alpha_i)\right]^2W^{(4)}(\alpha_i)}{2!(3!)^2 4!}
\la \left({\rm tr} \phi^3\right)^2\ {\rm tr} \phi^4\ra^{M_i}_G +
\frac{\left[W^{(3)}(\alpha_i)\right]^4}{4!(3!)^4}
\la \left({\rm tr} \phi^3\right)^4\ra^{M_i}_G
\right) + \right.\nn \\
\left. +
 \left(-\frac{W^{(8)}(\alpha_i)}{8!}\la {\rm tr} \phi^8\ra^{M_i}_G
+ \frac{W^{(3)}(\alpha_i)W^{(7)}(\alpha_i)}{3!7!}
\la {\rm tr} \phi^3\ {\rm tr} \phi^7\ra^{M_i}_G + \right.\right.\nn\\ \left.\left.
+ \frac{W^{(4)}(\alpha_i)W^{(6)}(\alpha_i)}{4!6!}
\la {\rm tr} \phi^4\ {\rm tr} \phi^6\ra^{M_i}_G
+ \frac{\left[W^{(5)}(\alpha_i)\right]^2}{2!(5!)^2}
\la \left({\rm tr} \phi^5\right)^2\ra^{M_i}_G   - \right.\right.\nn\\ \left.\left.
- \frac{\left[W^{(3)}(\alpha_i)\right]^2W^{(6)}(\alpha_i)}{2!(3!)^2 4!}
\la \left({\rm tr} \phi^3\right)^2\ {\rm tr} \phi^6\ra^{M_i}_G -
\frac{\left[W^{(4)}(\alpha_i)\right]^3}{3!(4!)^3}
\la \left({\rm tr} \phi^4\right)^3 \ra^{M_i}_G
+ \right.\right.\nn\\ \left.\left. +
\frac{\left[W^{(3)}(\alpha_i)\right]^3W^{(5)}(\alpha_i)}{3!(3!)^3 5!}
\la \left({\rm tr} \phi^3\right)^3\ {\rm tr} \phi^5\ra^{M_i}_G
\right) + \  \ldots\  \right]\left.Z_G^{M_i}(t|N_i)\right|_{t=0}
\label{Oi0}
\ee

\bigskip

$\bullet$
This formal series becomes ill-defined when some $\alpha_{ij} = 0$,
i.e. when we consider an expansion near the degenerate extremum $\alpha_i =
\alpha_j$. In such cases (for such $W(z)$) one should develop a
different, essentially non-Gaussian formalism, involving expressions
like (for $N=1$)\footnote{The contours in the sum depend on $p$ due to the
change of integration variable which depends on $p$.}
\be
\int_C d\phi \exp \left(-h\phi^p + \sum_{k=1}^\infty t_k\phi^k\right) =
\sum_{k\geq 0} \frac{Sh_k(t)}{p\ h^{\frac{k+1}{p}}}
\Gamma_{C(p)}\left(\frac{k+1}{p}\right)
\ee
where $Sh_k(t)$ are Shur polynomials,
\be
\exp \left(\sum_{k=1}^\infty t_k\phi^k\right) = \sum_{k=0}^\infty
Sh_k(t) \phi^k
\ee
and the Gamma-function for the chain $C$ is defined as
\be
\Gamma_{C}(z) = \int_{C} \phi^{z-1}e^{-\phi}d\phi
\ee

Generalization of this calculation for $N\neq 1$ is essentially
provided by the non-Gaussian multidensities $\rho$ for $W(z) =
W_p(z) = hz^p$.
Note only that association of chains $C$ with the function
$F_{W_p}$ and the polynomials $f_{W_p}$ is somewhat non-trivial:
for example, a naive choice of $f^{(0|1)} = \nu = const$ would imply that
the correlators
\be
\la {\rm tr}\phi^k \ra^{(0)}_{W_p} =
\oint_\infty z^k\rho^{(0|1)}_{W_p}(z|f = const) dz
= -\frac{1}{2}\oint_\infty z^k \sqrt{p^2z^{2p-2} - 4S} dz
\ee
do not vanish only if $k+1$ is an {\it odd} multiple of $p-1$,
i.e. if $k = (2n-1)(p-1)-1$, while for $N=1$ it should be
equal to $\Gamma_C\left(\frac{k+1}{p}\right)$.
(In the Gaussian case $p=2$, this requirement for $C$ is that
$\Gamma_C(z) \neq 0$ only for half-integer, but not for integer
arguments.) Note also that, for $p\neq 2$, the choice of
$f^{(0|1)} = \nu = const$ is in no way distinguished, since
there is no $M$ and thus no natural definition for $\nu$
(which is equal to $gN/M$ in the Gaussian case).

In what follows we consider the non-degenerate situation, when all
$\alpha_{ij}\neq 0$.

\bigskip

$\bullet$
Even if one needs the decomposition formula (\ref{factfor})
at $t=0$, the terms like $\frac{W^{(3)}(\alpha_i)}{3!}\frac{\partial}
{\partial t_3}$ in $\hat{\cal O}_W(\alpha_i)$, which do not contribute
to (\ref{Oi0}), should be kept, since they can combine with
$\hat{\cal O}_{ij}$ and give non-vanishing contributions even at $t=0$.

For explicit calculations, we introduce a {\it double} gradation
of the prepotentials: by powers of $g$ and $S$,
\be
{\cal F} = \sum_{p\geq 0}{\cal F}^{(p)} =
\sum_{\stackrel{p\geq 0}{m\geq 1}} {\cal F}^{(p|m)}+ \widetilde{\cal F} =
\sum_{m\geq 1}{\cal F}_m+ \widetilde{\cal F}
\ee
where ${\cal F}^{(p|m)}$ and ${\cal F}_m$
denote contributions with the $m$-th powers of $S$'s of
genus $p$ and of all genera respectively. $\widetilde{\cal F}$ is
the contribution singular in
$S$, which can be extracted from \ref{volume}.
Explicit expressions (with finitely many terms) can be
found for every particular $\left.{\cal F}^{(p|m)}\right|_{t=0}$.
The decomposition formulas (\ref{factfor}) express
$\left.{\cal F}_{DV}^{(p|m)}\right|_{t=0}$
through ${\cal F}_G^{(p'|m')}$ with $p'\leq p$, $m'\leq m$,
and some $t\neq 0$: at the end, one should put the lowest times
\be
t_k^{(i)} = -T^{(i)}_k = -\frac{M_i\sigma^{(i)}_{k-2}}{k}, \ \ \
3\leq k \leq n+1
\ee
but, before doing this, one should take some $t$-derivatives.

The decomposition formula (\ref{factfor}) states that
\be
e^{(1/g^2)\left.{\cal F}_{DV}\right|_{t=0}} =
e^{(1/g^2){\cal F}_{DV}^{pert}}
\left.
e^{-\sd} e^{g^2\cdd} e^{(1/g^2) F_G}
\right|_{t_k^{(i)} = -T^{(i)}_k}
\label{DVGa}
\ee
where the ``normalization" part of the prepotential
\be
e^{(1/g^2){\cal F}_{DV}^{pert}} = \frac{1}
{\prod_i {\rm Vol}\ U(S_i/g)} \prod_i
\frac{e^{-S_iW(\alpha_i)}}{M_i^{S_i^2/2g^2}}
\prod_{i<j} \alpha_{ij}^{2S_iS_j/g^2}
\ee
the sum of the Gaussian prepotentials
\be
F_G = \sum_{m\geq 1} F_m = \sum_{i=1}^n {\cal F}_G^{M_i}(t^{(i)})
\ee
and the operators
\be
\sd = \sum_{i\neq j}^n \left(\sum_{k\geq 1}^\infty
\frac{2S_i}{k\alpha_{ij}^k}
\frac{\partial}{\partial t^{(j)}_k}\right),
\nn \\
\cdd = \sum_{i\neq j}^n \left(\sum_{k,l\geq 1}^\infty
\frac{(-)^{k+1}(k+l-1)!}{k!l!\alpha_{ij}^{k+l}}
\frac{\partial}{\partial t^{(i)}_k}
\frac{\partial}{\partial t^{(j)}_l}\right)
\ee
Note that $\cdd$ annihilates any single $F_m$, it acts non-trivially
only on bilinear and higher-order combinations of $F$'s.

Substituting (see \cite{MMDV})
\be
\log \left({\rm Vol}\
U(S/g)\right)=-\frac{S^2}{2g^2}\log(\frac{S}{g})+\frac{1}{12}\log(\frac{S}{g})+\frac{3S^2}{4g^2}
-\frac{S}{2g}\log(2\pi)-\zeta'(-1)-
\sum_{k=2}^\infty\frac{B_{2k}}{4k(k-1)}\frac{g^{2k-2}}{S^{2k-2}}
\ee
one straightforwardly obtains from (\ref{DVGa})  (by
expanding exponentials at both sides and comparing the terms
with the same powers of $S$ and $g^2$)
\be
\left.{\cal F}^{DV}_1\right|_{t=0} = - \sum_i S_i W(\alpha_i) +
\left.F_1\right|_{t_k^{(i)} = -T^{(i)}_k}, \nn \\
\left.{\cal F}^{DV}_2\right|_{t=0} =
-\frac{1}{2}\sum_{i<j} (S_i^2 - 4S_iS_j + S_j^2)\log\alpha_{ij} -
\frac{3}{4}\sum_{i}S_i^2+\\
+\left[F_2 - \sd F_1 + \frac{1}{2}
\sum_{k\geq 1}^\infty
\frac{g^{2k-2}}{k!}\cdd^k F_1^2
\right]_{t_k^{(i)} = -T^{(i)}_k}, \\
\left.{\cal F}^{DV}_3\right|_{t=0} =
\left[F_3 - \sd F_2 + \frac{1}{2}\sd^2 F_1 +
\phantom{\sum_{k\geq 1}^\infty}
\right.\nn \\ \left.+
\sum_{k\geq 1}^\infty
\frac{g^{2k-2}}{k!}\cdd^k \left[F_1F_2 - F_1\sd F_1\right] +
\frac{1}{6}\sum_{k\geq 2}^\infty
\frac{g^{2k-4}}{k!}\left(\cdd^k F_1^3 - 3F_1\cdd^k F_1^2\right)
\right]_{t_k^{(i)} = -T^{(i)}_k}, \nn \\
\ldots
\label{Fexpan}
\ee
More terms of the expansion can be found in the Tables.

In the sums still remaining in (\ref{Fexpan}),
only finitely many terms contribute to
$\left.{\cal F}_{DV}^{(p|m)}\right|_{t=0}$,
with any given $p$ and $m$, and one should also take
into account the non-trivial $g^2$-dependence of
$F_m = \sum_i F_m(S_i|t^{(i)})$
\be
F_m(S|t) = S^m \sum_{p\geq 0}^\infty g^{2p} \left[
\sum_{s=1}^\infty \sum_{\stackrel{k_1<\ldots<k_s}{l_1,\ldots,l_s}}^\infty
\frac{t_{k_1}^{l_1}\ldots t_{k_s}^{l_s}}{l_1!\ldots l_s!}
\lla k_1^{l_1},\ldots,k_s^{l_s}\rra^{(p)} \right.\nn\\ \left.
\frac{1}{M^{\frac{1}{2}(l_1k_1 + \ldots l_sk_s)}}\
\delta\left(\frac{1}{2} \sum_{j=1}^s l_j(k_j-2) = m+2p-2\right)\right]
\label{Fm}
\ee
where we introduced a condensed notation for combinatorial quantities
(calculating the numbers of the relevant fat graphs)
\be
\lla k_1^{l_1},\ldots,k_s^{l_s}\rra^{(p)} =
M^{\frac{1}{2}(l_1k_1 + \ldots l_sk_s)}
\lla \left({\rm tr}\phi^{k_1}\right)^{l_1}
\ldots  \left({\rm tr}\phi^{k_s}\right)^{l_s}\rra_G^{(p)}
\label{cono}
\ee
The delta-function in (\ref{Fm}) restricts the possible
number of non-vanishing $t_k = T_k$ (with $3\leq k \leq n+1$)
for given $m,p$, $t_1$ and $t_2$.
However, since at the end of the day, we are going to put $t_1=0$ and
$t_2=0$, they should be eliminated by applying the operators
$\sd$ and $\cdd$, which contribute extra powers of $S$ and $g^2$
respectively. Thus, only finitely many terms in the expansion
(\ref{Fm}) contribute to a given
$\left.{\cal F}_{DV}^{(p|m)}\right|_{t=0}$.
A useful way to study these selection rules is to introduce
the notation ${\cal N}_{k_j}$ for $l_j$ in (\ref{cono}).
Then, the delta-function in (\ref{Fm}) imposes the constraint
\be
\sum_{k=3}^\infty (k-2){\cal N}_k = {\cal N}_1 + 2m+4p-4
\ee
for the correlator
\be
\lla\{{\cal N}_1,{\cal N}_2,\ldots\}\rra =
\lla \left({\rm tr}\phi\right)^{{\cal N}_1}
\left({\rm tr}\phi^2\right)^{{\cal N}_2} \ldots \rra
\ee
to be non-vanishing.

\bigskip

$\bullet$
Particular example of
${\bf m=1}:$

Since
\be
\left.{\cal F}^{DV}_1\right|_{t=0} + \sum_i S_i W(\alpha_i) =
\left.F_1\right|_{t_k^{(i)} = -T^{(i)}_k}
\ee
and in (\ref{Fm})
\be
\sum_{k=3}^\infty (k-2){\cal N}_k = {\cal N}_1 + 4p-2
\ee
there are no non-vanishing contributions to the r.h.s. for $p = 0$,
while for $p=1$ we have $2{\cal N}_4 + {\cal N}_3 = 2$
(i.e. either ${\cal N}_4 =1$ or ${\cal N}_3 = 2$),
and for $p=2$ we have $6{\cal N}_8 + 5{\cal N}_7 +
4{\cal N}_6 + 3{\cal N}_5 + 2{\cal N}_4 +  {\cal N}_3 = 6$:
\be
\left.{\cal F}_{DV}^{(0|1)}\right|_{t=0} = -\sum_i S_i W(\alpha_i), \nn \\ \nn \\
\left.{\cal F}_{DV}^{(1|1)}\right|_{t=0} =
\sum_i \frac{S_i}{M_i}\left(-\frac{\lla 4^1 \rra^{(1)}}{4}\ \sigma_2^{(i)} +
\frac{\lla 3^2 \rra^{(1)}}{2! \cdot 3^2}
\left[\sigma_1^{(i)}\right]^2\right) =
\nn \\ =
\sum_i \frac{S_i}{M_i}\left(-\frac{\sigma_2^{(i)}}{4} +
\frac{3\left[\sigma_1^{(i)}\right]^2}{2!\cdot 3^2}\right)
= \sum_i \frac{S_i}{12M_i}\left(2\left[\sigma_1^{(i)}\right]^2 -
3\sigma_2^{(i)}\right),\\
\ldots
\ee

See Tables for extra explicit expressions.

\newpage

\part{Tables}
\setcounter{section}{0}
\section{Gaussian partition function}
\setcounter{equation}{0}

\subsection{First generalized Catalan numbers}

The number of diagrams of different genera in the ``leaf" graph
$\langle {\rm Tr}\phi^{2k}\rangle$

\be
\begin{array}{ccccccccc}
2k & total \# & p & 0 & 1 & 2& 3& 4& 5 \\
 & (2k-1)!!   &         &   &   &  &  &  &   \\
&&&   &   &  &  &  &   \\
0 &  1                                                && 1 & - & - & - & - & - \\
2 &  1                                                && 1 & - & - & - & - & - \\
4 &  3                                                &&   2 & 1  & - & - & - & - \\
6 &  15                                               &&   5 & 10 & - & - & - & - \\
8 &  105                                              &&  14 & 70 & 21& - & - & - \\
10 &  945                                             &&  42 & 420 & 483 & - & - & - \\
12 &  945\cdot 11                                     && 132 & 2310 & 6468 & 1485  & - & - \\
14 &  945 \cdot 11\cdot 13                            && 429 & 12012 & 66066 & 56628 & - & - \\
16 &  945 \cdot 11 \cdot 13\cdot 15                   && 1430 & 60060 & 570570  & 1169740 & 225225  & - \\
18 &   945 \cdot 11 \cdot 13\cdot 15 \cdot 17         && 4862 & 291720 & 4390386  & 17454580 & 12317877  & - \\
20 &   945 \cdot 11 \cdot 13\cdot 15 \cdot 17\cdot 19 && 16796 & 1385670 & 31039008 & 211083730 & 351683046 & 59520825  \\
\end{array}\nn \\
\ldots
\ee

\subsection{First Gaussian densities (generating functions
of the generalized Catalan numbers); $\nu=gN/M$ \label{G1pf}}

$\bullet$
{\bf Genus $p=0$:}
\be
\frac{1}{M}\rho^{(0|1)}(z) = \frac{z - y(z)}{2} =
\sum_{k=0}^\infty \frac{4^k\Gamma(k+1/2)}{\Gamma(1/2)(k+1)!}
\frac{\nu^{k+1}}{z^{2k+1}}
\label{rho01}
\ee
where
\be
y^2(z) = y_G^2(z) = z^2 - 4\nu, \nn \\
\nu = \frac{gN}{M} = \frac{S}{M}
\label{yGa}
\ee

$\bullet$
{\bf Genus $p=1$:}
\be
M\rho^{(1|1)}(z) = \frac{\nu}{y^5(z)}  =
\sum_{k=0}^\infty \frac{4^k\Gamma(k+5/2)}{\Gamma(5/2)k!}
\frac{\nu^{k+1}}{z^{2k+5}}
\label{rho11}
\ee

$\bullet$
{\bf Genus $p=2$:}
\be
M^3\rho^{(2|1)}(z) = \frac{21 \nu\left(z^2 + \nu\right)}
{y^{11}(z)} =
21\sum_{k=0}^\infty \left(
\frac{4^k\Gamma(k+11/2)}{\Gamma(11/2)k!} +
\frac{4^{k-1}\Gamma(k+9/2)}{\Gamma(11/2)(k-1)!}
\right)\frac{\nu^{k+1}}{z^{2k+9}}=\nn \\ =
\frac{\nu }{z^9} + \frac{14}{3}\sum_{k=1}^\infty
\frac{4^{k-1}\Gamma(k+9/2)(5k+18)}{\Gamma(9/2)k!}
\frac{\nu^{k+1}}{z^{2k+9}}
\label{rho21}
\ee

$\bullet$
{\bf Genus $p=3$:}
\be
M^5\rho^{(3|1)}(z) = \frac{11\nu\left(135 z^4 + 558\nu z^2 + 158 \nu^2\right)}
{y^{17}(z)} = \nn \\ =
11\sum_{k=0}^\infty \left(
135 \cdot \frac{4^k\Gamma(k+17/2)}{\Gamma(17/2)k!} +
558 \cdot \frac{4^{k-1}\Gamma(k+15/2)}{\Gamma(17/2)(k-1)!}
+ 158 \cdot \frac{4^{k-2}\Gamma(k+13/2)}{\Gamma(17/2)(k-2)!}
\right)\frac{\nu^{k+1}}{z^{2k+13}}=\nn \\ =
\frac{\nu }{z^{13}} + 
\frac{\nu^2 }{z^{15}} +
\frac{44}{13\cdot 15}\sum_{k=2}^\infty
\frac{4^{k-2}\Gamma(k+13/2)\left(
ak^2+bk+c
\right)}{\Gamma(13/2)k!}\frac{\nu^{k+1}}{z^{2k+13}}
\label{rho31}
\ee
\be
ak^2+bk+c =
135\cdot 16 (k+13/2)(k+15/2) + 558\cdot 4 k(k+13/2) + 158 k(k-1)
= \nn \\
=130(35k^2+343 k+810)
\nn
\ee
\be \ldots \nn \ee

$\bullet$
{\bf Generic genus $p>0$:}

In general,
\be
M^{2p-1}\cdot\rho^{(p|1)}(z) = \frac{\nu Q_{p-1}(z^2|\nu)}
{y^{6p-1}(z)}
\label{rhop1}
\ee
where
\be
Q_{p-1}(z^2|\nu) = \frac{(4p-1)!!}{2p+1}z^{2p-2} + \ldots
\ee
is a polynomial of degree $p-1$ in $z^2$,
\be
Q_0(x^2|\nu)=\nu\nn\\
Q_1(x^2|\nu)=21(x^2+\nu)\nn\\
Q_2(x^2|\nu)=11(135x^4+558x^2\nu+158\nu^2)\nn\\
Q_3(x^2|\nu)=143(1575x^6+13689x^4\nu+18378x^2\nu^2+2339\nu^3)\nn\\
Q_4(x^2|\nu)=88179(675x^8+9660x^6\nu+28764x^4\nu^2+18908x^2\nu^3+1354\nu^4)\nn\\
Q_5(x^2|\nu)=111435(218295x^{10}+4534875x^8\nu+
23001156x^6\nu^2+34604118x^4\nu^3\nn\\+13447818x^2\nu^4+617926\nu^5)
\ee

\subsection{First recurrent relations for Gaussian {\it multi}densities
\label{Grecrel}}

Some recurrent relations for the Gaussian multidensities
\be
y(z_1)\hat\rho^{(0|2)}(z_1,z_2) =
\partial_{z_2}\frac{\hat\rho^{(0|1)}(z_1)-\hat\rho^{(0|1)}(z_2)}
{z_1-z_2}
\ee
\be
y(z_1)\hat\rho^{(1|1)}(z_1) =
\hat\rho^{(0|2)}(z_1,z_1)
\ee
\be \ldots \nn \ee
\be
y(z_1)\hat\rho^{(0|3)}(z_1,z_2,z_3) =
2\hat\rho^{(0|2)}(z_1,z_2)\hat\rho^{(0|2)}(z_1,z_3) + \nn \\ +
\partial_{z_2}\frac{\hat\rho^{(0|2)}(z_1,z_3)-\hat\rho^{(0|2)}(z_2,z_3)}
{z_1-z_2} +
\partial_{z_3}\frac{\hat\rho^{(0|2)}(z_1,z_2)-\hat\rho^{(0|2)}(z_2,z_3)}
{z_1-z_3}
\ee
\be
y(z_1)\hat\rho^{(1|2)}(z_1,z_2) =
2\hat\rho^{(0|2)}(z_1,z_2)\hat\rho^{(1|1)}(z_1) +
\hat\rho^{(0|3)}(z_1,z_1,z_2) +
\partial_{z_2}\frac{\hat\rho^{(1|1)}(z_1)-\hat\rho^{(1|1)}(z_2)}
{z_1-z_2}
\ee
\be
y(z_1)\hat\rho^{(2|1)}(z_1) =
\left(\hat\rho^{(1|1)}(z_1)\right)^2
+ \hat\rho^{(1|2)}(z_1,z_1)
\ee
\be \ldots \nn \ee

\subsection{First Gaussian {\it multi}densities: analytic expressions;
$\nu=gN/M$; $y(z)=z^2-4\nu$
\label{Gmpf}}

The first Gaussian {\it multi}densities (one-, two-, three-, \ldots point
functions) are:

$\bullet$
{\bf Genus $p=0$:}
\be
\frac{1}{M}\rho^{(0|1)}(z) = \frac{z - y(z)}{2}
\ee
\be
\rho^{(0|2)}(z_1,z_2) = \frac{1}{2(z_1-z_2)^2}\left(
\frac{z_1z_2-4\nu}{y(z_1)y(z_2)} - 1\right)
\label{rho02}
\ee
there is no singularity at $z_1=z_2$, for coinciding variables the genus
zero two-point function
\be
\rho^{(0|2)}(z,z) = \frac{\nu}{y^4(z)}
\ee
\be
M\rho^{(0|3)}(z_1,z_2,z_3) = \frac{2\nu}{y(z_1)y(z_2)y(z_3)}
\left(
\frac{1}{(z_1-z_2)(z_1-z_3)y^2(z_1)} + \right. \nn \\ \left. +
\frac{1}{(z_2-z_1)(z_2-z_3)y^2(z_2)} +
\frac{1}{(z_3-z_1)(z_3-z_2)y^2(z_3)}
\right) =
\frac{2\nu (z_1z_2 + z_2z_3 + z_3z_1 + 4\nu)
}{y^3(z_1)y^3(z_2)y^3(z_3)}
\label{rho03}
\ee
\be
\label{rho04}
M^2\rho^{(0|4)}(z_1,z_2,z_3,z_4)=\frac{\nu}{y(z_1)^5y(z_2)^5y(z_3)^5y(z_4)^5}
\times\\\times\Big(
2\big[4z_1^2z_2^2z_3^2z_4^2(z_1z_2+z_1z_3+z_1z_4+z_2z_3+z_2z_4+z_3z_4)\nn\\
+3z_1z_2z_3z_4(z_1^2z_2^2z_3^2+z_1^2z_2^2z_4^2+z_1^2z_3^2z_4^2+z_2^2z_3^2z_4^2)\big]\nn\\
-8\big[(z_1^2z_2^2z_3^2+z_1^2z_2^2z_4^2+z_1^2z_3^2z_4^2+z_2^2z_3^2z_4^2)(z_1z_2+z_1z_3+z_1z_4+z_2z_3+z_2z_4+z_3z_4)\nn\\
+6z_1z_2z_3z_4(z_1^2z_2^2+z_1^2z_3^2+z_1^2z_4^2+z_2^2z_3^2+z_2^2z_4^2+z_3^2z_4^2)-6z_1^3z_2^3z_3^3z_4^3\big]\nu\nn\\
-32\big[2(z_1^2z_2^2+z_1^2z_3^2+z_1^2z_4^2+z_2^2z_3^2+z_2^2z_4^2+z_3^2z_4^2)(z_1z_2+z_1z_3+z_1z_4+z_2z_3+z_2z_4+z_3z_4)\nn\\
+3(z_1^2z_2^2z_3^2+z_1^2z_2^2z_4^2+z_1^2z_3^2z_4^2+z_2^2z_3^2z_4^2)-9z_1z_2z_3z_4(z_1^2+z_2^2+z_3^2+z_4^2)\big]\nu^2\nn\\
+128\big[5(z_1^2+z_2^2+z_3^2+z_4^2)(z_1z_2+z_1z_3+z_1z_4+z_2z_3+z_2z_4+z_3z_4)-12z_1z_2z_3z_4\big]\nu^3\nn\\
+512\big[3(z_1^2+z_2^2+z_3^2+z_4^2)-8(z_1z_2+z_1z_3+z_1z_4+z_2z_3+z_2z_4+z_3z_4)\big]\nu^4-12288\nu^5\Big)
\ee

$\bullet$
{\bf Genus $p=1$:}
\be
M\rho^{(1|1)}(z) = \frac{\nu}{y^5(z)}
\ee
\be
M^2 \rho^{(1|2)}(z_1,z_2) = \frac{\nu}{y^7(z_1)y^7(z_2)}
\Big(
z_1z_2(5z_1^4 + 4z_1^3z_2 + 3z_1^2z_2^2 + 4z_1z_2^3 + 5z_2^4) +
 \nn \\  +
4\nu\left[z_1^4 - 13z_1z_2(z_1^2 + z_1z_2 + z_2^2) + z_2^4\right] +
16\nu^2(-z_1^2 + 13z_1z_2 - z_2^2) + 320\nu^3
\Big)
\label{rho12}
\ee
\be
M^3 \rho^{(1|3)}(z_1,z_2,z_3) =\frac{\nu}{y(z_1)^9y(z_2)^9y(z_3)^9}\nn\\
\Big(6\big[5z_1^2z_2^2z_3^2(z_1^5z_2^5+z_1^5z_3^5+z_2^5z_3^5+
z_1z_2z_3(z_1^3z_2^4+z_1^4z_2^3+z_1^3z_3^4+z_1^4z_3^3+z_2^3z_3^4+z_2^4z_3^3))\nn\\
+4z_1^4z_2^4z_3^4(z_1z_2^3+z_1^3z_2+z_1z_3^3+z_1^3z_3+z_2z_3^3+z_2^3z_3+z_1^2z_2^2+z_1^2z_3^2+z_2^2z_3^2)\nn\\
+3z_1^5z_2^5z_3^5(z_1+z_2+z_3)\big]
+4\big[5(z_1z_2+z_1z_3+z_2z_3)(z_1^6z_2^6+z_1^6z_3^6+z_2^6z_3^6)\nn\\
-108z_1^2z_2^2z_3^2(z_1^5z_2^3+z_1^3z_2^5+z_1^5z_3^3+z_1^3z_3^5+z_2^5z_3^3+z_2^3z_3^5+\nn\\
z_1z_2^2z_3^5+z_1z_2^5z_3^2+z_1^2z_2z_3^5+z_1^5z_2z_3^2+z_1^2z_2^5z_3+z_1^5z_2^2z_3)\nn\\
-96z_1^3z_2^3z_3^3(z_1z_2z_3^3+z_1z_2^3z_3+z_1^3z_2z_3+z_1^2z_2^3+z_1^3z_2^2+z_1^2z_3^3+z_1^3z_3^2+z_2^2z_3^3+z_2^3z_3^2)\nn\\
-180z_1^4z_2^4z_3^4(z_1z_2+z_1z_3+z_2z_3)+12z_1^2z_2^2z_3^2(z_1^4z_2^4+z_1^4z_3^4+z_2^4z_3^4)\big]\nu\nn\\
+16\big[-(z_1^6z_2^6+z_1^6z_3^6+z_2^6z_3^6)-21(z_1^5z_2^5(z_1^2+z_2^2)+z_1^5z_3^5(z_1^2+z_3^2)+z_2^5z_3^5(z_2^2+z_3^2)\nn\\
+z_1z_2z_3(z_1^3z_2^6+z_1^6z_2^3+z_1^3z_3^6+z_1^6z_3^3+z_2^3z_3^6+z_2^6z_3^3))\nn\\
-27z_1z_2z_3(z_1^4z_2^5+z_1^5z_2^4+z_1^4z_3^5+z_1^5z_3^4+z_2^4z_3^5+z_2^5z_3^4)\nn\\
-30z_1^2z_2^2z_3^2(z_1^4z_2^2+z_1^2z_2^4+z_1^4z_3^2+z_1^2z_3^4+z_2^4z_3^2+z_2^2z_3^4)+
132z_1^3z_2^3z_3^3(z_1^3+z_2^3+z_3^3)\nn\\
+156z_1^2z_2^2z_3^2(z_1z_2^5+z_1^5z_2+z_1z_3^5+z_1^5z_3+z_2z_3^5+z_2^5z_3)\nn\\
+378z_1^3z_2^3z_3^3(z_1z_2^2+z_1^2z_2+z_1z_3^2+z_1^2z_3+z_2z_3^2+z_2^2z_3)\nn\\
+396z_1^2z_2^2z_3^2(z_1^3z_2^3+z_1^3z_3^3+z_2^3z_3^3)
+234z_1^4z_2^4z_3^4\big]\nu^2\nn\\
+192\big[-(z_1^4z_2^4(z_1^2+z_2^2)+z_1^4z_3^4(z_1^2+z_3^2)+z_2^4z_3^4(z_2^2+z_3^2))\nn\\
+60z_1^2z_2^2z_3^2(z_1^2z_2^2+z_1^2z_3^2+z_2^2z_3^2)
+9(z_1^3z_2^3(z_1^4+z_2^4)+z_1^3z_3^3(z_1^4+z_3^4)+z_2^3z_3^3(z_2^4+z_3^4)\nn\\
+z_1z_2z_3(z_1z_2^6+z_1^6z_2+z_1z_3^6+z_1^6z_3+z_2z_3^6+z_2^6z_3))\nn\\
+27(z_1^5z_2^5+z_1^5z_3^5+z_2^5z_3^5)-156z_1^2z_2^2z_3^2(z_1z_2^3+z_1^3z_2+z_1z_3^3+z_1^3z_3+z_2z_3^3+z_2^3z_3)\nn\\
+48z_1^2z_2^2z_3^2(z_1^4+z_2^4+z_3^4)+13z_1z_2z_3(z_1^2z_2^5+z_1^5z_2^2+z_1^2z_3^5+z_1^5z_3^2+z_2^2z_3^5+z_2^5z_3^2)\nn\\
-144z_1^3z_2^3z_3^3(z_1+z_2+z_3)
+31z_1z_2z_3(z_1^4z_2^3+z_1^3z_2^4+z_1^4z_3^3+z_1^3z_3^4+z_2^4z_3^3+z_2^3z_3^4)\big]\nu^3\nn\\
-256\big[3(z_1^4z_2^4+z_1^4z_3^4+z_2^4z_3^4)+6(z_1^2z_2^2(z_1^4+z_2^4)+z_1^2z_3^2(z_1^4+z_3^4)+z_2^2z_3^2(z_2^4+z_3^4))\nn\\
+129(z_1^3z_2^3(z_1^2+z_2^2)+z_1^3z_3^3(z_1^2+z_3^2)+z_2^3z_3^3(z_2^2+z_3^2))\nn\\
+52z_1z_2z_3(z_1^5+z_2^5+z_3^5)+46(z_1z_2(z_1^6+z_2^6)+z_1z_3(z_1^6+z_3^6)+z_2z_3(z_2^6+z_3^6))\nn\\
+378z_1^2z_2^2z_3^2(z_1z_2+z_1z_3+z_2z_3)
+153z_1z_2z_3(z_1^2z_2^3+z_1^3z_2^2+z_1^2z_3^3+z_1^3z_3^2+z_2^2z_3^3+z_2^3z_3^2)\nn\\
+540z_1^2z_2^2z_3^2(z_1^2+z_2^2+z_3^2)
+135z_1z_2z_3(z_1z_2^4+z_1^4z_2+z_1z_3^4+z_1^4z_3+z_2z_3^4+z_2^4z_3)\big]\nu^4
\\+1024\big[-10(z_1^6+z_2^6+z_3^6)+75(z_1^2z_2^2(z_1^2+z_2^2)+z_1^2z_3^2(z_1^2+z_3^2)+z_2^2z_3^2(z_2^2+z_3^2))\nn\\
+174(z_1z_2(z_1^4+z_2^4)+z_1z_3(z_1^4+z_3^4)+z_2z_3(z_2^4+z_3^4))+237(z_1^3z_2^3+z_1^3z_3^3+z_2^3z_3^3)\nn\\
249z_1z_2z_3(z_1^2z_2+z_1z_2^2+z_1^2z_3+z_1z_3^2+z_2^2z_3+z_2z_3^2)+186z_1z_2z_3(z_1^3+z_2^3+z_3^3)+828z_1^2z_2^2z_3^2\big]\nu^5\nn\\
-12288\big[80(z_1z_2(z_1^2+z_2^2)+z_1z_3(z_1^2+z_3^2)+z_2z_3(z_2^2+z_3^2))-12(z_1^4+z_2^4+z_3^4)+
\ee
\be
+39(z_1^2z_2^2+z_1^2z_3^2+z_2^2z_3^2)+82z_1z_2z_3(z_1+z_2+z_3)\big]\nu^6\nn\\
+344064\big[-2(z_1^2+z_2^2+z_3^2)+7(z_1z_2+z_1z_3+z_2z_3)\big]\nu^7+3342336\nu^8\Big)
\label{rho13}
\ee

$\bullet$
{\bf Genus $p=2$:}
\be
M^3\rho^{(2|1)}(z) = \frac{21\nu\left(z^2 + \nu\right)}
{y^{11}(z)}
\ee
\be
M^4\rho^{(2|2)}(z_1,z_2) =
\frac{\nu}{y(z_1)^{13}y(z_2)^{13}}
\Big(
189z_1^{11}z_2^3+168z_1^{10}z_2^4+147z_1^9z_2^5+156z_1^8z_2^6+165z_1^7z_2^7+\nn\\
156z_1^6z_2^8+147z_1^5z_2^9+168z_1^4z_2^{10}+189z_1^3z_2^{11}\nn\\
+\big[399z_1^{11}z_2+462z_1^{10}z_2^2-4011z_1^9z_2^3-3664z_1^8z_2^4-3317z_1^7z_2^5-3342z_1^6z_2^6\nn\\
-3317z_1^5z_2^7-3664z_1^4z_2^8-4011z_1^3z_2^9+462z_1^2z_2^{10}+399z_1z_2^{11}\big]\nu\nn\\
+big[84z_1^{10}-9408z_1^9z_2-10516z_1^8z_2^2+33736z_1^7z_2^3+30980z_1^6z_2^4+27624z_1^5z_2^5\nn\\
+30980z_1^4z_2^6+33736z_1^3z_2^7-10516z_1^2z_2^8-9408z_1z_2^9+84z_2^{10}\big]\nu^2\nn\\
+\big[-1696z_1^8+92368z_1^7z_2+101120z_1^6z_2^2-129648z_1^5z_2^3-133248z_1^4z_2^4\nn\\
-129648z_1^3z_2^5+101120z_1^2z_2^6+92368z_1z_2^7-1696z_2^8\big]\nu^3\nn\\
+\big[12992z_1^6-487936z_1^5z_2-491072z_1^4z_2^2+202752z_1^3z_2^3-491072z_1^2z_2^4
-487936z_1z_2^5+12992z_2^6\big]\nu^4\nn\\
+\big[-77312z_1^4+1415936z_1^3z_2+2519552z_1^2z_2^2+1415936z_1z_2^3-77312z_2^4\big]\nu^5\nn\\
+\big[-560128z_1^2-2951168z_1z_2-560128z_2^2\big]\nu^6-1736704\nu^7\Big)
\ee

$\bullet$
{\bf Genus $p=3$:}
\be
M^5\rho^{(3|1)}(z) =
\frac{11\nu(135z_1^4+558z_1^2\nu+158\nu^2)}{y(z)^{17}}
\ee

Note that all these multidensities are defined for any finite value of $N$
and have nothing to do with the limit $N\rightarrow\infty$. See also
\cite{moments} for an alternative approach to the similar problem.

\subsection{First Gaussian {\it multi}densities: first terms of $N$-expansions
\label{Gmpfe}}

In this section $M=1$ and $g=1$ so that $\nu = N$.

$\bullet$
{\bf One-point functions:}
\be
\rho^{(0|1)}_G(x) =\frac{1}{x}N+\frac{1}{x^3}N^2+\frac{2}{x^5}N^3+\frac{5}{x^7}N^4+\frac{14}{x^9}N^5
+\frac{42}{x^{11}}N^6+\frac{132}{x^{13}}N^7+\frac{429}{x^{15}}N^8+\frac{1430}{x^{17}}N^9+O(N^{10})
\ee
\be
\rho^{(1|1)}_G(x)=\frac{1}{x^5}N+\frac{10}{x^7}N^2+\frac{70}{x^9}N^3+\frac{420}{x^{11}}N^4+\frac{2310}{x^{13}}N^5\nn\\
+\frac{12012}{x^{15}}N^6+\frac{60060}{x^{17}}N^7+\frac{291720}{x^{19}}N^8+\frac{1385670}{x^{21}}N^9+O(N^{10})
\ee
\be
\rho^{(2|1)}_G(x) = \frac{21}{x^9}N+\frac{483}{x^{11}}N^2+\frac{6468}{x^{13}}N^3+\frac{66066}{x^{15}}N^4+\frac{570570}{x^{17}}N^5\nn\\
+\frac{4390386}{x^{19}}N^6+\frac{31039008}{x^{21}}N^7+\frac{205633428}{x^{23}}N^8+\frac{1293938646}{x^{25}}N^9+O(N^{10})
\ee
\be
\rho^{(3|1)}_G(x) = \frac{1485}{x^{13}}N+\frac{56628}{x^{15}}N^2+\frac{1169740}{x^{17}}N^3+\frac{17454580}{x^{19}}N^4+\frac{211083730}{x^{21}}N^5\nn\\
+\frac{2198596400}{x^{23}}N^6+\frac{20465052608}{x^{25}}N^7+\frac{174437377400}{x^{27}}N^8+\frac{1384928666550}{x^{29}}N^9+O(N^{10})
\ee

$\bullet$
{\bf Two-point functions:}
\be
\rho^{(0|2)}_G(x,y) =
\frac{1}{2(x-y)^2}\left[-1 +
\left(1 - \frac{4N}{xy}\right)
\left(1+\frac{2N}{x^2} + \frac{6N^2}{x^4} + \frac{20N^3}{x^6} +
\frac{70N^4}{x^8} + \frac{252N^5}{x^{10}} + \ldots \right)
\times \right.\nn \\ \left. \times
\left(1+\frac{2N}{y^2} + \frac{6N^2}{y^4} + \frac{20N^3}{y^6} +
\frac{70N^4}{y^8} + \frac{252N^5}{y^{10}} + \ldots \right)
\right] = \nn \\ \nn =
\frac{N}{x^2y^2} +
N^2\left(\frac{3}{x^4y^2} + \frac{2}{x^3y^3} + \frac{3}{x^2y^4}\right)
+
N^3\left(\frac{10}{x^6y^2} + \frac{8}{x^5y^3} + \frac{12}{x^4y^4}
+ \frac{8}{x^3y^5} + \frac{10}{x^2y^6}\right)
+ \nn \\ +
N^4\left(\frac{35}{x^8y^2} + \frac{30}{x^7y^3} + \frac{45}{x^6y^4}
+ \frac{36}{x^5y^5}
+ \frac{45}{x^4y^6} + \frac{30}{x^3y^7} + \frac{35}{x^2y^8}
\right)
+ \nn \\ +
N^5\left(\frac{126}{x^{10}y^2} + \frac{112}{x^9y^3} + \frac{168}{x^8y^4}
 + \frac{144}{x^7y^5}  + \frac{180}{x^6y^6}  + \frac{144}{x^5y^7}
+ \frac{168}{x^4y^8} + \frac{112}{x^3y^9} + \frac{126}{x^2y^{10}}
\right)
+ O(N^6)
\ee
\be
\rho^{(1|2)}_G(x,y) =
\Big(\frac{5}{x^6y^2}+\frac{4}{x^5y^3}+\frac{3}{x^4y^4}+
\frac{4}{x^3y^5}+\frac{5}{x^2y^6}\Big)N\nn\\
+\Big(\frac{70}{x^8y^2}+\frac{60}{x^7y^3}+\frac{60}{x^6y^4}+\frac{60}{x^5y^5}+
\frac{60}{x^4y^6}+\frac{60}{x^3y^7}+\frac{70}{x^2y^8}\Big)N^2\nn\\
+\Big(\frac{630}{x^{10}y^2}+\frac{560}{x^9y^3}+\frac{630}{x^8y^4}+
\frac{600}{x^7y^5}+\frac{600}{x^6y^6}+\frac{600}{x^5y^7}+
\frac{630}{x^4y^8}+\frac{560}{x^3y^9}+\frac{630}{x^2y^{10}}\Big)N^3\nn\\
+\Big(\frac{4620}{x^{12}y^2}+
\frac{4200}{x^{11}y^3}+\frac{5040}{x^{10}y^4}+\frac{4760}{x^9y^5}+
\frac{4900}{x^8y^6}+\frac{4800}{x^7y^7}+\nn\\
+\frac{4900}{x^6y^8}+\frac{4760}{x^5y^9}+\frac{5040}{x^4y^{10}}+
\frac{4200}{x^3y^{11}}+\frac{4620}{x^2y^{12}}\Big)N^4\nn\\
+\Big(\frac{30030}{x^{14}y^2}+\frac{27720}{x^{13}y^3}+
\frac{34650}{x^{12}y^4}+\frac{32760}{x^{11}y^5}+\frac{34650}{x^{10}y^6}
+\frac{33600}{x^9y^7}+\frac{34300}{y^8x^8}+\nn\\
+\frac{33600}{x^7y^9}+\frac{34650}{x^6y^{10}}+\frac{32760}{x^5y^{11}}
+\frac{34650}{x^4y^{12}}+\frac{27720}{x^3y^{13}}
+\frac{30030}{x^2y^{14}}\Big)N^5+O(N^6)
\ee
\be
\rho^{(2|2)}_G(x,y) =
(\frac{189}{xy^9}+\frac{168}{x^8y^2}+\frac{147}{x^7y^3}+\frac{156}{x^6y^4}+\frac{165}{x^5y^5}
+\frac{156}{x^4y^6}+\frac{147}{x^3y^7}+\frac{168}{x^2y^8}+\frac{189}{x^9y})N\nn\\
+(\frac{5313}{x^{11}y}+\frac{4830}{x^{10}y^2}+\frac{4725}{x^9y^3}+\frac{4760}{x^8y^4}+\frac{4795}{x^7y^5}
+\frac{4770}{x^6y^6}\nn\\+\frac{4795}{x^5y^7}+\frac{4760}{x^4y^8}+\frac{4725}{x^3y^9}+\frac{4830}{x^2y^{10}}
+\frac{5313}{xy^{11}})N^2\nn\\
+(\frac{84084}{x^{13}y}+\frac{77616}{x^{12}y^2}+\frac{81774}{x^{11}y^3}+\frac{80640}{x^{10}y^4}+\frac{80640}{x^9y^5}
+\frac{80640}{x^8y^6}\nn\\+\frac{81340}{x^7y^7}+\frac{80640}{x^6y^8}+\frac{80640}{x^5y^9}+\frac{80640}{x^4y^{10}}
+\frac{81774}{x^3y^{11}}+\frac{77616}{x^2y^{12}}+\frac{84084}{xy^{13}})N^3+O(N^4)
\ee
$$\ldots$$

$\bullet$
{\bf Three-point functions:}
\be
\rho^{(0|3)}_G(x,y,z) =
\Big(\frac{2}{x^2y^2z^3}+\frac{2}{x^2y^3z^2}+\frac{2}{x^3y^2z^2}\Big)N\nn\\
+\Big(\frac{12}{x^2y^2z^5}+\frac{12}{x^2y^5z^2}+\frac{12}{x^5y^2z^2}\ \
+\ \ \frac{12}{x^2y^3z^4}+\frac{12}{x^3y^4z^2}+\frac{12}{x^4y^3z^2}\ \ +\nn\\
+\ \ \frac{12}{x^2y^4z^3}+\frac{12}{x^3y^2z^4}+\frac{12}{x^4y^2z^3}\ \
+\ \ \frac{8}{x^3y^3z^3}\Big)N^2\\
+\Big(\frac{60}{x^3y^6z^2}+\frac{60}{x^6y^3z^2}+\frac{60}{z^6x^3y^2}
+\frac{60}{z^3x^2y^6}+\frac{60}{z^3x^6y^2}+\frac{60}{z^6x^2y^3}+
\ee
\be
+\frac{60}{x^7y^2z^2}+\frac{60}{z^7x^2y^2}+\frac{60}{x^2y^7z^2}\ \
+\ \ \frac{72}{z^3x^4y^4}+\frac{72}{z^4x^4y^3}+\frac{72}{z^4x^3y^4}\nn\\
+\frac{72}{x^5y^4z^2}+\frac{72}{z^5x^4y^2}+\frac{72}{z^4x^5y^2}
+\frac{72}{z^4x^2y^5}+\frac{72}{z^5x^2y^4}+\frac{72}{x^4y^5z^2}\nn\\
+\frac{48}{z^3x^3y^5}+\frac{48}{z^5x^3y^3}+\frac{48}{z^3x^5y^3}\Big)N^3\nn\\
+\Big(\frac{280}{z^9x^2y^2}+\frac{280}{x^9y^2z^2}+\frac{280}{x^2y^9z^2}
+\frac{280}{z^3x^8y^2}+\frac{280}{z^8x^3y^2}+\frac{280}{z^3x^2y^8}+\frac{280}{z^8x^2y^3}+\frac{280}{x^8y^3z^2}
+\frac{280}{x^3y^8z^2}\nn\\
+\frac{360}{z^4x^7y^2}+\frac{360}{z^7x^4y^2}+\frac{360}{z^4x^2y^7}+\frac{360}{z^7x^2y^4}+
\frac{360}{x^7y^4z^2}+\frac{360}{x^4y^7z^2}\nn\\
+\frac{360}{z^5x^6y^2}+\frac{360}{z^6x^5y^2}+\frac{360}{z^5x^2y^6}+\frac{360}{z^6x^2y^5}
+\frac{360}{x^5y^6z^2}+\frac{360}{x^6y^5z^2}\nn\\
+\frac{360}{z^4x^6y^3}+\frac{360}{z^3x^6y^4}+\frac{360}{z^6x^4y^3}+\frac{360}{z^3x^4y^6}
+\frac{360}{z^6x^3y^4}+\frac{360}{z^4x^3y^6}\nn\\
+\frac{240}{z^3x^7y^3}+\frac{240}{z^3x^3y^7}+\frac{240}{z^7x^3y^3}
+\frac{288}{z^3x^5y^5}+\frac{288}{z^5x^3y^5}+\frac{288}{z^5x^5y^3}
+\frac{432}{z^4x^5y^4}+\frac{432}{z^4x^4y^5}+\frac{432}{z^5x^4y^4}\Big)N^4\nn\\
+\Big(\frac{1680}{z^3x^8y^4}+\frac{1680}{z^4x^8y^3}+\frac{1680}{z^4x^3y^8}
+\frac{1680}{z^8x^3y^4}+\frac{1680}{z^8x^4y^3}+\frac{1680}{z^3x^4y^8}\nn\\
+\frac{1680}{x^4y^9z^2}+\frac{1680}{z^9x^4y^2}+\frac{1680}{z^4x^2y^9}
+\frac{1680}{z^9x^2y^4}+\frac{1680}{x^9y^4z^2}+\frac{1680}{z^4x^9y^2}\nn\\
+\frac{1680}{z^8x^5y^2}+\frac{1680}{z^5x^2y^8}+\frac{1680}{z^5x^8y^2}
+\frac{1680}{z^8x^2y^5}+\frac{1680}{x^8y^5z^2}+\frac{1680}{x^5y^8z^2}\nn\\
+\frac{1440}{z^3x^7y^5}+\frac{1440}{z^5x^7y^3}+\frac{1440}{z^5x^3y^7}
+\frac{1440}{z^7x^3y^5}+\frac{1440}{z^3x^5y^7}+\frac{1440}{z^7x^5y^3}\nn\\
+\frac{2160}{z^4x^5y^6}+\frac{2160}{z^6x^5y^4}+\frac{2160}{z^5x^6y^4}
+\frac{2160}{z^6x^4y^5}+\frac{2160}{z^4x^6y^5}+\frac{2160}{z^5x^4y^6}\nn\\
+\frac{2160}{z^4x^7y^4}+\frac{2160}{z^4x^4y^7}+\frac{2160}{z^7x^4y^4}\ \
+\ \ \frac{1120}{z^3x^3y^9}+\frac{1120}{z^9x^3y^3}+\frac{1120}{z^3x^9y^3}\nn\\
+\frac{1800}{z^7x^6y^2}+\frac{1800}{z^6x^7y^2}+\frac{1800}{x^7y^6z^2}
+\frac{1800}{z^7x^2y^6}+\frac{1800}{z^6x^2y^7}+\frac{1800}{x^6y^7z^2}\nn\\
+\frac{1800}{z^6x^3y^6}+\frac{1800}{z^3x^6y^6}+\frac{1800}{z^6x^6y^3}\ \
+\frac{1260}{z^{11}x^2y^2}+\frac{1260}{x^{11}y^2z^2}+\frac{1260}{x^2y^{11}z^2}\nn\\
+\frac{1260}{z^3x^{10}y^2}+\frac{1260}{z^{10}x^3y^2}+\frac{1260}{z^3x^2y^{10}}
+\frac{1260}{z^{10}x^2y^3}+\frac{1260}{x^{10}y^3z^2}+\frac{1260}{x^3y^{10}z^2}
+\ \ \frac{1728}{z^5x^5y^5}\Big)N^5+O(N^6)
\ee
\be
\rho^{(1|3)}_G(x,y,z) =
(\frac{24}{x^3zy^4}+\frac{24}{x^4zy^3}+\frac{24}{z^4xy^3}+\frac{24}{z^3x^4y}+\frac{24}{z^4x^3y}+
\frac{24}{z^3xy^4}\nn\\
+\frac{24}{z^2x^2y^4}+\frac{24}{z^2x^4y^2}+\frac{24}{z^4x^2y^2}\ \
+\ \ \frac{18}{z^2x^3y^3}+\frac{18}{z^3x^2y^3}+\frac{18}{z^3x^3y^2}\nn\\
+\frac{30}{x^2zy^5}+\frac{30}{z^5x^2y}+\frac{30}{z^2xy^5}+
\frac{30}{z^5xy^2}+\frac{30}{z^2x^5y}+\frac{30}{x^5zy^2}
+\frac{30}{x^6zy}+\frac{30}{xzy^6}+\frac{30}{z^6xy}
)N\nn\\
+(\frac{540}{x^5zy^4}+\frac{540}{z^4x^5y}+\frac{540}{z^5x^4y}
+\frac{540}{z^5xy^4}+\frac{540}{x^4zy^5}+\frac{540}{z^4xy^5}\nn\\
+\frac{480}{z^2x^5y^3}+\frac{480}{z^3x^5y^2}+\frac{480}{z^5x^3y^2}+
\frac{480}{z^5x^2y^3}+\frac{480}{z^3x^2y^5}+\frac{480}{z^2x^3y^5}+\\
+\frac{540}{z^6x^3y}+\frac{540}{x^3zy^6}+\frac{540}{z^3xy^6}
+\frac{540}{x^6zy^3}+\frac{540}{z^3x^6y}+\frac{540}{z^6xy^3}+
\ee
\be
+\frac{560}{z^7xy^2}+\frac{560}{x^2zy^7}+\frac{560}{z^2xy^7}+\frac{560}{z^2x^7y}+
\frac{560}{z^7x^2y}+\frac{560}{x^7zy^2}\\
+\frac{480}{z^4x^2y^4}+\frac{480}{z^4x^4y^2}+\frac{480}{z^2x^4y^4}\ \
+\ \ \frac{560}{z^8xy}+\frac{560}{xzy^8}+\frac{560}{x^8zy}\nn\\
+\frac{468}{z^4x^3y^3}+\frac{468}{z^3x^3y^4}+\frac{468}{z^3x^4y^3}\ \
+\ \ \frac{480}{z^2x^2y^6}+\frac{480}{z^6x^2y^2}+\frac{480}{z^2x^6y^2})N^2\nn\\
+(\frac{6300}{z^3x^7y^2}+\frac{6300}{z^7x^3y^2}+\frac{6300}{z^2x^7y^3}
+\frac{6300}{z^7x^2y^3}+\frac{6300}{z^3x^2y^7}+\frac{6300}{z^2x^3y^7}\ \
+\ \ \frac{6300}{x^{10}zy}+\frac{6300}{z^{10}xy}+\frac{6300}{xzy^{10}}\nn\\
+\frac{6000}{z^4x^2y^6}+\frac{6000}{z^2x^6y^4}+\frac{6000}{z^6x^2y^4}+
\frac{6000}{z^4x^6y^2}+\frac{6000}{z^2x^4y^6}+\frac{6000}{z^6x^4y^2}\ \
+\ \ \frac{6000}{z^2x^5y^5}+\frac{6000}{z^5x^2y^5}+\frac{6000}{z^5x^5y^2}\nn\\
+\frac{6300}{z^2x^9y}+\frac{6300}{x^9zy^2}+\frac{6300}{x^2zy^9}
+\frac{6300}{z^9xy^2}+\frac{6300}{z^2xy^9}+\frac{6300}{z^9x^2y}\ \
+\ \ \frac{6660}{z^6x^3y^3}+\frac{6660}{z^3x^3y^6}+\frac{6660}{z^3x^6y^3}\nn\\
+\frac{6720}{z^4x^7y}+\frac{6720}{x^7zy^4}+\frac{6720}{z^7xy^4}
+\frac{6720}{z^4xy^7}+\frac{6720}{z^7x^4y}+\frac{6720}{x^4zy^7}\ \
+\ \ \frac{5600}{z^8x^2y^2}+\frac{5600}{z^2x^2y^8}+\frac{5600}{z^2x^8y^2}\nn\\
\frac{6300}{z^2x^9y}+\frac{6300}{x^9zy^2}+\frac{6300}{x^2zy^9}
+\frac{6300}{z^9xy^2}+\frac{6300}{z^2xy^9}+\frac{6300}{z^9x^2y}\ \
+\ \ \frac{5600}{z^8x^2y^2}+\frac{5600}{z^2x^2y^8}+\frac{5600}{z^2x^8y^2}\nn\\
+\frac{6720}{z^8x^3y}+\frac{6720}{x^3zy^8}+\frac{6720}{x^8zy^3}+
\frac{6720}{z^3xy^8}+\frac{6720}{z^8xy^3}+\frac{6720}{z^3x^8y}\nn\\
+\frac{6600}{z^5x^6y}+\frac{6600}{z^6xy^5}+\frac{6600}{z^6x^5y}+
\frac{6600}{z^5xy^6}+\frac{6600}{x^6zy^5}+\frac{6600}{x^5zy^6}\nn\\
+\frac{6480}{z^4x^3y^5}+\frac{6480}{z^4x^5y^3}+\frac{6480}{z^3x^5y^4}+
\frac{6480}{z^3x^4y^5}+\frac{6480}{z^5x^4y^3}+\frac{6480}{z^5x^3y^4}\ \
+\ \ \frac{6336}{z^4x^4y^4}+)N^3+O(N^4)
\ee
$$\ldots$$

$\bullet$
{\bf Four-point functions:}
\be
\rho^{(0|4)}_G(x,y,z,w) =
(
\frac{8}{w^2xy^2z}+\frac{8}{x^2ywz^2}+\frac{8}{w^2x^2yz}+\frac{8}{xy^2wz^2}+\frac{8}{w^2xyz^2}+\frac{8}{x^2y^2wz}\nn\\
+\frac{6}{xy^3wz}+\frac{6}{xywz^3}+\frac{6}{w^3xyz}+\frac{6}{yx^3wz})N\nn\\
+(
\frac{72}{w^3x^2yz^2}+\frac{72}{w^2xy^3z^2}+\frac{72}{w^2xy^2z^3}+\frac{72}{w^3xy^2z^2}+\frac{72}{w^2x^3y^2z}
+\frac{72}{x^3y^2wz^2}\nn\\
+\frac{72}{w^2x^3yz^2}+\frac{72}{w^2x^2y^3z}+\frac{72}{x^2y^3wz^2}+\frac{72}{w^3x^2y^2z}
+\frac{72}{x^2y^2wz^3}+\frac{72}{w^2x^2yz^3}\nn\\
+\frac{72}{x^2ywz^4}+\frac{72}{xy^4wz^2}+\frac{72}{w^2xy^4z}+\frac{72}{w^4xyz^2}+\frac{72}{w^2xyz^4}+\frac{72}{w^2x^4yz}\nn\\
+\frac{72}{w^4xy^2z}+\frac{72}{xy^2wz^4}+\frac{72}{x^4y^2wz}+\frac{72}{x^2y^4wz}+\frac{72}{w^4x^2yz}+\frac{72}{x^4ywz^2}\nn\\
+\frac{72}{w^3xy^3z}+\frac{72}{xy^3wz^3}+\frac{72}{w^3xyz^3}+\frac{72}{x^3y^3wz}+\frac{72}{w^3x^3yz}+\frac{72}{x^3ywz^3}\nn\\
+\frac{60}{yx^5wz}+\frac{60}{xywz^5}+\frac{60}{w^5xyz}+\frac{60}{xy^5wz}\ \
+\ \ \frac{48}{w^2x^2y^2z^2})N^2\\
+(
\frac{576}{w^3x^3y^2z^2}+\frac{576}{w^2x^3y^3z^2}+\frac{576}{w^3x^2y^3z^2}+\frac{576}{w^2x^2y^3z^3}
+\frac{576}{w^2x^3y^2z^3}+\frac{576}{w^3x^2y^2z^3}\nn\\
+\frac{576}{w^3x^2yz^4}+\frac{576}{w^4x^3y^2z}+\frac{576}{x^3y^2wz^4}+\frac{576}{x^4y^2wz^3}
+\frac{576}{w^3x^4y^2z}+\frac{576}{w^2x^4y^3z}+\frac{576}{x^4y^3wz^2}+\frac{576}{w^4xy^2z^3}\nn\\
+\frac{576}{w^4x^3yz^2}+\frac{576}{w^2xy^4z^3}+\frac{576}{x^2y^4wz^3}+\frac{576}{w^3xy^4z^2}
+\frac{576}{w^3x^4yz^2}+\frac{576}{w^2x^4yz^3}+\frac{576}{w^2xy^3z^4}+\frac{576}{w^4xy^3z^2}+
\ee
\be
+\frac{576}{w^3xy^2z^4}+\frac{576}{w^4x^2y^3z}+\frac{576}{w^4x^2yz^3}+\frac{576}{w^2x^3yz^4}
+\frac{576}{w^2x^3y^4z}+\frac{576}{x^3y^4wz^2}+\frac{576}{w^3x^2y^4z}+\frac{576}{x^2y^3wz^4}\nn\\
+\frac{480}{w^2x^6yz}+\frac{480}{xy^6wz^2}+\frac{480}{w^6xy^2z}+\frac{480}{w^6xyz^2}
+\frac{480}{x^6ywz^2}+\frac{480}{x^6y^2wz}+\frac{480}{w^2xyz^6}+\frac{480}{xy^2wz^6}\nn\\
+\frac{480}{w^6x^2yz}+\frac{480}{x^2ywz^6}+\frac{480}{w^2xy^6z}+\frac{480}{x^2y^6wz}\ \
+\ \ \frac{480}{w^5xy^2z^2}+\frac{480}{w^5x^2yz^2}+\frac{480}{x^2y^2wz^5}+\frac{480}{x^2y^5wz^2}\nn\\
+\frac{480}{w^2x^2yz^5}+\frac{480}{w^2x^5yz^2}+\frac{480}{w^5x^2y^2z}+\frac{480}{w^2x^2y^5z}
+\frac{480}{w^2x^5y^2z}+\frac{480}{x^5y^2wz^2}+\frac{480}{w^2xy^5z^2}+\frac{480}{w^2xy^2z^5}\nn\\
+\frac{540}{x^3ywz^5}+\frac{540}{xy^5wz^3}+\frac{540}{w^5x^3yz}+\frac{540}{w^5xyz^3}
+\frac{540}{w^3xy^5z}+\frac{540}{x^5ywz^3}+\frac{540}{x^5y^3wz}+\frac{540}{w^3x^5yz}\nn\\
+\frac{540}{w^5xy^3z}+\frac{540}{x^3y^5wz}+\frac{540}{xy^3wz^5}+\frac{540}{w^3xyz^5}\ \
+\ \ \frac{420}{xy^7wz}+\frac{420}{yx^7wz}+\frac{420}{xywz^7}+\frac{420}{w^7xyz}\nn\\
+\frac{648}{w^3x^3y^3z}+\frac{648}{w^3xy^3z^3}+\frac{648}{x^3y^3wz^3}+\frac{648}{w^3x^3yz^3}\ \
+\ \ \frac{384}{w^2x^2y^2z^4}+\frac{384}{w^4x^2y^2z^2}+\frac{384}{w^2x^4y^2z^2}+\frac{384}{w^2x^2y^4z^2}\nn\\
+\frac{576}{x^4y^4wz}+\frac{576}{w^4x^4yz}+\frac{576}{x^4ywz^4}+\frac{576}{w^4xy^4z}
+\frac{576}{w^4xyz^4}+\frac{576}{xy^4wz^4})N^3+O(N^4)
\ee
$$\ldots$$

\subsection{Recurrent formulas for Gaussian correlators}


$\bullet$
{\bf General recurrent formula for connected correlators
with ${\rm tr}\phi$:}\footnote{
In applications of this formula one should be careful: when some
$i_\alpha-a_\alpha = 0$, while some other $i_\beta-a_\beta>0$,
the corresponding connected correlator at the r.h.s.
$\ldots\la\la {\rm tr}I\ldots {\rm tr}\phi^{i_\beta-a_\beta}\ldots \ra\ra_G
= 0$.
}
\be
\la\la \left({\rm tr}\phi\right)^{k}\
{\rm tr}\phi^{i_1}\ldots {\rm tr}\phi^{i_m} \ra\ra_G = \nn \\ = k!
\sum_{\stackrel{0\leq a_1,\ldots,a_m\leq k}{a_1+\ldots+a_m = k}}
\frac{i_1!}{a_1!(i_1-a_1)!}\ldots
\frac{i_m!}{a_m!(i_m-a_m)!}\
\la\la {\rm tr}\phi^{i_1-a_1}\ldots {\rm tr}\phi^{i_m-a_m} \ra\ra_G
\ee
{\bf Particular examples:}
\be
\la\la \left({\rm tr}\phi\right)^{k} \ra\ra_G = \delta_{k,0} +
N\delta_{k,2}
\ee
\be
\la\la \left({\rm tr}\phi\right)^{k}\ {\rm tr}\phi^2 \ra\ra_G = N^2\delta_{k,0} +
2N\delta_{k,2}
\ee
\be
\la\la \left({\rm tr}\phi\right)^{k}\ {\rm tr}\phi^3 \ra\ra_G = 3N^2\delta_{k,1} +
6N\delta_{k,3}
\ee
\be
\la\la \left({\rm tr}\phi\right)^{k}\ {\rm tr}\phi^4 \ra\ra_G =
(2N^3+N)\delta_{k,0} + 12N^2\delta_{k,2} + 24N\delta_{k,4}
\ee
$$\ldots$$
\be
\la\la \left({\rm tr}\phi\right)^{k}\ {\rm tr}\phi^{k} \ra\ra_G = Nk!
\ee
\be
\la\la \left({\rm tr}\phi\right)^{k}\ {\rm tr}\phi^{m} \ra\ra_G = 0,
\ \ {\rm for} \ k>m
\ee

In general, for the $k+1$-point correlators
\be
\la\la \left({\rm tr}\phi\right)^{k}\ {\rm tr}\phi^{2m} \ra\ra_G =
c_{2m}\delta_{k,0} +
2m(2m-1)c_{2m-2}\delta_{k,2} + \ldots + (2m)!N\delta_{k,2m}
\ee
\be
\la\la \left({\rm tr}\phi\right)^{k}\ {\rm tr}\phi^{2m+1} \ra\ra_G =
(2m+1)c_{2m}\delta_{k,1} +
2m(4m^2-1)c_{2m-2}\delta_{k,3} + \ldots +
(2m+1)!N\delta_{k,2m+1}
\ee
where
$$c_{2m}= \la\la tr\phi^{2k}\ra\ra_G$$

The disconnected correlators
\be
\la \left({\rm tr}\phi\right)^{2k}\ra_G = (2k-1)!!N^k
\ee

$\bullet$
{\bf Gaussian correlators of ${\rm tr}\phi^2$ only}
\be
\la\la \left({\rm tr}\phi^2\right)^k \ra\ra_G =
4(k-1)(k-2)\la\la \left({\rm tr}\phi^2\right)^{k-2} \ra\ra_G =
2^{k-1}(k-1)!N^2
\ee
\be
\la \left({\rm tr}\phi^2\right)^k \ra_G =
N^2\la \left({\rm tr}\phi^2\right)^{k-1} \ra_G
+ 2(k-1)\left[N^2+2(k-2)\right]
\la \left({\rm tr}\phi^2\right)^{k-2} \ra_G = \nn \\ =
\sum_{j=1}^k \frac{1}{j!}
\sum_{\stackrel{0<l_1,\ldots,l_j<k}{l_1+\ldots+l_j=k}}
\frac{k!}{l_1!\ldots l_j!}
\la\la \left({\rm tr}\phi^2\right)^{l_1} \ra\ra_G \ldots
\la\la \left({\rm tr}\phi^2\right)^{l_j} \ra\ra_G =\\=
\sum_{j=1}^k \frac{2^{k-j}k!}{j!} N^{2j}
\sum_{\stackrel{0<l_1,\ldots,l_j<k}{l_1+\ldots+l_j=k}}
\frac{1}{l_1\cdot\ldots\cdot l_j}
\ee
{\bf Particular examples:}
\be
\la {\rm tr}\phi^2 \ra_G = N^2
\ee
\be
\la \left({\rm tr}\phi^2\right)^2 \ra_G = N^4 + 2N^2
\ee
\be
\la \left({\rm tr}\phi^2\right)^3 \ra_G = N^6 + 6N^4 + 8N^2
\ee
\be
\la \left({\rm tr}\phi^2\right)^4 \ra_G =
N^8 + 12N^6 + 44N^4 + 48N^2
\ee
\be
\la \left({\rm tr}\phi^2\right)^5 \ra_G =
N^{10} + 20N^8 + 140N^6 + 400N^4 + 384N^2
\ee
$$\ldots$$

$\bullet$
{\bf Recurrent formulas for correlators with ${\rm tr}\phi^k$
with $k\geq 2$}

These are more sophisticated, from the point of view of $N$
dependence (genus expansion). See (\ref{224}) and (\ref{235})
for particular examples with ${\rm tr}\phi^2$.

\subsection{Lowest Gaussian correlators,
connected and disconnected
\label{correx}}


To restore the $g$ and $M$ dependencies below, each correlator
$$\la{\rm tr}\phi^{k_1}\ldots{\rm tr}\phi^{k_m}\ra$$
should be multiplied by
$$\left(\frac{g}{M}\right)^{\frac{r_1+\ldots+k_m}{2}}$$

$\bullet$
{\bf Selection rules:}

{\bf a)}
for {\it disconnected} correlators for $N=1$
\be
\la {\rm tr}\phi^{i_1}\ldots {\rm tr}\phi^{i_k}\ra_G =
(i_1 + \ldots + i_k - 1)!!
\ee

{\bf b)} \ \ \
$
i_1 + \ldots + i_k = {\rm even}
$

\bigskip

$\bullet$
{\bf Single-point correlators (the generalized Catalan numbers):}
\be
\la {\rm tr}\phi^2 \ra_G =
\la\la {\rm tr}\phi^2 \ra\ra_G = N^2
\ee
\be
\la {\rm tr}\phi^4 \ra_G =
\la\la {\rm tr}\phi^4 \ra\ra_G = 2N^3 + N
\ee
\be
\la {\rm tr}\phi^6 \ra_G =
\la\la {\rm tr}\phi^6 \ra\ra_G = 5N^4 + 10N^2
\ee
\be
\la {\rm tr}\phi^8 \ra_G =
\la\la {\rm tr}\phi^8 \ra\ra_G = 14N^5 + 70N^3 + 21N
\ee
\be
\la {\rm tr}\phi^{10} \ra_G =
\la\la {\rm tr}\phi^{10} \ra\ra_G = 42N^6 + 420N^4 + 483N^2
\ee
$$\ldots$$

$\bullet$
{\bf Two-point correlators:}
\be
\la\la \left({\rm tr}\phi\right)^2 \ra\ra_G =
\la \left({\rm tr}\phi\right)^2 \ra_G = N
\ee
\be
\la\la \left({\rm tr}\phi^2\right)^2 \ra\ra_G = 2N^2
\ee
\be
\la \left({\rm tr}\phi^2\right)^2 \ra_G = N^4 + 2N^2
\ee
\be
\la\la \left({\rm tr}\phi^3\right)^2 \ra\ra_G =
\la \left({\rm tr}\phi^3\right)^2 \ra_G = 12N^3 + 3N
\ee
\be
\la\la \left({\rm tr}\phi^4\right)^2 \ra\ra_G = 36N^4 + 60N^2
\ee
\be
\la \left({\rm tr}\phi^4\right)^2 \ra_G = 4N^6 + 40N^4 + 61N^2
\ee
\be
\la\la \left({\rm tr}\phi^5\right)^2 \ra\ra_G =
\la \left({\rm tr}\phi^5\right)^2 \ra_G = 180N^5 + 600N^3 + 165N
\ee
$$\ldots$$
\be
\la\la {\rm tr}\phi\ {\rm tr}\phi^3 \ra\ra_G =
\la {\rm tr}\phi\ {\rm tr}\phi^3 \ra_G = 3N^2
\ee
\be
\la\la {\rm tr}\phi\ {\rm tr}\phi^5 \ra\ra_G =
\la {\rm tr}\phi\ {\rm tr}\phi^5 \ra_G = 10N^3 + 5N
\ee
\be
\la\la {\rm tr}\phi\ {\rm tr}\phi^7 \ra\ra_G =
\la {\rm tr}\phi\ {\rm tr}\phi^7 \ra_G = 35N^4 + 70N^2
\ee
\be
\la\la {\rm tr}\phi\ {\rm tr}\phi^9 \ra\ra_G =
\la {\rm tr}\phi\ {\rm tr}\phi^9 \ra_G = 126N^5 + 630N^3 + 189N
\ee
$$\ldots$$
\be
\la\la {\rm tr}\phi^2\ {\rm tr}\phi^4 \ra\ra_G = 8N^3 + 4N
\ee
\be
\la {\rm tr}\phi^2\ {\rm tr}\phi^4 \ra_G = 2N^5 + 9N^3 + 4N
\ee
\be
\la\la {\rm tr}\phi^2\ {\rm tr}\phi^6 \ra\ra_G = 30N^4 + 60N^2
\ee
\be
\la {\rm tr}\phi^2\ {\rm tr}\phi^6 \ra_G = 5N^6 + 40N^4 + 60N^2
\ee
\be
\la\la {\rm tr}\phi^2\ {\rm tr}\phi^8 \ra\ra_G = 112N^5 + 560N^3 + 178N
\ee
\be
\la {\rm tr}\phi^2\ {\rm tr}\phi^8 \ra_G = 14N^7 +
182N^5 + 581N^3 + 178N
\ee
$$\ldots$$
\be
\la\la {\rm tr}\phi^3\ {\rm tr}\phi^5 \ra\ra_G =
\la {\rm tr}\phi^3\ {\rm tr}\phi^5 \ra_G = 45N^4 + 60N^2
\ee
\be
\la\la {\rm tr}\phi^3\ {\rm tr}\phi^7 \ra\ra_G =
\la {\rm tr}\phi^3\ {\rm tr}\phi^7 \ra_G = 168N^5 + 630N^3 + 147N
\ee
$$\ldots$$
\be
\la\la {\rm tr}\phi^4\ {\rm tr}\phi^6 \ra\ra_G = 144N^5 + 600N^3 + 156N
\ee
\be
\la {\rm tr}\phi^4\ {\rm tr}\phi^6 \ra_G = 10N^7 + 169N^5 + 610N^3 + 156N
\ee
$$\ldots$$

$\bullet$
{\bf Three-point correlators:}
\be
\la\la \left({\rm tr}\phi^2\right)^3\ra\ra_G = 8N^2
\ee
\be
\la \left({\rm tr}\phi^2\right)^3 \ra_G = N^6 + 6N^4 + 8N^2
\ee
$$\ldots$$
\be
\la\la \left({\rm tr}\phi\right)^2\ {\rm tr}\phi^2 \ra\ra_G = 2N
\ee
\be
\la \left({\rm tr}\phi\right)^2\ {\rm tr}\phi^2 \ra_G = N^3+2N
\ee
\be
\la\la \left({\rm tr}\phi\right)^2\ {\rm tr}\phi^4 \ra\ra_G = 12N^2
\ee
\be
\la \left({\rm tr}\phi\right)^2\ {\rm tr}\phi^4 \ra_G = 2N^4+13N^2
\ee
\be
\la\la \left({\rm tr}\phi\right)^2\ {\rm tr}\phi^6 \ra\ra_G = 60N^3+30N
\ee
\be
\la \left({\rm tr}\phi\right)^2\ {\rm tr}\phi^6 \ra_G = 5N^5+70N^3+30N
\ee
\be
\la\la \left({\rm tr}\phi\right)^2\ {\rm tr}\phi^8 \ra\ra_G = 280N^4+560N^2
\ee
\be
\la \left({\rm tr}\phi\right)^2\ {\rm tr}\phi^8 \ra_G = 14N^6+350N^4+581N^2
\ee
$$\ldots$$
\be
\la\la \left({\rm tr}\phi^2\right)^2\ {\rm tr}\phi^4 \ra\ra_G = 48N^3
+ 24N
\ee
\be
\la \left({\rm tr}\phi^2\right)^2\ {\rm tr}\phi^4 \ra_G =
2N^7 + 21N^5 + 58N^3 + 24N
\ee
\be
\la\la \left({\rm tr}\phi^2\right)^2\ {\rm tr}\phi^6 \ra\ra_G =
240N^4 + 480N^2
\ee
\be
\la \left({\rm tr}\phi^2\right)^2\ {\rm tr}\phi^6 \ra_G =
5N^8 + 80N^6 + 380N^4 + 480N^2
\ee
$$\ldots$$
\be
\la\la \left({\rm tr}\phi^3\right)^2\ {\rm tr}\phi^2 \ra\ra_G =
72N^3 + 18N
\ee
\be
\la \left({\rm tr}\phi^3\right)^2\ {\rm tr}\phi^2 \ra_G =
12N^5 + 75N^3 + 18N
\ee
\be
\la\la \left({\rm tr}\phi^3\right)^2\ {\rm tr}\phi^4 \ra\ra_G =
432N^4 + 468N^2
\ee
\be
\la \left({\rm tr}\phi^3\right)^2\ {\rm tr}\phi^4 \ra_G =
24N^6 + 450N^4 + 471N^2
\ee
$$\ldots$$
\be
\la\la \left({\rm tr}\phi^4\right)^2\ {\rm tr}\phi^2 \ra\ra_G = \nn \\ =
2\left(8N \la\la {\rm tr}\phi^2\ {\rm tr}\phi^4 \ra\ra_G +
4\la\la \left({\rm tr}\phi\right)^2\ {\rm tr}\phi^4 \ra\ra_G\right)
+ 2\cdot 16 \la\la {\rm tr}\phi^6 \ra\ra_G = \nn \\ =
288N^4 + 480N^2
\label{224}
\ee
\be
\la \left({\rm tr}\phi^4\right)^2\ {\rm tr}\phi^2 \ra_G =
36N^6 + 360N^4 + 549N^2
\ee
$$\ldots$$
\be
\la\la {\rm tr}\phi\  {\rm tr}\phi^2\ {\rm tr}\phi^3 \ra\ra_G = 12N^2
\ee
\be
\la {\rm tr}\phi\  {\rm tr}\phi^2\ {\rm tr}\phi^3 \ra_G = 3N^4+12N^2
\ee
\be
\la\la {\rm tr}\phi\  {\rm tr}\phi^2\ {\rm tr}\phi^5 \ra\ra_G = 60N^3+30N
\ee
\be
\la {\rm tr}\phi\  {\rm tr}\phi^2\ {\rm tr}\phi^5 \ra_G = 10N^5+65N^3+30N
\ee
\be
\la\la {\rm tr}\phi\  {\rm tr}\phi^2\ {\rm tr}\phi^7 \ra\ra_G = 280N^4+560N^2
\ee
\be
\la {\rm tr}\phi\  {\rm tr}\phi^2\ {\rm tr}\phi^7 \ra_G = 35N^6+350N^4+560N^2
\ee
$$\ldots$$
\be
\la\la {\rm tr}\phi\  {\rm tr}\phi^3\ {\rm tr}\phi^4 \ra\ra_G = 72N^3+24N
\ee
\be
\la {\rm tr}\phi\  {\rm tr}\phi^3\ {\rm tr}\phi^4 \ra_G = 6N^5+75N^3+24N
\ee
\be
\la\la {\rm tr}\phi\  {\rm tr}\phi^3\ {\rm tr}\phi^6 \ra\ra_G = 360N^4+540N^2
\ee
\be
\la {\rm tr}\phi\  {\rm tr}\phi^3\ {\rm tr}\phi^6 \ra_G = 15N^6+390N^4+540N^2
\ee
$$\ldots$$
\be
\la\la {\rm tr}\phi\  {\rm tr}\phi^4\ {\rm tr}\phi^5 \ra\ra_G = 360N^4+540N^2
\ee
\be
\la {\rm tr}\phi\  {\rm tr}\phi^4\ {\rm tr}\phi^5 \ra_G = 20N^6+380N^4+545N^2
\ee
$$\ldots$$
\be
\la\la {\rm tr}\phi^2\  {\rm tr}\phi^3\ {\rm tr}\phi^5 \ra\ra_G
= 6N\la\la {\rm tr}\phi\ {\rm tr}\phi^5 \ra\ra_G + \nn \\ +
\left(10N \la\la \left({\rm tr}\phi^3\right)^2 \ra\ra_G +
10\la\la {\rm tr}\phi\ {\rm tr}\phi^2\ {\rm tr}\phi^3 \ra\ra_G\right)
+ 2\cdot 3\cdot 5 \la\la {\rm tr}\phi^6 \ra\ra_G = \nn \\ =
360N^4 + 480N^2
\label{235}
\ee
\be
\la {\rm tr}\phi^2\  {\rm tr}\phi^3\ {\rm tr}\phi^5 \ra_G =
45N^6 + 420N^4 + 480N^2
\ee
$$\ldots$$

$\bullet$
{\bf Four-point correlators:}
\be
\la\la \left({\rm tr}\phi\right)^4 \ra\ra_G = 0
\ee
\be
\la \left({\rm tr}\phi\right)^4 \ra_G = 3N^2
\ee
\be
\la\la \left({\rm tr}\phi^2\right)^4 \ra\ra_G = 48N^2
\ee
\be
\la \left({\rm tr}\phi^2\right)^4 \ra_G = N^8+12N^6+44N^4+48N^2
\ee
$$\ldots$$
\be
\la\la \left({\rm tr}\phi\right)^3\ {\rm tr}\phi^3 \ra\ra_G = 6N
\ee
\be
\la \left({\rm tr}\phi\right)^3\ {\rm tr}\phi^3 \ra_G = 9N^3+6N
\ee
\be
\la\la \left({\rm tr}\phi\right)^3\ {\rm tr}\phi^5 \ra\ra_G = 60N^2
\ee
\be
\la \left({\rm tr}\phi\right)^3\ {\rm tr}\phi^5 \ra_G = 30N^4+75N^2
\ee
\be
\la\la \left({\rm tr}\phi\right)^3\ {\rm tr}\phi^7 \ra\ra_G = 420N^3+210N
\ee
\be
\la \left({\rm tr}\phi\right)^3\ {\rm tr}\phi^7 \ra_G = 105N^5+630N^3+210N
\ee
$$\ldots$$
\be
\la\la \left({\rm tr}\phi^2\right)^3\ {\rm tr}\phi^4 \ra\ra_G =
384N^3+192N
\ee
\be
\la \left({\rm tr}\phi^2\right)^3\ {\rm tr}\phi^4 \ra_G =
2N^9 + 37N^7 + 226N^5 + 488N^3 + 192N
\ee
\be
\la\la \left({\rm tr}\phi^3\right)^3\ {\rm tr}\phi \ra\ra_G =
648 N^3+162 N
\ee
\be
\la \left({\rm tr}\phi^3\right)^3\ {\rm tr}\phi \ra_G =
108 N^5+675 N^3 +162N
\ee
$$\ldots$$
\be
\la\la \left({\rm tr}\phi\right)^2\ \left({\rm tr}\phi^2\right)^2 \ra\ra_G = 8N
\ee
\be
\la  \left({\rm tr}\phi\right)^2\ \left({\rm tr}\phi^2\right)^2 \ra_G =
N^5+6N^3+8N
\ee
\be
\la\la \left({\rm tr}\phi\right)^2\ \left({\rm tr}\phi^3\right)^2 \ra\ra_G = 72N^2
\ee
\be
\la  \left({\rm tr}\phi\right)^2\ \left({\rm tr}\phi^3\right)^2 \ra_G =
30N^4+75N^2
\ee
\be
\la\la \left({\rm tr}\phi\right)^2\ \left({\rm tr}\phi^4\right)^2 \ra\ra_G =
576N^3 + 192N
\ee
\be
\la  \left({\rm tr}\phi\right)^2\ \left({\rm tr}\phi^4\right)^2 \ra_G =
4N^7 + 88N^5 + 661N^3 + 192N
\ee
$$\ldots$$
\be
\la\la \left({\rm tr}\phi^2\right)^2\ \left({\rm tr}\phi^3\right)^2 \ra\ra_G =
576N^3+144N
\ee
\be
\la  \left({\rm tr}\phi^2\right)^2\ \left({\rm tr}\phi^3\right)^2 \ra_G =
12N^7 + 171N^5 + 618N^3 + 144N
\ee
$$\ldots$$
\be
\la\la \left({\rm tr}\phi\right)^2\ {\rm tr}\phi^2\ {\rm tr}\phi^4 \ra\ra_G =
72N^2
\ee
\be
\la  \left({\rm tr}\phi\right)^2\ {\rm tr}\phi^2\ {\rm tr}\phi^4 \ra_G =
2N^6+25N^4+78N^2
\ee
\be
\la\la \left({\rm tr}\phi\right)^2\ {\rm tr}\phi^2\ {\rm tr}\phi^6 \ra\ra_G =
480N^3 + 240N
\ee
\be
\la  \left({\rm tr}\phi\right)^2\ {\rm tr}\phi^2\ {\rm tr}\phi^6 \ra_G =
5N^7+110N^5+590N^3+240N
\ee
$$\ldots$$
\be
\la\la {\rm tr}\phi\ \left({\rm tr}\phi^2\right)^2\ {\rm tr}\phi^3\ra\ra_G =
72N^2
\ee
\be
\la {\rm tr}\phi\ \left({\rm tr}\phi^2\right)^2\ {\rm tr}\phi^3\ra_G =
3N^6+30N^4+72N^2
\ee
\be
\la\la {\rm tr}\phi\ \left({\rm tr}\phi^2\right)^2\ {\rm tr}\phi^5\ra\ra_G =
480N^3 + 240N
\ee
\be
\la {\rm tr}\phi\ \left({\rm tr}\phi^2\right)^2\ {\rm tr}\phi^5\ra_G =
10N^7+145N^5+550N^3+240N
\ee
$$\ldots$$
\be
\la\la \left({\rm tr}\phi\right)^2\ {\rm tr}\phi^3\ {\rm tr}\phi^5 \ra\ra_G =
540N^3 + 210N
\ee
\be
\la  \left({\rm tr}\phi\right)^2\ {\rm tr}\phi^3\ {\rm tr}\phi^5 \ra_G =
105N^5 + 630N^3 +210N
\ee
$$\ldots$$
\be
\la\la {\rm tr}\phi\ {\rm tr}\phi^2\ {\rm tr}\phi^3\ {\rm tr}\phi^4\ra\ra_G =
576N^3+192N
\ee
\be
\la {\rm tr}\phi\ {\rm tr}\phi^2\ {\rm tr}\phi^3\ {\rm tr}\phi^4 \ra_G =
126N^5+627N^3+192N
\ee
$$\ldots$$

$\bullet$
{\bf Five-point correlators:}
\be
\la\la \left({\rm tr}\phi^2\right)^5 \ra\ra_G = 384N^2
\ee
\be
\la \left({\rm tr}\phi^2\right)^5 \ra_G =
N^{10} + 20N^8 + 140N^6 + 400N^4 + 384N^2
\ee
$$\ldots$$
\be
\la\la \left({\rm tr}\phi\right)^4\ {\rm tr}\phi^2 \ra\ra_G = 0
\ee
\be
\la \left({\rm tr}\phi\right)^4\ {\rm tr}\phi^2 \ra_G = 3N^4+12N^2
\ee
\be
\la\la \left({\rm tr}\phi\right)^4\ {\rm tr}\phi^4 \ra\ra_G = 24N
\ee
\be
\la \left({\rm tr}\phi\right)^4\ {\rm tr}\phi^4 \ra_G = 6N^5+75N^3+24N
\ee
$$\ldots$$
\be
\la\la \left({\rm tr}\phi\right)^3\ {\rm tr}\phi^2\ {\rm tr}\phi^3 \ra\ra_G = 36N
\ee
\be
\la \left({\rm tr}\phi\right)^3\ {\rm tr}\phi^2\ {\rm tr}\phi^3 \ra_G = 9N^5+60N^3+36N
\ee
\be
\la\la \left({\rm tr}\phi\right)^3\ {\rm tr}\phi^2\ {\rm tr}\phi^5 \ra\ra_G = 480N^2
\ee
\be
\la \left({\rm tr}\phi\right)^3\ {\rm tr}\phi^2\ {\rm tr}\phi^5 \ra_G = 30N^6+315N^4+600N^2
\ee
$$\ldots$$
\be
\la\la \left({\rm tr}\phi\right)^3\ {\rm tr}\phi^3\ {\rm tr}\phi^4 \ra\ra_G = 504N^2
\ee
\be
\la \left({\rm tr}\phi\right)^3\ {\rm tr}\phi^3\ {\rm tr}\phi^4 \ra_G = 18N^6+345N^4+582N^2
\ee
$$\ldots$$
\be
\la\la \left({\rm tr}\phi^2\right)^3\ \left({\rm tr}\phi\right)^2\ra\ra_G = 48N
\ee
\be
\la \left({\rm tr}\phi^2\right)^3\ \left({\rm tr}\phi\right)^2\ra_G = N^7+12N^5+44N^3+48N
\ee
$$\ldots$$
\be
\la\la \left({\rm tr}\phi\right)^2\  \left({\rm tr}\phi^2\right)^2\
{\rm tr}\phi^4 \ra\ra_G = 384N^2
\ee
\be
\la \left({\rm tr}\phi\right)^2\  \left({\rm tr}\phi^2\right)^2\
{\rm tr}\phi^4 \ra_G = 2N^8 + 41N^6 + 278N^4 + 624N^2
\ee
$$\ldots$$

$\bullet$
{\bf Six-point correlators:}
\be
\la\la \left({\rm tr}\phi\right)^6 \ra\ra_G = 0
\ee
\be
\la \left({\rm tr}\phi\right)^6 \ra_G = 15N^3
\ee
$$\ldots$$
\be
\la\la \left({\rm tr}\phi\right)^5\ {\rm tr}\phi^3 \ra\ra_G = 0
\ee
\be
\la \left({\rm tr}\phi\right)^5\ {\rm tr}\phi^3 \ra_G = 45N^4 + 60N
\ee
\be
\la\la \left({\rm tr}\phi\right)^5\ {\rm tr}\phi^5 \ra\ra_G = 120N
\ee
\be
\la \left({\rm tr}\phi\right)^5\ {\rm tr}\phi^5 \ra_G =
150N^5+675N^3+120N
\ee
$$\ldots$$
\be
\la\la \left({\rm tr}\phi\right)^4\ \left({\rm tr}\phi^2\right)^2 \ra\ra_G = 0
\ee
\be
\la \left({\rm tr}\phi\right)^4\ \left({\rm tr}\phi^2\right)^2 \ra_G =
3N^6+30N^4+72N^2
\ee
$$\ldots$$
\be
\la\la \left({\rm tr}\phi\right)^4\ \left({\rm tr}\phi^3\right)^2 \ra\ra_G =
216N
\ee
\be
\la \left({\rm tr}\phi\right)^4\ \left({\rm tr}\phi^3\right)^2 \ra_G =
144N^5+585N^3+216N
\ee
$$\ldots$$
\be
\la\la \left({\rm tr}\phi^2\right)^4\ \left({\rm tr}\phi\right)^2 \ra\ra_G =
384N
\ee
\be
\la \left({\rm tr}\phi^2\right)^4\ \left({\rm tr}\phi\right)^2 \ra_G =
N^9+20N^7+140N^5+400N^3+384N
\ee
$$\ldots$$
\be
\la\la \left({\rm tr}\phi\right)^3\ \left({\rm tr}\phi^2\right)^2\
{\rm tr}\phi^3 \ra\ra_G = 288N
\ee
\be
\la \left({\rm tr}\phi\right)^3\ \left({\rm tr}\phi^2\right)^2\
{\rm tr}\phi^3 \ra_G = 9N^7+132N^5+516N^3+288N
\ee
$$\ldots$$

$\bullet$
{\bf Seven-point correlators:}
\be
\la\la \left({\rm tr}\phi\right)^6\ {\rm tr}\phi^2 \ra\ra_G = 0
\ee
\be
\la \left({\rm tr}\phi\right)^6\ {\rm tr}\phi^2 \ra_G = 15N^5+90N^3
\ee
\be
\la\la \left({\rm tr}\phi\right)^6\ {\rm tr}\phi^4 \ra\ra_G = 0
\ee
\be
\la \left({\rm tr}\phi\right)^6\ {\rm tr}\phi^4 \ra_G = 30N^6+555N^4+360N^2
\ee
$$\ldots$$
\be
\la\la \left({\rm tr}\phi\right)^5\ {\rm tr}\phi^2\ {\rm tr}\phi^3 \ra\ra_G = 0
\ee
\be
\la \left({\rm tr}\phi\right)^5\ {\rm tr}\phi^2\ {\rm tr}\phi^3 \ra_G =
45N^6+420N^4+480N^2
\ee
$$\ldots$$
\be
\la\la \left({\rm tr}\phi\right)^4\ \left({\rm tr}\phi^2\right)^3 \ra\ra_G = 0
\ee
\be
\la \left({\rm tr}\phi\right)^4\ \left({\rm tr}\phi^2\right)^3 \ra_G =
3N^8+54N^6+312N^4+576N^2
\ee
$$\ldots$$

$\bullet$
{\bf Eight-point correlators:}
\be
\la\la \left({\rm tr}\phi\right)^8 \ra\ra_G = 0
\ee
\be
\la \left({\rm tr}\phi\right)^8 \ra_G = 105N
\ee
$$\ldots$$
\be
\la\la \left({\rm tr}\phi\right)^7\ {\rm tr}\phi^3 \ra\ra_G = 0
\ee
\be
\la \left({\rm tr}\phi\right)^7\ {\rm tr}\phi^3 \ra_G = 315N^5+630N^3
\ee
$$\ldots$$
\be
\la\la \left({\rm tr}\phi\right)^6\ \left({\rm tr}\phi^2\right)^2 \ra\ra_G = 0
\ee
\be
\la \left({\rm tr}\phi\right)^6\ \left({\rm tr}\phi^2\right)^2 \ra_G =
15N^7+210N^5+720N^3
\ee
$$\ldots$$

$\bullet$
{\bf Nine-point correlators:}
\be
\la\la \left({\rm tr}\phi\right)^8 {\rm tr}\phi^2 \ra\ra_G = 0
\ee
\be
\la \left({\rm tr}\phi\right)^8 {\rm tr}\phi^2 \ra_G = 105N^6+840N^4
\ee
$$\ldots$$

$\bullet$
{\bf Ten-point correlators:}
\be
\la\la \left({\rm tr}\phi\right)^{10} \ra\ra_G = 0
\ee
\be
\la \left({\rm tr}\phi\right)^{10} \ra_G = 945N
\ee

\subsection{Lowest prepotentials ${\cal F}^{(p)}$:
first terms of $t$-expansions}

The prepotential is the generating function of the connected correlators
\be
{\cal F}_G = Nt_0 + \sum_{k=0}^\infty t_k \la\la{\rm tr}\phi^k\ra\ra_G +
\frac{1}{2!}\sum_{k_1,k_2=0}^\infty t_{k_1}t_{k_2}
\la\la{\rm tr}\phi^{k_1}\ {\rm tr}\phi^{k_2}\ra\ra_G + \nn
\\ +
\frac{1}{6!}\sum_{k_1,k_2,k_3=0}^\infty t_{k_1}t_{k_2}t_{k_3}
\la\la{\rm tr}\phi^{k_1}\ {\rm tr}\phi^{k_2}\ {\rm tr}\phi^{k_3}\ra\ra_G
+ \ldots
\ee
If the correlators are taken from Table \ref{correx}, one should restore
$g$ and $M$ dependencies and substitute
\be
N = \frac{MS}{g}
\ee
so that
\be
{\cal F}_G(t|S) =
\sum_{p\geq 0} g^{2p-2}{\cal F}_G^{(p)}(t|S)
= \nn \\ =
\sum_{m=1}^{\infty} \frac{1}{m!}
\sum_{k_1,\ldots,k_m=0}^\infty \frac{t_{k_1}\ldots t_{k_m}}{g^m}
\left(\frac{g}{M}\right)^{\frac{k_1+\ldots+k_m}{2}}
\la\la{\rm tr}\phi^{k_1}\ldots {\rm tr}\phi^{k_m}\ra\ra_G
\la {\rm tr}\phi^2 \ra_G
= \nn \\ \nn \\ = \frac{Nt_0}{g}
\frac{N^2 t_2}{M} + \frac{g(2N^3 + N)t_4}{M^2} + \frac{g^2(5N^4 + 10N^2)t_6}{M^3} +
\frac{g^3(14N^5 + 70N^3 + 21N)t_8}{M^4} + \nn \\ +
\frac{g^4(42N^6 + 420N^4 + 483N^2)t_{10}}{M^5}
+ \ldots + \nn \\ \\ +
\frac{Nt_1^2}{2gM} + \frac{N^2t_2^2}{M^2} + \frac{g(12N^3 + 3N)t_3^2}{2M^3}
+  \frac{g^2(36N^4 + 60N^2)t_4^2}{2M^4} +
\frac{g^3(180N^5 + 600N^3 + 165N)t_5^2}{2M^5}
+ \ldots + \nn \\ \nn \\ +
\frac{3N^2t_1t_3}{M^2} + \frac{g(10N^3 + 5N)t_1t_5}{M^3}
+ \frac{g^2(35N^4 + 70N^2)t_1t_7}{M^4} +
\frac{g^3(126N^5 + 630N^3 + 189N)t_1t_9}{M^5} +
 \ldots + \nn \\ +
\frac{g(8N^3 + 4N)t_2t_4}{M^3} + \frac{g^2(30N^4 + 60N^2)t_2t_6}{M^4} +
\frac{g^3(112N^5 + 560N^3 + 178N)t_2t_8}{M^5} +
 \ldots + \nn \\ +
\frac{g^2(45N^4+60N^2)t_3t_5}{M^4} +
\frac{g^3(168N^5 + 630N^3 + 147N)t_3t_7}{M^5} +
 \ldots + \nn \\ +
\frac{g^3(144N^5 + 600N^3 + 156N)t_4t_6}{M^5}
+ \ldots +
\frac{8N^2t_2^3}{6M^3} + \ldots +\\
+\frac{2Nt_1^2t_2}{2gM^2} + \frac{12N^2t_1^2t_4}{2M^3} +
\frac{g(60N^3+30N)t_1^2t_6}{2M^4} + \frac{g^2(280N^4+560N^2)t_1^2t_8}{2M^5}
+ \ldots + \nn \\ +
\frac{g(48N^3+24N)t_2^2t_4}{2M^4} +
\frac{g^2(240N^4 + 480N^2)t_2^2t_6}{2M^5} + \ldots + \nn \\ +
\frac{g(72N^3 + 18N)t_2t_3^2}{2M^4} +
\frac{g^2(432N^4 + 468N^2)t_3^2t_4}{2M^5} + \ldots + \nn \\ +
\frac{g^2(288N^4 + 480N^2)t_2t_4^2}{2M^5} + \ldots + \nn \\ +
\frac{12N^2t_1t_2t_3}{M^3} +
\frac{g(60N^3+30N)t_1t_2t_5}{M^4} +
\frac{g^2(280N^4+560N^2)t_1t_2t_7}{M^5} + \ldots +
\ee
\be
+\frac{g(72N^3+24N)t_1t_3t_4}{M^4} +
\frac{g^2(360N^4+540N^2)t_1t_3t_6}{M^5} + \ldots +
\\+\frac{g^2(360N^4+540N^2)t_1t_4t_5}{M^5} + \ldots + \nn \\ +
\frac{g^2(360N^4 + 480N^2)t_2t_3t_5}{M^5}
+ \ldots +
\frac{48N^2t_2^4}{24M^4} + \ldots + \nn \\ +
\frac{6Nt_1^3t_3}{6gM^3} +
\frac{60N^2t_1^3t_5}{6M^4} +
\frac{g(420N^3+210N)t_1^3t_7}{6M^5} + \ldots +
\frac{g(384N^3+192N)t_2^3t_4}{6M^5} + \ldots + \nn \\ +
\frac{8Nt_1^2t_2^2}{4gM^3} + \frac{72N^2t_1^2t_3^2}{4M^4} +
\frac{g(576N^3 + 192N)t_1^2t_4^2}{4M^5} + \ldots + \nn \\ +
\frac{g(576N^3+144N)t_2^2t_3^2}{4M^5} + \ldots + \nn \\ +
\frac{72N^2 t_1^2t_2t_4}{2M^4} +
\frac{g(540N^3+210N)t_1^2t_3t_5}{2M^5} +
\frac{g(480N^3+240N)t_1^2t_2t_6}{2M^5} + \ldots + \nn \\ +
\frac{72N^2t_1t_2^2t_3}{2M^4} +
\frac{g(480N^3+240N)t_1t_2^2t_5}{2M^5} + \ldots + \nn \\ +
\frac{g(576N^3+192N)t_1t_2t_3t_4}{M^5}
+ \ldots + \\ +
\frac{384N^2t_2^5}{120M^5} + \ldots+\\
\frac{24Nt_1^4t_4}{24M^5} + \ldots + \nn \\ +
\frac{36Nt_1^3t_2t_3}{6gM^4} +
\frac{480N^2t_1^3t_2t_5}{6M^5} +
\frac{504N^2t_1^3t_3t_4}{6M^5} + \ldots + \nn \\ +
\frac{48Nt_1^2t_2^3}{12gM^4} + \ldots + \nn \\ +
\frac{384N^2t_1^2t_2^2t_4}{4M^5}
+ \ldots + \nn \\ +
\frac{Nt_1^5t_5}{gM^5} + \ldots + \nn \\ +
\frac{216Nt_1^4t_3^2}{48gM^5} + \ldots + \nn \\ +
\frac{384Nt_1^2t_2^4}{48gM^5} + \ldots + \nn \\ +
\frac{288Nt_1^3t_2^2t_3}{16gM^5}
+ \ldots + \nn \\
\ee
$$\ldots$$

\subsection{Functions $Z_G(t|N)$:
first terms of $t$-expansion}
\be
e^{-Nt_0} Z_G(t|N) = 1 + \nn \\ +
\frac{N^2 t_2}{M} + \frac{g(2N^3 + N)t_4}{M^2} + \frac{g^2(5N^4 + 10N^2)t_6}{M^3} +
\frac{g^3(14N^5 + 70N^3 + 21N)t_8}{M^4} + \nn \\ +
\frac{g^4(42N^6 + 420N^4 + 483N^2)t_{10}}{M^5}
+ \ldots + \\ +
\frac{Nt_1^2}{2gM} + \frac{(N^4+2N^2)t_2^2}{2M^2} + \frac{g(12N^3 + 3N)t_3^2}{2M^3}
+  \frac{g^2(4N^6+40N^4 + 61N^2)t_4^2}{2M^4} +
\nn \\ +
\frac{g^3(180N^5 + 600N^3 + 165N)t_5^2}{2M^5}
+ \ldots + \\ +
\frac{3N^2t_1t_3}{M^2} + \frac{g(10N^3 + 5N)t_1t_5}{M^3}
+ \frac{g^2(35N^4 + 70N^2)t_1t_7}{M^4} + \nn \\ +
\frac{g^3(126N^5 + 630N^3 + 189N)t_1t_9}{M^5} +
 \ldots + \nn \\ +
\frac{g(2N^5+9N^3 + 4N)t_2t_4}{M^3} +
\frac{g^2(5N^6+40N^4 + 60N^2)t_2t_6}{M^4} + \nn \\ +
\frac{g^3(14N^7+182N^5 + 581N^3 + 178N)t_2t_8}{M^5} +
 \ldots + \nn \\ +
\frac{g^2(45N^4+60N^2)t_3t_5}{M^4} +
\frac{g^3(168N^5 + 630N^3 + 147N)t_3t_7}{M^5} +
 \ldots + \nn \\ +
\frac{g^3(10N^7+169N^5 + 610N^3 + 156N)t_4t_6}{M^5}
+ \ldots + \\ +
\frac{(N^6+6N^4+8N^2)t_2^3}{6M^3} + \ldots + \nn \\ +
\frac{(N^3+2N)t_1^2t_2}{2gM^2} +
\frac{(2N^4+13N^2)t_1^2t_4}{2M^3} +
\frac{g(5N^5+70N^3+30N)t_1^2t_6}{2M^4} + \nn \\ +
\frac{g^2(14N^6+350N^4+581N^2)t_1^2t_8}{2M^5}
+ \ldots + \nn \\ +
\frac{g(2N^7+21N^5+58N^3+24N)t_2^2t_4}{2M^4} +
\frac{g^2(5N^8+80N^6+380N^4 + 480N^2)t_2^2t_6}{2M^5} + \ldots + \nn \\ +
\frac{g(12N^5+75N^3 + 18N)t_2t_3^2}{2M^4} +
\frac{g^2(24N^6+450N^4 + 471N^2)t_3^2t_4}{2M^5} + \ldots + \nn \\ +
\frac{g^2(36N^6+360N^4 + 549N^2)t_2t_4^2}{2M^5} + \ldots + \nn \\ +
\frac{(3N^4+12N^2)t_1t_2t_3}{M^3} +
\frac{g(10N^5+65N^3+30N)t_1t_2t_5}{M^4} + \nn \\ +
\frac{g^2(35N^6+350N^4+560N^2)t_1t_2t_7}{M^5} + \ldots + \nn \\ +
\frac{g(6N^5+75N^3+24N)t_1t_3t_4}{M^4} +
\frac{g^2(15N^6+390N^4+540N^2)t_1t_3t_6}{M^5} + \ldots + \nn \\ +
\frac{g^2(20N^6+380N^4+545N^2)t_1t_4t_5}{M^5} + \ldots + \nn \\ +
\frac{g^2(45N^6+420N^4 + 480N^2)t_2t_3t_5}{M^5}
+ \ldots +
\ee
\be +
\frac{3N^2t_1^4}{24g^2M^2} +
\frac{N^8+12N^6+44N^4+48N^2)t_2^4}{24M^4} + \ldots + \nn \\ +
\frac{(9N^3+6N)t_1^3t_3}{6gM^3} +
\frac{(30N^4+75N^2)t_1^3t_5}{6M^4} +
\frac{g(105N^5+630N^3+210N)t_1^3t_7}{6M^5} + \ldots + \nn \\ +
\frac{g(2N^9+37N^7+226N^5+488N^3+192N)t_2^3t_4}{6M^5} + \ldots + \nn \\ +
\frac{(N^5+6N^3+8N)t_1^2t_2^2}{4gM^3} +
\frac{(30N^4+75N^2)t_1^2t_3^2}{4M^4} +
\frac{g(4N^7+88N^5+661N^3 + 192N)t_1^2t_4^2}{4M^5} + \ldots + \nn \\ +
\frac{g(12N^7+171N^5+618N^3+144N)t_2^2t_3^2}{4M^5} + \ldots + \nn \\ +
\frac{(2N^6+25N^4+78N^2) t_1^2t_2t_4}{2M^4} + \nn \\ +
\frac{g(105N^5+630N^3+210N)t_1^2t_3t_5}{2M^5} +
\frac{g(5N^7+110N^5+590N^3+240N)t_1^2t_2t_6}{2M^5} + \ldots + \nn \\ +
\frac{(3N^6+30N^4+72N^2)t_1t_2^2t_3}{2M^4} +
\frac{g(10N^7+145N^5+550N^3+240N)t_1t_2^2t_5}{2M^5} + \ldots + \nn \\ +
\frac{g(126N^5+627N^3+192N)t_1t_2t_3t_4}{M^5}
+ \ldots +
\\ +
\frac{(N^{10} + 20N^8 + 140N^6 + 400N^4 + 384N^2)t_2^5}{120M^5} + \ldots + \nn \\ +
\frac{(3N^4+12N^2)t_1^4t_2}{24gM^4} +
\frac{(6N^5+75N^3+24N)t_1^4t_4}{24M^5} + \ldots + \nn \\ +
\frac{(9N^5+60N^3+36N)t_1^3t_2t_3}{6gM^4} +
\frac{(30N^6+315N^4+600N^2)t_1^3t_2t_5}{6M^5} + \nn \\ +
\frac{(18N^6+345N^4+582N^2)t_1^3t_3t_4}{6M^5} + \ldots + \nn \\ +
\frac{(N^7+12N^5+44N^3+48N)t_1^2t_2^3}{12gM^4} + \ldots + \nn \\ +
\frac{(2N^8+41N^6+278N^4+624N^2)t_1^2t_2^2t_4}{4M^5}
+ \ldots +
\\ +
\frac{15N^3t_1^6}{720g^3M^3} + \ldots + \nn \\ +
\frac{(45N^4+60N)t_1^5t_3}{120g^2M^4} +
\frac{(150N^5+675N^3+120N)t_1^5t_5}{120gM^5} + \ldots + \nn \\ +
\frac{(3N^6+30N^4+72N^2)t_1^4t_2^2}{48g^2M^4} +
\frac{(144N^5+585N^3+216N)t_1^4t_3^2}{48gM^5} + \ldots + \nn \\ +
\frac{(N^9+20N^7+140N^5+400N^3+384N)t_1^2t_2^4}{48gM^5} + \ldots + \nn \\ +
\frac{(9N^7+132N^5+516N^3+288N)t_1^3t_2^2t_3}{16gM^5}
+ \ldots +\\ +
\frac{(15N^5+90N^3)t_1^6t_2}{720g^3M^4} +
\frac{(30N^6+555N^4+360N^2)t_1^6t_4}{720g^2M^5} + \ldots + \nn \\ +
\frac{(45N^6+420N^4 + 480N^2)t_1^5t_2t_3}{60g^2M^5} + \ldots + \nn \\ +
\frac{(3N^8+54N^6+312N^4+576N^2)t_1^4t_2^3}{240g^2M^5}
+ \ldots +
\ee
\be +
\frac{105N^4t_1^8}{5760g^4M^4} + \ldots +
\frac{(315N^5+630N^3)t_1^7t_3}{720g^3M^5} + \ldots + \nn \\ +
\frac{(15N^7+210N^5+720N^3)t_1^6t_2^2}{180g^3M^5}
+ \ldots + \\ +
\frac{(105N^6+840N^4)t_1^8t_2}{5760g^4M^5}
+ \ldots +  \\ +
\frac{945N^5t_1^{10}}{10!g^5M^5}
+ \ldots + \\ + \ldots
\ee

The MAPLE generating program for the expansion of the integral
\be\label{tttr}
{\int d\phi e^{-{1\over 2}\Tr\phi^2 +\sum_{k=1}g^kt_k\Tr\phi^k}
\over \int d\phi e{-{1\over 2}\Tr\phi^2}}
\ee
in $g$ and $t_k$ written by V.Pestun, is available at
http://wwwth.itep.ru/progs/giv.mws or .../giv.tar.
It generates the function ``tttr", which is the
integral (\ref{tttr}) and requires, at least, MAPLE 7.

\section{Non-Gaussian partition functions\label{NGd}}
\setcounter{equation}{0}

From the Virasoro constraints in the case of non-Gaussian potential one can
get non-Gaussian densities and multidensities, for instance, the simplest
genus zero one-point density is as follows
\be
\rho_W^{(0|1)}(z)=\frac{W'(z)-y(z)}{2}
\ee
Here
\be
y_W^2(z)=W'(z)^2-4f_W^{(0|1)}(z)
\ee
From equation (\ref{recrel1}) follows expressions for all non-Gaussian
(multi)densities, for instance,
\be
\rho_W^{(0|2)}(z,x)=\frac{1}{y(z)}\Big(\p_x\Big(\frac{\rho_W^{(0|1)}(x)-\rho_W^{(0|1)}(z)}{x-z}\Big)+f_W^{(0|2)}(z|x)\Big)
\ee
and
\be
\rho_W^{(1|1)}(z)=\frac{1}{y(z)}\Big(\rho_W^{(0|2)}(z|z)+f_W^{(1|1)}(z)\Big)
\ee
It can be shown explicitly that the function $\rho_W^{(0|2)}(z,x)$ is symmetric
in $z$ and $x$, if $f_W^{(0|2)}(z|x)$ is defined as in (\ref{fthroughrho})
\be
f_W^{(0|2)}(z|x) = \check{R}_W(z)\rho^{(0|1)}_W(x)
\ee
The two point function can be presented in form explicitly symmetric in
$x$ and $z$, namely,
\be
\rho_W^{(0|2)}(z,x)=\frac{1}{2y(x)y(z)(x-z)^2}\Big(W'(x)W'(z)-y(x)y(z)+
(x-z)^2\check{K}_W(x,z)F^{(0)}+\check{A}_W(x,z)F^{(0)}\Big)
\ee
where
\be
\check{K}_W(x,z)=\sum_{n,m=0}nmT_nT_m\sum_{i=0}^{n-2}\sum_{j=0}^{m-2}x^iz^j\frac{\p}{\p T_{n-i-2}}\frac{\p}{\p T_{m-j-2}}
\ee
and
\be
\check{A}_W(x,z)=2\sum nT_n\Big(2\frac{\p}{\p T_{n-2}}+
\sum_{i=1}^{n-2}[(n-i)(x-z)(x^{i-1}-z^{i-1})+(zx^{i-1}+xz^{i-1})]\frac{\p}{\p T_{n-i-2}}\Big)
\ee
Thus, the two-point function has a smooth limit when $x\to z$
\be
\rho_W^{(0|2)}(x,x)=\frac{1}{2y(x)^2}\Big(
\frac{1}{8}(W^{(3)}W'-2W''^2)\\
-\frac{1}{4y(x)^2}\Big(((W'')^2+W'W^{(3)}-2\check
{R}''_W(z)F^{(0)})(W'^2-4\check{R}'_W(z)F^{(0)})-2(W'W''-2\check{R}'_W(z)F^{(0)})^2\Big)\\
+\check{K}_W(x,x)F^{(0)}+\check{B}(x)F^{(0)}\Big)
\Big)
\ee
\be
\check{B}(x)=\frac{1}{2}\sum
nT_n\sum_{i=1}^{n-2}\Big(4(n-i)(i-2)+(i-1)(i-4)\Big)x^{i-2}\frac{\p}{\p T_{n-i-2}}
\ee
\be
\la\la {\rm tr} I \ra\ra_W^{(0)} = \sum_{i=1}^n \tilde S_i,
\nn \\
\la\la {\rm tr} \phi^k \ra\ra_W^{(0)} = \sum_{i=1}^n \tilde S_i\alpha_i^k, \
\ {\rm for} \ \ 0\leq k\leq n,
\nn \\
\la\la {\rm tr} \phi^{n+1} \ra\ra_W^{(0)} = \sum_{i=1}^n \tilde
S_i\alpha_i^{n+1} + \frac{1}{(n+1)T_{n+1}}\left(\sum_{i=1}^n \tilde
S_i\right)^2
\ee

\section{CIV-DV prepotential:
$\left.{\cal F}^{DV}_m\right|_{t=0}$ through Gaussian
prepotentials}
\setcounter{equation}{0}

From (\ref{factfor}) we obtain here (as a generalization of the results of
\cite{MMDV}) the first terms of expansion for the prepotential
$\left.{\cal F}^{DV}_m\right|_{t=0}$ through Gaussian
prepotentials. They coincide with the results from
Seiberg-Witten theory (see \cite{IM4-5},\cite{IM6},\cite{IM7}):
\be
\left.{\cal F}^{DV}_1\right|_{t=0} = - \sum_i S_i W(\alpha_i) +
\left.F_1\right|_{t_k^{(i)} = -T^{(i)}_k}, \nn \\
\left.{\cal F}^{DV}_2\right|_{t=0} =
-\frac{1}{2}\sum_{i<j} (S_i^2 - 4S_iS_j + S_j^2)\log\alpha_{ij} -
\frac{3}{4}\sum_{i}S_i^2+\\
+\left[F_2 - \sd F_1 + \frac{1}{2}
\sum_{k\geq 1}^\infty
\frac{g^{2k-2}}{k!}\cdd^k F_1^2
\right]_{t_k^{(i)} = -T^{(i)}_k}, \nn \\
\left.{\cal F}^{DV}_3\right|_{t=0} =
\left[F_3 - \sd F_2 + \frac{1}{2}\sd^2 F_1 +
\phantom{\sum_{k\geq 1}^\infty}
\right.\nn \\ \left.+
\sum_{k\geq 1}^\infty
\frac{g^{2k-2}}{k!}\cdd^k \left[F_1F_2 - F_1\sd F_1\right] +
\frac{1}{6}\sum_{k\geq 2}^\infty
\frac{g^{2k-4}}{k!}\left(\cdd^k F_1^3 - 3F_1\cdd^k F_1^2\right)
\right]_{t_k^{(i)} = -T^{(i)}_k}, \nn \\
\left.{\cal F}^{DV}_4\right|_{t=0} =
\left[F_4 - \sd F_3 + \frac{1}{2}\sd^2 F_2 - \frac{1}{6}\sd^3 F_1 +
\phantom{\sum_{k\geq 1}^\infty}
\right.\nn \\ \left. +
\sum_{k\geq 1}^\infty
\frac{g^{2k-2}}{k!}\cdd^k
\left[\frac{1}{2}F_2^2 + F_1F_3 - F_2\sd F_1 - F_1\sd F_2 +
\frac{1}{2}F_1\sd^2F_1 + \frac{1}{2}\left[\sd F_1\right]^2
\right] + \right.\nn \\ \left. +
\frac{1}{2}\sum_{k\geq 2}^\infty
\frac{g^{2k-4}}{k!}\left[\cdd^k F_1^2F_2 - 2F_1\cdd^k F_1F_2 -
F_2\cdd^kF_1^2\right] +
\right.\nn \\ \left. +
\frac{1}{2}\sum_{k\geq 2}^\infty
\frac{g^{2k-4}}{k!}\left[\cdd^k F_1^2\sd F_1 -2F_1\cdd^k F_1\sd F_1 -
\sd F_1 \cdd^k F_1^2\right] +
\right. \ee\be \left. +
\sum_{k\geq 3}^\infty
\frac{1}{24} \frac{g^{2k-6}}{k!}\left[\cdd^kF_1^4
+ 12F_1^2\cdd^k F_1^2 - 4F_1\cdd^k F_1^3 -
\phantom{\sum}
\right.\right.
\\ \left.\left. -
3\left(\cdd \otimes I \phantom{\sum} + \phantom{\sum} I\otimes \cdd
\right)^k F_1^2\otimes F_1^2\right]
\phantom{\sum_{k\geq 1}^\infty}
\right]_{t_k^{(i)} = -T^{(i)}_k},
\nn \\
\ldots
\ee

\subsection{Explicit formulas for $\left.{\cal F}_{DV}^{(p|m)}\right|_{t=0}$
}

${\bf m=1}:$
\be
\left.{\cal F}_{DV}^{(0|1)}\right|_{t=0} = -\sum_i S_i W(\alpha_i)
\ee
\be
\left.{\cal F}_{DV}^{(1|1)}\right|_{t=0} =
\sum_i \frac{S_i}{M_i}\left(-\frac{\lla 4^1 \rra^{(1)}}{4}\ \sigma_2^{(i)} +
\frac{\lla 3^2 \rra^{(1)}}{2! \cdot 3^2}
\left[\sigma_1^{(i)}\right]^2\right) =
\nn \\ =
\sum_i \frac{S_i}{M_i}\left(-\frac{\sigma_2^{(i)}}{4} +
\frac{3\left[\sigma_1^{(i)}\right]^2}{2!\cdot 3^2}\right)
= \sum_i \frac{S_i}{12M_i}\left(2\left[\sigma_1^{(i)}\right]^2 -
3\sigma_2^{(i)}\right)
\ee
\be
\left.{\cal F}_{DV}^{(2|1)}\right|_{t=0} = \sum_i \frac{S_i}{M_i^3}
\left(-\frac{\lla 8^1 \rra^{(2)}}{8}\ \sigma_6^{(i)} + \right.\nn \\ \left.  +
\frac{\lla 3^1,7^1 \rra^{(2)}}{3\cdot 7}\ \sigma_1^{(i)}\sigma_5^{(i)} +
\frac{\lla 4^1,6^1 \rra^{(2)}}{4\cdot 6}\ \sigma_2^{(i)}\sigma_4^{(i)} +
\frac{\lla 5^2 \rra^{(2)}}{2!\cdot 5^2}\left[\sigma_3^{(i)}\right]^2 -
\right.\nn \\ \left.  -
\frac{\lla 3^2,6^1 \rra^{(2)}}{2!\cdot 3^2\cdot 6}
\left[\sigma_1^{(i)}\right]^2\sigma_4^{(i)} -
\frac{\lla 3^1,4^1,5^1 \rra^{(2)}}{3\cdot 4\cdot 5}\
\sigma_1^{(i)}\sigma_2^{(i)}\sigma_3^{(i)} -
\frac{\lla 4^3 \rra^{(2)}}{3!\cdot 4^3}
\left[\sigma_2^{(i)}\right]^3 + \right.\nn \\ \left.  +
\frac{\lla 3^2,4^2 \rra^{(2)}}{(2!)^2\cdot 3^2\cdot 4^2}
\left[\sigma_1^{(i)}\right]^2\left[\sigma_2^{(i)}\right]^2 -
\frac{\lla 3^4,4^1 \rra^{(2)}}{4!\cdot 3^4\cdot 4}
\left[\sigma_1^{(i)}\right]^4\sigma_2^{(i)} +
\frac{\lla 3^6 \rra^{(2)}}{6!\cdot 3^6}\left[\sigma_1^{(i)}\right]^6
\right)
\ee
\be
\ldots \nn
\ee

${\bf m=2}:$
\be
\left.{\cal F}_{DV}^{(0|2)}\right|_{t=0} =
-\frac{1}{2}\sum_{i<j} (S_i^2 - 4S_iS_j + S_j^2)\log\alpha_{ij} -
\frac{3}{4}\sum_{i}S_i^2
\ee
\be
\left.{\cal F}_{DV}^{(1|2)}\right|_{t=0} =
\sum_i \frac{S_i^2}{M_i^2} \left(
-\frac{\lla 6^1\rra^{(1)}}{6}\sigma_4^{(i)} + \frac{\lla
4^2\rra^{(1)}}{2!\cdot 4^2} \left[\sigma_2^{(i)}\right]^2 +
\frac{\lla 3,5 \rra^{(1)}}{3\cdot 5} \sigma_1^{(i)}\sigma_3^{(i)} -
\right. \nn \\ \left. -
\frac{\lla 3^2,4\rra^{(1)}}{2!\cdot 3^2\cdot 4}\left[\sigma_1^{(i)}\right]^2
\sigma_2^{(i)} + \frac{\lla 3^4 \rra^{(1)}}{4!\cdot
3^4}\left[\sigma_1^{(i)}\right]^4\right) - \nn \\ -
\sum_{i\neq j}\frac{S_iS_j}{M_i^2} \left(
\frac{\lla 4^1\rra^{(1)}}{2\alpha_{ij}^4} +
\frac{2\lla 3^2\rra^{(1)}}{9\alpha_{ij}^3}\sigma_1^{(i)}  -
 \frac{\lla 2,4
\rra^{(1)}}{4\alpha_{ij}^2}\sigma_2^{(i)} +
\frac{\lla 2^1,3^2\rra^{(1)}}{2!\cdot
3^2\alpha_{ij}^2}\left[\sigma_1^{(i)}\right]^2
+ \right. \nn \\ \left. +
\frac{2\lla 1,5 \rra^{(1)}}{5\alpha_{ij}}\sigma_3^{(i)}
+ \frac{2\lla 1,3,4\rra^{(1)}}{3\cdot 4\alpha_{ij}}
\sigma_1^{(i)}\sigma_2^{(i)} +
\frac{2\lla 1,3^3\rra^{(1)}}{3!\cdot 3^3\alpha_{ij}}
\left[\sigma_1^{(i)}\right]^3
\right) +
\sum_{i\neq j}\frac{S_iS_j}{M_iM_j\alpha_{ij}^4} \left[
\lla 1^2 \rra^{(0)}\right]^2
\ee
\be
\ldots \nn
\ee

${\bf m=3}:$
\be
\left.{\cal F}_{DV}^{(0|3)}\right|_{t=0} =
\left[F_3 - \sd F_2 + \frac{1}{2}\sd^2 F_1 +
\phantom{\sum_{k\geq 1}^\infty}
\right.\nn \\ \left.+
\sum_{k\geq 1}^\infty
\frac{g^{2k-2}}{k!}\cdd^k \left[F_1F_2 - F_1\sd F_1\right] +
\frac{1}{6}\sum_{k\geq 2}^\infty
\frac{g^{2k-4}}{k!}\left(\cdd^kF_1^3 - 3F_1\cdd^k F_1^2\right)
\right]_{t_k^{(i)} = -T^{(i)}_k} = \nn \\ =
\sum_{i=1}^n \left[S_i^3 \left(\frac{2t_4^{(i)}}{M_i^2} +
\frac{6\left[t_3^{(i)}\right]^2}{M_i^3}\right) -
\sd S_i^2 \left(\frac{t_2^{(i)}}{M_i} +
\frac{3t_1^{(i)}t_3^{(i)}}{M_i^2}\right)
+ \frac{1}{2}\sd^2 S_i\frac{\left[t_1^{(i)}\right]^2}{2M_i}
\right]_{t_k^{(i)} = -T^{(i)}_k}
= \nn \\ =
\sum_i \left(\frac{S_i^3}{M_i} \left(-\frac{2\sigma^{(i)}_2}{4} +
\frac{6\left[\sigma^{(i)}_1\right]^2}{3^2}\right) -
\sum_{j\neq i} \frac{2S_i^2S_j}{M_i} \left(\frac{1}{2\alpha_{ij}^2}
+ \frac{\sigma^{(i)}_1}{\alpha_{ij}}\right)
+ \sum_{j,k\neq i} \frac{2S_iS_jS_k}{M_i\alpha_{ij}\alpha_{ik}}
\right) = \nn \\ =
\frac{2}{3} \sum_i \frac{S_i^3}{M_i} \left(\sum_{j\neq i}
\frac{1}{\alpha_{ij}^2} + \frac{5}{4}\sum_{\stackrel{j<k}{j,k\neq i}}
\frac{1}{\alpha_{ij}\alpha_{ik}}\right) +
\sum_{i\neq j} S_i^2S_j\left(\frac{2}{M_j\alpha_{ij}^2} -
\frac{3}{M_i\alpha_{ij}^2} - \frac{2}{\alpha_{ij}}\sum_{k\neq i,j}
\frac{1}{\alpha_{ik}}\right) + \nn \\ +
4\sum_{i<j<k}S_iS_jS_k \left(\frac{1}{M_i\alpha_{ij}\alpha_{ik}} +
\frac{1}{M_j\alpha_{ji}\alpha_{jk}} +
\frac{1}{M_k\alpha_{ki}\alpha_{kj}}\right)
\label{F03}
\ee
As conjectured by R.Dijkgraaf and C.Vafa \cite{DV}, this expression coincides
(up to rescalings $S_{\cite{IM6}} = 2S$, ${\cal F}_{\cite{IM6}}=2{\cal F}$)
with the result of calculation \cite{IM6} in the framework of
Seiberg-Witten theory \cite{SW}-\cite{IM7}.\footnote{According
to \cite{IM6},
\be
\left.{\cal F}_{DV}^{(0|3)}\right|_{t=0} =
 \sum_i u_iS_i^3 + \sum_{i\neq j} u_{i;j}S_i^2S_j +
\sum_{i<j<k}u_{ijk}S_iS_jS_k,
\nn \\
u_i = \frac{1}{6}\left(-\sum_{j\neq i}\frac{1}{M_j\alpha_{ij}^2} +
\frac{1}{4M_i}\sum_{j<k} \frac{1}{\alpha_{ij}\alpha_{ik}}\right) =
\frac{1}{6M_i} \left(\sum_{j\neq i}
\frac{1}{\alpha_{ij}^2} + \frac{5}{4}\sum_{\stackrel{j<k}{j,k\neq i}}
\frac{1}{\alpha_{ij}\alpha_{ik}}\right),
\nn \\
u_{i;j} = \frac{1}{4} \left(-\frac{3}{M_i\alpha_{ij}^2} +
\frac{2}{M_j\alpha_{ij}^2} - \frac{2}{M_i\alpha_{ij}}\sum_{k\neq
i,j}\frac{1}{\alpha_{ik}}\right),
\nn \\
u_{ijk} = \frac{1}{M_i\alpha_{ij}\alpha_{ik}} +
\frac{1}{M_j\alpha_{ji}\alpha_{jk}} + \frac{1}{M_k\alpha_{ki}\alpha_{kj}}
\label{IM6f}
\ee
and the two expressions for $u_i$ coincide because of the sum rule
\cite{IM7},
$$
\sum_{j\neq i}\frac{1}{\alpha_{ij}^2M_j} = -
\frac{1}{M_i}\sum_{j\neq i}\frac{1}{\alpha_{ij}^2} -
\frac{1}{M_i}\sum_{\stackrel{j<k}{j,k\neq i}}\frac{1}{\alpha_{ij}\alpha_{ik}}
= 0
$$
obtained by summation over all the residues $\alpha_i$ of the integral
$$
\oint_{\alpha_i} \frac{dx}{(x-\alpha_i)^2P_n(x)} =
\oint_0 \frac{dz}{z^3\prod_{j\neq i}(\alpha_{ij}+z)}
= \frac{1}{M_i}\left(\sum_{j\neq i}\frac{1}{\alpha_{ij}^2} +
\sum_{\stackrel{j<k}{j,k\neq i}}\frac{1}{\alpha_{ij}\alpha_{ik}}\right)
$$
The identity between (\ref{F03}) and (\ref{IM6f}) was first
demonstrated by V.Pestun ({\it January 2003, unpublished}).
}

\subsection{Example of $n=2$ (cubic potential $W(z)$,
$3T_3 = \kappa$, $M_1 = -M_2 = \kappa \alpha_{12}$)}

In the case of ${\bf n=2}$, the only non-vanishing $\sigma$-parameter is
$\sigma_1^{(1)} = -\sigma_1^{(2)} = 1/\alpha_{12}$, and
from above formulas we easily reproduce the old results \cite{MMDV}:
\be
\left.{\cal F}_{DV}^{(0|1)}\right|_{t=0} = -S_1W(\alpha_1) - S_2W(\alpha_2)
\ee
\be
\left.{\cal F}_{DV}^{(1|1)}\right|_{t=0} =
\frac{S_1-S_2}{6\kappa\alpha_{12}^3}
\ee
\be
\left.{\cal F}_{DV}^{(2|1)}\right|_{t=0} =
\frac{(S_1-S_2)\lla 3^6\rra^{(2)}}{6!\cdot 3^6 \kappa^3\alpha_{12}^9}=\frac{35(S_1-S_2)}{6\kappa^3\alpha_{12}^9}
\ee
\be
\left.{\cal F}_{DV}^{(0|2)}\right|_{t=0} =
-\frac{1}{2}(S_1^2-4S_1S_2+S_2^2)\log\alpha_{12}
\ee
\be
\left.{\cal F}_{DV}^{(1|2)}\right|_{t=0} =
\frac{(S_1^2+S_2^2)\lla 3^4\rra^{(1)}}{4!\cdot 3^4\cdot
\kappa^2\alpha_{12}^6} - \nn \\ -
\frac{2S_1S_2}{\kappa^2\alpha_{12}^6}
\left(\frac{\lla 4^1\rra^{(1)}}{2} + \frac{2\lla 3^2\rra^{(1)}}{9}
+ \frac{\lla 2, 3^2 \rra^{(1)}}{2!\cdot 3^2}
+ \frac{2\lla 1^1,3^3\rra^{(1)}}{3!\cdot 3^3} +
\left[\lla 1^2 \rra^{(0)}\right]^2\right) = \\ =
\frac{1}{\kappa^2\alpha_{12}^6}
\left(\frac{4536}{8\cdot 3^5}(S_1^2+S_2^2) - \left(
1+{4\cdot 3\over 9}+{2\cdot 18\over 18}+
{4\cdot 162\over 3^4}+2\right)S_1S_2\right)=\\
=\frac{1}{\kappa^2\alpha_{12}^6}
\left(\frac{7}{3}(S_1^2+S_2^2) - \frac{31}{3}S_1S_2\right)
\ee
\be
\left.{\cal F}_{DV}^{(0|3)}\right|_{t=0} =
\frac{6}{3^2}\frac{S_1^3-S_2^3}{\kappa\alpha_{12}^3} -
\frac{5S_1S_2(S_1-S_2)}{\kappa\alpha_{12}^3} =
\frac{2}{3\kappa\alpha_{12}^3}\left(S_1^3-S_2^3 -
\frac{15}{2}S_1S_2(S_1-S_2)\right)
\ee

\newpage

\part{Comments}
\setcounter{section}{0}

\section{Generalized sum rule (\ref{sumrule2}): derivation and applications \label{sumrul}}
\setcounter{equation}{0}

The generalized sum rule (\ref{sumrule2}) can be derived in many different ways
(see, for example, \cite{sumru1} for various derivations).
Here we present yet another possible derivation.
For the sake of simplicity, we put in this section
the coefficient $M$ equal to $1$.
The dependence on $M$ can be easily restored.

From the Toda chain
equation (\ref{Toda}), one can obtain the recursion relation for the
density
\be
\rho_{N+1}(x)+\rho_{N-1}(x)-2\rho_N(x)=\frac{\p_x^2\rho_N(x)}{N}
\label{recre}
\ee
All the densities with $N>1$ can be found directly from this equation, but we
use another approach. That is, we introduce the generation function for
the Gaussian densities,
\be
I_\lambda(x)=\sum_{N=1}^{\infty}\rho_N(x) \lambda^N
\ee
It is easy to see that equation (\ref{recre}) leads to the differential
equation for this generation function. More precisely, one gets
\be
2 \sinh^2(\varphi)\p_\varphi J_{e^{2\varphi}}(x)=\p_x^2
J_{e^{2\varphi}}(x)
\ee
where
\be
J_{e^{2\varphi}}(x)=\sinh^2(\varphi)I_{e^{2\varphi}}(x)
\ee
Making the inverse Laplace transformation for $J$, one gets
\be
2 \sinh^2(\varphi)\p_\varphi \widehat{J}(p)=p^2
\widehat{J}(p)
\ee
where $\widehat{J}(p)$ is the inverse Laplace image
\be
\widehat{J}(x)=\int_0^{\infty}e^{-xp}\widehat{J}(p)dp
\ee
This differential equation can be solved, and one
gets the general solution
\be
\widehat{I}_{e^{2\varphi}}(p)=\frac{\exp(-{\frac{p^2
\coth(\varphi)}{2}})}{\sinh^2(\varphi)}f(p)
\ee
To find the integration constant $f(p)$, one can determine the generation
function for $N=1$ and compare it with the explicitly calculation.
This gives $f(p)=\frac{1}{4}$.
To get the generalized sum rule (\ref{sumrule2}), one should consider the
Laplace image of $I_\lambda$,
\be\label{lapI}
I_\lambda(p)=\frac{\lambda}{(1-\lambda)^2}e^{\frac{p^2}{2}\frac{1+\lambda}{1-\lambda}}
\ee
i.e. the integral
\be
\frac{\lambda}{(1-\lambda)^2}\int_0^\infty e^{-px}
e^{\frac{p^2}{2}\frac{1+\lambda}{1-\lambda}}
\ee
which is divergent. To deal with this integral, one should
treat it as a formal series in $1/x$.
Then, after simple calculations, one gets
\be
I_\lambda(x)=
\frac{1}{\sqrt{2\pi}}\frac{x
\lambda}{(1-\lambda)^2}\int\frac{1}{x^2-t^2\frac{1+\lambda}{1-\lambda}}e^{-\frac{t^2}{2}}dt
=\nn\\
\frac{1}{\sqrt{2\pi}}\sum_{N=1}^\infty x\int\Big(\Big[\frac{x^2+t^2}{x^2-t^2}\Big]^N-1\Big)
e^{-\frac{t^2}{2}}\frac{dt}{2t^2}\lambda^N
\ee
and the coefficients of expansion in $\lambda$ give the
desired sum rule. Unfortunately, we know any generalizations of the
derivation
neither for the Gaussian multidensities, nor for the non-Gaussian
(multi)densities.

Using the sum rule, one can easily restore the genus expansion.
Introducing the variable $\widetilde{x}$ such that
$\widetilde{x}^2=\frac{x^2}{N}$,
one gets the genus expansion as the $1/N^2$-expansion
\be
\rho_N(x)=\frac{x}{\sqrt{2\pi}}\int\Big(\Big[\frac{x^2+t^2}{x^2-t^2}\Big]^N-1\Big)e^{-\frac{t^2}{2}}\frac{dt}{2t^2}=\nn\\
\frac{x}{\sqrt{2\pi}}\int\Big(e^{\frac{2Nt^2}{x^2}}\exp\Big(N
\sum_{k=1}^\infty\Big(\frac{t^2}{x^2}\Big)^{2k+1}\frac{2}{(2k+1)}\Big)
-1\Big)e^{-\frac{t^2}{2}}\frac{dt}{2t^2}\nn\\
=\frac{x}{\sqrt{2\pi}}\int\Big(e^{\frac{2t^2}{\widetilde{x}^2}}\exp(\sum_{k=1}^\infty
\Big(\frac{t^2}{\widetilde{x}^2}-1\Big)^{2k+1}\frac{2}{(2k+1)N^{2k}})
\Big)e^{-\frac{t^2}{2}}\frac{dt}{2t^2}=\nn\\
\frac{x\sqrt{\widetilde{x}^2-4}}{2\widetilde{x}\sqrt{2\pi}}\int\Big(\exp\Big(\sum_{k=1}^\infty
\Big(\frac{t^2}{\widetilde{x}^2-4}\Big)^{2k+1}\frac{2}{(2k+1)N^{2k}}\Big)-e^{-\frac{2t^2}{\widetilde{x}^2-4}}\Big)
e^{-\frac{t^2}{2}}\frac{dt}{t^2}
\ee
Expanding this formula into the $1/N^2$-series gives the genus
expansion, and one can get all the expressions from table (\ref{G1pf}).

One can find the expression for the leading term of the polynomial $Q$
in (\ref{rhop1}). For genus $k$, one has
\be
\rho^{(1|k)}=\frac{N\widetilde{x}\sqrt{\widetilde{x}^2-4}}{\widetilde{x}}\Big(\frac{(4k-1)!!}
{(2k+1)(\widetilde{x}^2-4)^{2k+1}}+
\frac{(4k+1)!!}{(\widetilde{x}^2-4)^{2k+2}}\sum_{i=1}^{k}\frac{1}{(2i+1)(2k-2i+1)}+\cdots\Big)\nn\\
=\frac{N\widetilde{x}}{\widetilde{x}y^{6k-1}}\Big(\frac{(4k-1)!!y^{2k-2}}{(2k+1)}+
y^{2k-4}(4k+1)!!\sum_{i=1}^{k}\frac{1}{(2i+1)(2k-2i+1)}+\cdots\Big)=\nn\\
\frac{N\widetilde{x}}{\widetilde{x}y^{6k-1}}\Big(\frac{(4k-1)!!\widetilde{x}^{2k-2}}{(2k+1)}+\nn\\
\widetilde{x}^{2k-4}\Big[-\frac{4(k-1)(4k-1)!!}{(2k+1)}+(4k+1)!!\sum_{i=1}^{k}\frac{1}{(2i+1)(2k-2i+1)}\Big]
+O(\widetilde{x}^{2k-6})\Big)
\ee
where $y^2=\widetilde{x}^2-4$.

\section{ Evaluation of multidensities in Gaussian model \label{mdsec}}
\setcounter{equation}{0}

In principle, all the multidensities in the Gaussian model can be derived
step-by-step from the Virasoro constraints, as are the first few in
s.\ref{Gmpf}. But it is desirable to get expressions like
(\ref{sumrule2}) which unify all genera. Here we consider only the two-point
function.

$\bullet$
There is a sum rule for the two-point function for $N=1$
\be
\oint\oint\rho^{(2)}(z_1,z_2)z_1^{k_1}z_2^{k_2}dz_1dz_2=\\
\left\{
\begin{array}{l}
 (2k_1+2k_2-1)!!\quad k_1,k_2\, odd,\\
 (2k_1+2k_2-1)!!-(2k_1-1)!!-(2k_2-1)!!\quad k_1,k_2\, even,\\
 0\quad else
\end{array}
\right.
\ee

$\bullet$
The Virasoro constraints for the Gaussian potential can be combined as follows
\be
(x\widehat{\nabla}_x -\p_0)Z_N=\widehat{\nabla}_x^2 Z_N
\ee
Thus, for the disconnected two-point resolvent, one gets
\be
\rho^{(2)}(x,x)+\rho^{(1)}(x)^2=\frac{x^2}{\sqrt{2\pi}}\int\Big(\Big[\frac{x^2+t^2}{x^2-t^2}\Big]^N
-1-2N\frac{t^2}{x^2}\Big)e^{-\frac{t^2}{2}}\frac{dt}{2t^2}
\ee

$\bullet$
From the Toda chain equation (\ref{Toda}), it is easy to derive the following
equation for the two-point resolvent
\be
\rho^{(2)}_{N+1}(x,y)+\rho^{(2)}_{N-1}(x,y)-2\rho^{(2)}_N(x,y)=\frac{(\p_x+\p_y)^2\rho^{(2)}_N(x,y)}{N}-
\frac{\p_x^2 \rho^{(1)}_N(x) \p_y^2\rho^{(1)}_N(y)}{N^2}
\ee
This equation is consistent with known low genus densities, that is,
they satisfy the equation in the leading order in $1/N$.

$\bullet$
Using the harmonic oscillator calculation as in \cite{sumru1}, one gets
\be
\eta_N^{(2)}(y,z)=\eta_N^{(1)}(y+z)-G_N(y,z)
\ee
where
\be
G_N(y,z)=\exp{\big(\frac{z^2+y^2}{2}\big)}\Big[\sum_{n=0}^{N-1} B(n,n;y)B(n,n;z)+
2\sum_{n=0}^{N-1}\sum_{s=n+1}^{N-1}  B(n,s;y)B(n,s;z)
\Big]
\label {GG}
\ee
Here $B(n,s;x)$ is some (Hypergeometric) function, a proper matrix element for
the harmonic oscillator
\be
B(n,s;x)=\sqrt{n!s!}\sum_{i=0}^n\frac{x^{n+s-2i}}{i!(n-i)!(s-i)!}
\ee
Using the Laplace transformation, one can get the two-point resolvent
\be
\rho^{(2)}(s,t)=\int_{0}^{\infty}\int_{0}^{\infty}e^{-(ys+zt)}\eta^{(2)}(y,z)dydz
\ee

$\bullet$
Using the integral representation of the Hermite polynomials $H_n$
\be
H_n(x)=\frac{(-1)^n n!}{2\pi i}\oint\frac{d z}{z^{n+1}}e^{-z^2/2-zx}
\ee
one can represent the function $G_N(y,z)$ as the four-fold contour integral
\be
G_N(y,z)=\exp{\Big(\frac{y^2+z^2}{2}\Big)}(\oint dz)^4
\sum_{s,n=0}^{N-1}\frac{s!n!}{(z_1z_3)^{n+1}(z_2z_4)^{s+1}}e^{z_1z_2+z_3z_4-(z_1+z_2)y-(z_3+z_4)z}
\ee
For the generation function, one gets the expression
\be
K(\lambda)=\exp{\Big(-\frac{y^2+z^2}{2}\Big)}\sum_{N=1}^{\infty}G_N \lambda^N=
(\oint dz)^4\int_{0}^\infty\int_0^\infty dp_1
dp_2e^{-p_1-p_2}\frac{1}{p_1-z_1z_3}\frac{1}{p_2-z_2z_4}\nn\\
\Big[\frac{1}{1-\frac{\lambda p_1p_2}{z_1z_2z_3z_4}}-\frac{1}{1-\frac{\lambda p_2}{z_2z_4}}-\frac{1}{1-\frac{\lambda p_1}{z_1z_3}}+
\frac{1}{1-\lambda}\Big]e^{z_1z_2+z_3z_4-(z_1+z_2)y-(z_3+z_4)z}
\ee

The full generation function for the two-point correlation function is
\be
\sum_{N=1}^{\infty}\eta_N^{(2)}\lambda^N=
\frac{\lambda}{(1-\lambda)^2}e^{\frac{(y+z)^2}{2}\frac{1+\lambda}{1-\lambda}}-e^{\frac{y^2+z^2}{2}}K(\lambda)=\nn\\
\frac{\lambda}{(1-\lambda)^2}e^{(y^2+z^2)\frac{1+\lambda}{2(1-\lambda)}}
\Big(
\sum_{k=0}^{\infty}\frac{1}{k!}(\frac{zy}{1-\lambda})^k\sum_{m=0}^{k}(\frac{\lambda yz}{1-\lambda})^m\frac{1}{m!}
-\sum_{k=0}^{\infty}\frac{1}{k!}(\frac{\lambda zy}{1-\lambda})^k\sum_{m=0}^{k}(\frac{
yz}{1-\lambda})^m\frac{1}{m!}
\Big)
\ee
It is easy to check that this formula gives an answer consistent with
(\ref{GG}) for $N$ from 1 to 3.

This formula is much more involved than the corresponding expression
(\ref{lapI}) for
the one-point densities, and this is not immediate to get the genus
expansion for the two-point function from it.

\section{Continuous limits}
\setcounter{equation}{0}

There is an infinite variety of continuum limits, and only
very few have been partly investigated.
Among these are:
\begin{itemize}
\item
Naive large-$N$ (planar or t'Hooft') limit \cite{t'Hooft,largeN}
\item
Double-scaling limit \cite{ds,vircref,MMMM}
\item
BMN limit \cite{BMN}
\item
DV limit \cite{DV,SWthDV1,DVfollowup}
\item
...
\end{itemize}
In all these limits the FSHMM converts into some other matrix model.
Thus it is natural first to study these other models by themselves
and only after that perform a detailed and justified analysis of the
variety of continuum limits. This is a subject of separate publications.

Also, particular representatives of these other matrix model families
are identically equivalent to FSHMM, what allows to analyze the limiting
procedures inside these families: various continuum limits are associated
with different generalizations of the FSHMM.
One of such examples is provided by the family of Generalized Kontsevich
models \cite{GKM,UFN3}, in which the {\it Gaussian Kontsevich model}
coincides with the FSHMM.

\section{Acknowledgements}

We thank Vasily Pestun for collaboration
at early stages of this project,
Yuri Makeenko for important comments about the status
of the problem and Nataly Amburg, Leonid Chekhov and Anton Khoroshkin
for useful discussions.

Our work is partly supported by Federal Program of the Russian Ministry of
Industry, Science and Technology No 40.052.1.1.1112 and by the grants:
INTAS 00-334, RFBR 03-02-17373
and the RFBR grant for support of young scientists (A.A.),
Volkswagen Stiftung (A.Mir. and A.Mor.),
INTAS 00-561, RFBR 01-02-17682a, Grant of Support for the Scientific
Schools 1936.2003.02 (A.Mir.),
INTAS 00-561, RFBR 01-02-17488 and Grant of Support for the
Scientific Schools (A.Mor.).


\begin{thebibliography}{12}

\bibitem{ST} A.Polyakov, {\sl Gauge Fields and Strings},
Chur, Switzerland: Harwood (1987) 301 p. (Contemporary Concepts in Physics,
3)\\
M.Green, J.Shwarz and E.Witten, {\sl Superstring Theory}, v.1,2,
Cambridge, Uk: Univ. Press (1987)
(Cambridge Monographs On Mathematical Physics) \\
A.Morozov, Sov.Phys.Usp. {\bf 35} (1992) 671-714
(Usp.Fiz.Nauk {\bf 162} (1992) 83-176\\
J.Polchinski, {\sl String Theory}, v.1,2, Cambridge, UK: Univ. Press (1998)\\
A.Marshakov, Phys.Usp. {\bf 45} (2002) 915-954 (Usp.Fiz.Nauk
{\bf 45} (2002) 977-1020)

\bibitem{GKLMM} A.Gerasimov, S.Khoroshkin, D.Lebedev, A.Mironov and
A.Morozov,
Int.J.Mod.Phys. A10 (1995) 2589-2614, hep-th/9405011\\
A.Mironov, A.Morozov and L.Vinet, Theor.Math.Phys. {\bf 100} (1995) 890-899
(Teor.Mat.Fiz. {\bf 100} (1994) 119-131)\\
S.Kharchev, A.Mironov and A.Morozov, Theor.Math.Phys.
{\bf 104} (1995) 129-143\\
A.Mironov, Theor.Math.Phys. {\bf 114} (1998) 127\\
A.Mironov and A.Morozov, Phys.Lett. {\bf B524} (2002) 217-226

\bibitem{HMM} Mehta M.L., {\sl Random matrices},
(2nd ed., Academic Press, New York, 1991)\\
Br\'ezin E., Itzykson C., Parisi G., and Zuber J.B.,
Commun.Math.Phys. {\bf 59} (1978) 35\\
Bessis D.,
Commun.Math.Phys. {\bf 69} (1979) 147\\
Bessis D., Itzykson C., and Zuber J.B.,
Adv.Appl.Math. {\bf 1} (1980) 109\\
Itzykson C. and Zuber J.-B.,
J.~Math.Phys. {\bf 21} (1980) 411

\bibitem{GMMMO} A.Gerasimov, A.Marshakov, A.Mironov, A.Morozov and
A.Orlov, Nucl.Phys. {\bf B357} (1991) 565-618

\bibitem{detrep}
S.Kharchev, A.Marshakov, A.Mironov, A.Orlov and A.Zabrodin, Nucl.Phys.
{\bf B366} (1991) 569-601

\bibitem{UFN3} A.Morozov, Phys.Usp. {\bf 37} (1994) 1-55;
hep-th/9502091

\bibitem{revs} A.Mironov, Int.J.Mod.Phys. {\bf A9} (1994) 4355-4405;
Phys.Part.Nucl. {\bf 33} (2002) 537-582 (Fiz.Elem.Chast.Atom.Yadra
{\bf 33} (2002) 1051-1145\\
I.Kostov, hep-th/9907060

\bibitem{KMMMP} A.Marshakov, A.Mironov and A.Morozov,
Phys.Lett. {\bf B265} (1991) 99-107\\
S.Kharchev, A.Marshakov, A.Mironov, A.Morozov and S.Pakuliak,
Nucl.Phys. {\bf B404} (1993) 717-750\\
A.Mironov, S.Pakuliak,
Theor.Math.Phys. {\bf 95} (1993) 604-625 (Teor.Mat.Fiz. {\bf 95} (1993)
317-340)

\bibitem{Kos} I.Kostov, Phys.Lett. {\bf B297} (1992) 74-81

\bibitem{kost} J.Alfaro and I.Kostov, hep-th/9604011

\bibitem{zab}
M.Mineev-Weinstein, P.B.Wiegmann and A.Zabrodin,
Phys.Rev.Lett. {\bf 84} (2000) 5106-5109\\
I.Kostov, I.Krichever,
M.Mineev-Weinstein, P.B.Wiegmann and A.Zabrodin, hep-th/0005259\\
A. Boyarsky, A. Marshakov, O. Ruchayskiy, P. Wiegmann and A. Zabrodin,
Phys.Lett. {\bf B515} (2001) 483-492\\
I.Krichever, A.Marshakov and A.Zabrodin, hep-th/0309010

\bibitem{K} Kontsevich M.L.,
Funk.Anal.Prilozh. {\bf 25} (1991) v.~2, p.~50 (in Russian)

\bibitem{WK} E.Witten, in: New York 1991 Proc., Differential geometric
methods in theoretical physics, v.1, pp.176-216\\
A.Marshakov, A.Mironov and A.Morozov, Phys.Lett. {\bf B274} (1992) 280

\bibitem{GKM} S.Kharchev, A.Marshakov, A.Mironov, A.Morozov and A.Zabrodin,
Nucl.Phys. {\bf B380} (1992) 181-240; Phys.Lett. {\bf B275} (1992) 311-314

\bibitem{versus} S.Kharchev, A.Marshakov, A.Mironov and A.Morozov, Nucl.Phys.
{\bf B397} (1993) 339-378

\bibitem{LGM}
S.Kharchev, A.Marshakov, A.Mironov and A.Morozov,
Mod.Phys.Lett. {\bf A8} (1993) 1047-1061

\bibitem{GKMfollowup} L.Chekhov and Yu.Makeenko, Phys.Lett.
{\bf B278} (1992) 271-278\\
L.Chekhov, hep-th/9509001\\
S.Kharchev, hep-th/9810091

\bibitem{Wig} Wigner E.P.,
Ann.Math. {\bf 53} (1951) 36

\bibitem{gMM} Dyson F.J.,
J.Math.Phys. {\bf 3} (1962) 140\\
Gross D. and Witten E.,
Phys.Rev. {\bf D21} (1980) 446\\
Eguchi T. and Kawai H.,
Phys.Rev.Lett. {\bf 48} (1982) 1063\\
Voiculescu D.V., Dykema K.J., and Nica A.,
{\sl Free random variables}  (AMS, Providence, 1992)\\
Di Francesco P., Ginsparg P., and Zinn-Justin J.,
Phys.Rep. {\bf 254} (1995) 1

\bibitem{MMs} David F.,
Nucl.Phys. {\bf B257[FS14]} (1985) 45\\
Kazakov V.A., Kostov I.K., and Migdal A.A.,
Phys.~Lett. {\bf B157} (1985) 295

\bibitem{Giv} A.Givental, math.AG/0008067

\bibitem{Givfollowup}
J.S.Song and Y.S.Song, hep-th/0103254\\
A.Alexandrov, hep-th/0205261, to be published in J.Math.Phys.

\bibitem{MMDV}
F.~David,
Phys.Lett. B302 (1993) 403-410, hep-th/9212106;\\
G.~Bonnet, F.~David, B.~Eynard,
J.Phys. {\bf A33} (2000) 6739-6768, cond-mat/0003324\\
A.Klemm, M.Mari\~no and S.Theisen, JHEP {\bf 0303} (2003) 051

\bibitem{GR} I.S.Gradshteyn and I.M.Ryzhik, Table of Integrals,
Series, and Products, Academic Press, Fifth Edition

\bibitem{DV} R.~ Dijkgraaf and C.~ Vafa,
Nucl.Phys. {\bf B644} (2002) 3-20; Nucl.Phys. {\bf B644} (2002) 21-39;
hep-th/0208048

\bibitem{loopeqn} Migdal A.A.,
Phys.Rep. {\bf 102} (1983) 199\\
Ambj{\o}rn J., Jurkiewicz J., and Makeenko Yu.,
Phys.Lett. {\bf B251} (1990) 517

\bibitem{Wbr} E.Witten, Nucl.Phys. {\bf B460} (1996) 335-350

\bibitem{BV} I.A.Batalin and G.A.Vilkovisky, Phys.Lett. {\bf B102} (1981)
27-31; Phys.Lett. {\bf B120} (1983) 166-170; Phys.Rev. {\bf D28} (1983)
2567-2582; Nucl.Phys. {\bf B234} (1984) 106-124; J.Math.Phys. {\bf 26}
(1985) 172-184

\bibitem{BVfollowup} See for a review:\\
M.Henneaux and C.Teitelboim, {\sl Quantization of Gauge Systems}, Princeton
University Press, 1992\\
J.Gomis, J.Par\'is and S.Samuel, Phys.Rept. {\bf 259} (1995) 1-145

\bibitem{BFV} E.S.Fradkin and G.A.Vilkovisky, Phys.Lett. {\bf B55} (1975)
224\\
I.A.Batalin and G.A.Vilkovisky, Phys.Lett. {\bf B69} (1977) 309-312

\bibitem{BFVfollowup} See for a review:\\
M.Henneaux, Phys.Rept. {\bf 126} (1985) 1

\bibitem{WBV} E.Witten, Mod.Phys.Lett. {\bf A5} (1990) 487

\bibitem{WBVfollowup} A.S.Schwarz, Comm.Math.Phys. {\bf 155} (1993) 249-260

\bibitem{MMRG}
J.Polchinski, Nucl.Phys. {\bf B231} (1984) 269\\
A.Mironov and A.Morozov, Phys.Lett. {\bf B490} (2000) 173-179

\bibitem{SW} N.Seiberg and E.Witten, Nucl.Phys. {\bf B426} (1994) 19

\bibitem{MM}
David F.,
Mod.Phys.Lett. {\bf A5} (1990) 1019\\
A.Mironov and A.Morozov, Phys.Lett. {\bf B252} (1990) 47-52\\
Ambj{\o}rn J. and Makeenko Yu.,
Mod.Phys.Lett. {\bf A5} (1990) 1753\\
H.Itoyama and Y.Matsuo,
Phys.Lett. {\bf 255B} (1991) 202

\bibitem{vircref} Fukuma M., Kawai H., and Nakayama R.,
Int.J.Mod.Phys. {\bf A6} (1991) 1385\\
Dijkgraaf R., Verlinde H., and Verlinde E.,
Nucl.Phys. {\bf B348} (1991) 435

\bibitem{MMMM} Makeenko Yu., Marshakov A., Mironov A., and Morozov A.,
Nucl.Phys. {\bf B356} (1991) 574

\bibitem{IM7} H.Itoyama and A.Morozov, hep-th/0301136

\bibitem{sumru} J.Harer and D.Zagier, Invent.Math. {\bf 85} (1986) 457-485\\
S.K.Lando and A.K.Zvonkin, {\sl Embedded graphs}, Max-Plank-Institut f\"r
Mathematik, Preprint Series 2001 ({\bf 63})

\bibitem{sumru1} C.Itzykson and J.-B.Zuber, Comm.Math.Phys.
{\bf 134} (1990) 197-208

\bibitem{WDVV} E.~Witten, Nucl. Phys. {\bf B340} (1990) 281\\
R.~Dijkgraaf, H.~Verlinde and E.~Verlinde,
Nucl. Phys. {\bf B352} (1991) 59\\
B. Dubrovin, Lecture Notes in Math. {\bf 1620},
Springer, Berlin, 1996, 120-348

\bibitem{WDVVMMM} A.~Marshakov, A.~Mironov and A.~Morozov,
Phys. Lett. {\bf B389} (1996) 43;
Mod.Phys.Lett. {\bf A12} (1997) 773-787;
Int.J.Mod.Phys. {\bf A15} (2000) 1157-1206

\bibitem{WDVVf} A.Losev, JETP Lett. {\bf 65} (1997) 374\\
K.Ito and S.-K.Yang, Phys.Lett. {\bf B433} (1998) 56-62\\
G.Bertoldi and M.Matone, Phys.Rev. {\bf D57}
(1998) 6483-6485\\
A.Morozov, Phys.Lett. {\bf B427} (1998) 93-96\\
A.Mironov and A.Morozov, Phys.Lett., {\bf B424} (1998) 48-52\\
H.W. Braden, A.Marshakov, A.Mironov and A.Morozov, Phys.Lett.
 {\bf B448} (1999) 195\\
A.Veselov, Phys.Lett. {\bf A261} (1999) 297-302\\
J.M.Isidro, Nucl.Phys., {\bf B539} (1999) 379-402\\
See also reviews in:\\
A. Mironov, hep-th/9903088;\\
A.~Marshakov, Theor.Math.Phys. {\bf 132} (2002) 895
(Teor.Mat.Fiz. {\bf 132} (2002) 3)

\bibitem{CIV} F. Cachazo, K. Intriligator and C. Vafa, Nucl.Phys.
{\bf B603} (2001) 3-41\\
F. Cachazo and C. Vafa, hep-th/020601

\bibitem{DVfollowup1}
N.~Dorey, T.~J.~Hollowood, S.~Prem Kumar and A.~Sinkovics,
JHEP {\bf 0211} (2002) 039; {\it ibid.}, 040; {\it ibid.},
{\bf 0212} (2002) 003\\
F. Ferrara, Nucl.Phys. {\bf B648} (2003) 161-173; Phys.Rev.
{\bf D67} (2003) 085013\\
D. Berenstein, Phys.Lett. {\bf B552} (2003) 255-264\\
R. Dijkgraaf, S. Gukov, V. Kazakov and C. Vafa, Phys.Rev.
{\bf D68} (2003) 045007\\
A. Gorsky, Phys.Lett. {\bf B554} (2003) 185-189\\
R. Dijkgraaf, M.T. Grisaru, C.S. Lam, C. Vafa and D. Zanon,
hep-th/0211017;\\
Bo Feng, hep-th/0211202; Nucl.Phys. {\bf B661} (2003) 113-138\\
F.~Cachazo, M.~R.~Douglas, N.~Seiberg and E.~Witten, JHEP {\bf 0212}
(2002) 071\\
F.~Cachazo, N.~Seiberg and E.~Witten, JHEP {\bf 0302} (2003) 042;
JHEP {\bf 0304} (2003) 018\\
A. Dymarsky and V. Pestun, Phys.Rev. {\bf D67} (2003) 125001\\
R.Boels, Jan de Boer, R.Duivenvoorden, J.Wijnhout, hep-th/0305189

\bibitem{DVfollowup}
G. Bonelli, hep-th/0209225\\
H. Fuji and Y. Ookouchi, hep-th/0210148\\
R. Argurio, V.L. Campos, G. Ferretti and R. Heise, hep-th/0210291,
hep-th/0211249\\
J. McGreevy, hep-th/0211009\\
H. Suzuki, hep-th/0211052, hep-th/0212121\\
I. Bena and R. Roiban, hep-th/0211075\\
Y.~Demasure, R.~A.~Janik, hep-th/0211082\\
R. Gopakumar, hep-th/0211100\\
I.~Bena, R.~Roiban and R.~Tatar, hep-th/0211271\\
Y. Tachikawa, hep-th/0211274, hep-th/0211189\\
Y. Ookouchi, hep-th/0211287\\
S. Ashok, R. Corrado, N. Halmagyi, K. Kennaway and C. Romelsberger,
hep-th/0211291\\
K.~Ohta, hep-th/0212025\\
R.A. Janik and N.A. Obers, hep-th/0212069\\
S. Seki, hep-th/0212079\\
C. Hofman, hep-th/0212095\\
C. Ahn, S. Nam, hep-th/0212231\\
C. Ahn, hep-th/0301011\\
S.Aoyama and T.Masuda, hep-th/0309232

\bibitem{SWth}
A.Gorsky, I.Krichever, A.Marshakov, A.Mironov and A.Morozov,
Phys.Lett. {\bf B355} (1995) 466-477\\
E.Martinec and N.Warner,
Nucl.Phys. {\bf 459} (1996) 97\\
R.Donagi and E.Witten,
Nucl.Phys. {\bf B460} (1996) 299-334\\
A.Gorsky, A.Mironov, A.Marshakov and A.Morozov,
Nucl.Phys. {\bf B527} (1998) 690-716\\
H.Itoyama and A.Morozov,
Nucl.Phys., {\bf B477} (1996) 855-877;
Nucl.Phys., {\bf B491} (1997) 529-573; hep-th/9601168\\
E. D'Hoker and D. H. Phong, hep-th/9912271\\
A.Marshakov, {\sl Seiberg-Witten Theory and Integrable Systems},
Singapore, Singapore: World Scientific (1999) 253 p.\\
H.W.Braden and I.M.Krichever (Eds.), {\sl Integrability:
The Seiberg-Witten and Whitham Equations}, Gordon and Beach, 2000\\
A.Gorsky and A.Mironov, hep-th/0011197

\bibitem{SWthDV1} L.Chekhov and A.Mironov, Phys.Lett. {\bf B552}
(2003) 293-302\\
V.Kazakov and A.Marshakov, J.Phys. {\bf A36} (2003) 3107-3136

\bibitem{IM4-5}
H. Itoyama and A. Morozov, Nucl.Phys. {\bf B657} (2003) 53-78;
Phys.Lett. {\bf B555} (2003) 287-295;

\bibitem{IM6}
H. Itoyama and A. Morozov, Prog.Theor.Phys. {\bf 109} (2003) 433-463

\bibitem{SWthDV} L.Chekhov, A.Marshakov, A.Mironov and D.Vasiliev,
Phys.Lett. {\bf B562} (2003) 323-338\\
A.Mironov, Fortsch.Phys. {\bf 51} (2003) 781-786

\bibitem{moments} Ambj{\o}rn J., Chekhov L., Kristjansen C.F.,
and Makeenko Yu.,
Nucl.Phys. {\bf B404} (1993) 127\\
Ambj{\o}rn J., Chekhov L., and Makeenko Yu.,
Phys.Lett. {\bf B282} (1992) 341\\
G.Akemann, Nucl.Phys. {\bf B482} (1996) 403-430

\bibitem{t'Hooft} 't~Hooft G.,
Nucl.Phys. {\bf B72} (1974) 461

\bibitem{largeN} Veneziano G.,
Nucl.Phys. {\bf B117} (1976) 519\\
D. De Wit and G.~'t~Hooft, Phys.Lett. {\bf 69} (1977) 61\\
Witten E.,
eds.\ G.~'t~Hooft {\sl et al.} (Plenum, New York, 1980) p.~403\\
Wadia S.R.,
Phys.Rev. {\bf D24} (1981) 970\\
A.Mironov, A.Morozov and G.Semenoff, Int.J.Mod.Phys. {\bf A11} (1996)
5031-5080\\
B.Eynard, JHEP {\bf 0301} (2003) 051; hep-th/0309036

\bibitem{ds} Br\'ezin E. and Kazakov V.A.,
Phys.Lett. {\bf B236} (1990) 144\\
Gross  D. and Migdal A.A.,
Phys.Rev.Lett. {\bf 64} (1990) 127;
Nucl.Phys. {\bf B340} (1990) 333\\
Douglas M. and Shenker S.,
Nucl.Phys. {\bf B335} (1990) 635

\bibitem{BMN} D.Berenstein, J.Maldacena and H.Nastase,
JHEP {\bf 0204} (2002) 013\\
N.R.Constable, D.Z.Freedman, M.Headrick and S.Minwalla, JHEP {\bf 0210}
(2002) 068\\
D.J.Gross, A.Mikhailov and R.Roiban, Annals Phys. {\bf 301} (2002) 31-52;
JHEP {\bf 0305} (2003) 025\\
N.Beisert, C.Kristjansen, J.Plefka, G.W.Semenoff and M.Staudacher,
Nucl.Phys. {\bf B643} (2002) 3-30; {\it ibid.}, {\bf B650} (2003) 125-161

\end{thebibliography}
\end{document}